\documentclass[%
 reprint,
superscriptaddress,
nofootinbib,
amsmath,amssymb,
aps,
prd,
]{revtex4-2}

\usepackage{graphicx}
\usepackage{dcolumn}
\usepackage{bm}
\usepackage{amsmath}
\usepackage{lipsum}
\usepackage{tabularx}
\usepackage[table]{xcolor}
\usepackage{enumitem}

\definecolor{DarkGreen}{RGB}{40,150,60}
\definecolor{NicoBlue}{RGB}{45, 104, 196}
\definecolor{darkpink}{RGB}{200,55,113}
\usepackage[
  colorlinks=true,
  linkcolor=darkpink,
  citecolor=green,
  urlcolor=darkpink
]{hyperref}

\begin{document}

\title{Highly-accurate neutron star modeling in the Hartle-Thorne Approximation}

\author{Carlos Conde-Ocazionez}
\email{\textcolor{darkpink}{carlosc7@illinois.edu}}
\affiliation{The Grainger College of Engineering, Illinois Center for Advanced Studies of the Universe, Department of Physics, University of Illinois at Urbana-Champaign, Urbana, IL 61801, USA}

\author{Tuojin Yin}
\email{tuojinyi@usc.edu}
\affiliation{Department of Physics, University of Southern California, Los Angeles, CA 90089, USA}

\author{Jaquelyn Noronha-Hostler\vspace{0.1cm}}
\email{jnorhos@illinois.edu}
\affiliation{The Grainger College of Engineering, Illinois Center for Advanced Studies of the Universe, Department of Physics, University of Illinois at Urbana-Champaign, Urbana, IL 61801, USA}

\author{Nicolás Yunes\vspace{0.1cm}}
\email{nyunes@illinois.edu}
\affiliation{The Grainger College of Engineering, Illinois Center for Advanced Studies of the Universe, Department of Physics, University of Illinois at Urbana-Champaign, Urbana, IL 61801, USA}

\begin{abstract}
Future X-ray missions, such as NICER and LOFT, together with gravitational-wave observations from ground-based detectors, will provide new insights into neutron stars. Interpreting accurate observations in the future will require accurate models of their gravitational fields.
In this first paper of a two-part series, we construct the perturbation equations for slowly-rotating, isolated, and unmagnetized neutron stars, extending the Hartle-Thorne approximation to seventh order in a slow-rotation expansion. We obtain exact, closed-form, analytical solutions for the exterior metric at each order in spin. From these solutions, we derive expressions for the mass and mass-current scalar multipole moments, $M_{\ell}$ and $S_{\ell}$, respectively, up to seventh order in spin frequency, using two distinct methods.
This high-order expansion allows us to calculate second-, fourth-, and sixth-order relative spin corrections to the observed mass and moment of inertia; second- and fourth-order relative spin corrections to the quadrupole and octopole moments; second-order relative spin corrections to the hexadecapole and dotriacontapole moments; and leading-order-in-spin expressions for the hexacontatetrapole and hectoicosaoctapole moments. 
Going to seventh order in the spin-frequency approximation will enable very precise calculations of X-ray pulse profiles, as well as the I-Love-Q and three-hair relations for slowly-rotating neutron stars. These results will be valuable for breaking parameter degeneracies in future multimessenger observations.
\end{abstract}
                            
\maketitle

\section{\label{sec:level1}Introduction}

\textit{The Neutron Star Interior Composition Explorer} (NICER) and the \textit{Large Observatory for X-ray Timing} (LOFT) are dedicated to studying the properties of neutron stars with unprecedented precision \cite{gendreau2016neutron, Mignani:2012vc}. A key scientific objective of these missions is to accurately estimate neutron star masses and radii, which requires detailed and realistic modeling of rotating neutron stars that accounts for the effects of strong gravity, high densities, and rapid rotation. Gravitational-wave (GW) detectors, such as advanced LIGO (aLIGO)~\cite{LIGOScientific:2007fwp, LIGOScientific:2014pky}, Virgo~\cite{VIRGO:2012dcp} and KAGRA~\cite{Aso:2013eba}, have also begun to detect the GWs emitted when neutron stars inspiral and merge~\cite{LIGOScientific:2016aoc}. A key goal of these detectors is to infer the mass and radius of these merging neutron stars to obtain properties of the equation of state of dense matter, which can be aided through certain quasi-universal relations between the multipole moments of rotating stars.  

Neutron stars are among the most compact objects in the universe, with gravitational fields so strong that their structure and dynamics must be described within the framework of general relativity (GR). For non-rotating, spherically-symmetric stars, the spacetime inside the star is determined by solving the Tolman–Oppenheimer–Volkoff (TOV) equations \cite{Tolman:1939jz,Oppenheimer:1939ne}, while the exterior gravitational field is described by the Schwarzschild solution, as guaranteed by Birkhoff's theorem \cite{birkhoff1927relativity}. The exterior gravitational field, i.e.~the metric tensor exterior to the radius of the star, is then fully prescribed by just the mass of the neutron star.

When a neutron star is rotating, the modeling of its structure and gravitational field becomes significantly more complex. Rotation introduces additional relativistic effects, such as frame dragging and centrifugal deformation, which cause the star to lose its spherical symmetry \cite{Hartle:1967he}. The resulting spacetime is stationary and axisymmetric, and the gravitational field depends not only on the mass distribution but also on the star’s rotation rate. In contrast to the static case, where Birkhoff’s theorem guarantees that the exterior solution is uniquely given by the Schwarzschild metric, no such general result exists for rotating stars. Although the Kerr solution describes the spacetime around a rotating black hole, it does not apply to rotating neutron stars, which possess matter and a physical surface \cite{Wiltshire:2009zza}. Therefore, the full spacetime—both interior and exterior—must be computed as a solution to the Einstein field equations for a rotating, self-gravitating fluid. This is typically done using one of two approaches: perturbative methods, which expand the equations in powers of the star’s angular velocity $\Omega$ \cite{Hartle:1967he, Hartle:1968si, Hartle:1973zza, Benhar:2005gi, Yagi:2014bxa}, or full numerical-relativity simulations, which solve the equations on a grid without approximations \cite{Stergioulas:2003yp, Friedman:2013xza, Bonazzola:1993zz,Bonazzola:1998qx, Komatsu:1989zz, Cook:1993qj}. 

To model the spacetime around a rotating neutron star, one widely used method is the Hartle-Thorne approximation, developed in the late 1960s by James Hartle and Kip Thorne \cite{Hartle:1967he, Hartle:1968si}. This approach uses a perturbative expansion in the star’s angular velocity $\Omega$ to compute corrections to both the fluid distribution and the spacetime geometry. The central assumption is that the star rotates slowly, so that deviations from spherical symmetry can be treated as small perturbations. Despite this simplifying assumption, the method captures essential relativistic effects, such as frame dragging and rotational flattening of the poles, and it provides analytic control over the structure of the spacetime up to a chosen order in the small spin expansion.

In contrast, numerical relativity offers a way to model rapidly-rotating neutron stars by directly solving the full, nonlinear Einstein field equations. Several numerical schemes exist, and a detailed comparison between them can be found in \cite{Stergioulas:2003yp, Friedman:2013xza}. Two of the most widely used modern codes are LORENE/rotstar \cite{Bonazzola:1993zz,Bonazzola:1998qx} and RNS \cite{Stergioulas:1994ea}. LORENE is based on a 3+1 decomposition of spacetime using spectral methods, while RNS is based on the method developed by Komatsu, Eriguchi, and Hachisu (KEH) \cite{Komatsu:1989zz}, with modifications introduced by Cook, Shapiro, and Teukolsky \cite{Cook:1993qj}. The KEH method relies on an integral representation of the independent components of the field equations using Green's functions. These codes allow for the modeling of neutron stars without the slow-rotation approximation and can incorporate realistic features such as magnetic fields, differential rotation, and various equations of state. 
However, the use of numerical relativity generality comes at a cost: the simulations are computationally intensive and require highly-accurate grids when extracting global spacetime properties in certain scenarios, such as high-order multipole moments for very slowly-rotating stars~\cite{Yagi:2014bxa}. 

One of the main goals in modeling rotating neutron stars is to describe their multipolar structure, as the spacetime multipole moments are closely tied to observable properties, such as pulse profiles \cite{Psaltis:2013zja}, quasi-periodic oscillations \cite{Pappas:2012nt}, and gravitational-wave signatures \cite{Ryan:1997hg}. In particular, the mass quadrupole $M_2$ and higher-order multipole moments encode information about the star’s internal structure and rotational state. In 2014, Yagi et al.~\cite{Yagi:2014bxa} computed the multipole moments of neutron stars up to the mass hexadecapole moment $M_4$, using both slow-rotation perturbative methods and full numerical simulations with the LORENE and RNS codes. However, extracting higher multipole moments directly from numerical-relativity simulations proved extremely challenging. The multipole moments are defined asymptotically at spatial infinity, and their accurate extraction requires extremely high-resolution grids to overcome numerical noise. This sensitivity makes it difficult to obtain reliable results beyond the hexadecapole moment (e.g.~$S_5$, $M_6$ and $S_7$), especially for stars that do not rotate rapidly.

Given these limitations, the Hartle-Thorne perturbative approach is particularly well-suited for studying the multipolar structure of neutron star spacetimes. This approach provides a systematic framework for computing higher-order moments, especially in the slow-rotation regime, where it remains accurate and computationally efficient. A further advantage lies in the mathematical structure of the method: the angular dependence of the spacetime is expressed analytically in terms of Legendre polynomials, while the radial dependence is determined by solving numerically a set of linear ordinary differential equations at each perturbative order. Notably, comparisons with numerical relativity simulations in full GR have demonstrated that the Hartle-Thorne approximation remains remarkably accurate for modeling even the fastest observed pulsars \cite{Berti:2004ny, Yagi:2014bxa}. For instance, the mass quadrupole moment $M_2$ exhibits, at most, a 20$\%$ relative fractional error for the fastest pulsars when computed to leading order in the slow-rotation expansion \cite{Berti:2004ny}. However, the slow-rotation approximation begins to break down as the star’s rotational frequency approaches the mass-shedding frequency limit, where centrifugal forces become strong enough to unbind matter from the stellar surface. In such extreme regimes, the first few orders of the perturbative expansion cannot fully reproduce the dynamics \cite{Benhar:2005gi}, whereas full numerical relativity can capture them in detail, as demonstrated in \cite{Cook:1993qr}.

The Hartle-Thorne approximation has a long history of successful applications in modeling neutron star observables. One of the earliest results came from Hartle and Thorne themselves in 1968 \cite{Hartle:1968si}, where they computed the first rotational corrections to the mass of a neutron star. By constructing sequences of slowly-rotating models for a variety of equations of state and angular velocities, they showed that rotational corrections to the mass can reach up to $30\%$ relative to the non-rotating TOV mass $M_*$, depending on the stiffness of the equation of state and the spin frequency. 
In 1973, Hartle extended the formalism to third order in the spin to study the moment of inertia to second order in angular velocity \cite{Hartle:1973zza}. This work was later revisited by Benhar et al.~\cite{Benhar:2005gi}, who performed a detailed comparison between the slow-rotation expansion and full numerical relativity results \cite{Cook:1993qr, Berti:2003nb}. The authors of \cite{Benhar:2005gi} concluded that third-order corrections are essential to accurately reproduce the moment of inertia of neutron stars that rotate at frequencies comparable to the fastest observed pulsars. For example, for the APR2 equation of state~\cite{Akmal:1998cf} and a (linear) spin frequency of $f=641$ Hz, second-order corrections to the moment of inertia can be as large as 13\%.
Further extending the method, Yagi et al.~\cite{Yagi:2014bxa} pushed the slow-rotation expansion to fourth order, computing higher multipole moments, such as the mass-current octupole, $S_{3}$, and mass hexadecapole, $M_{4}$. By comparing their results to numerical relativity simulations, the authors of~\cite{Yagi:2014bxa} showed that fourth-order corrections to the TOV mass, moment of inertia, and mass quadrupole moment become significant for spin frequencies above 100–450 Hz, depending on the equation of state. These studies highlight the necessity of including higher-order corrections in the slow-rotation expansion when modeling rapidly-rotating neutron stars.

In this work, we aim to systematically assess the accuracy and applicability of the slow-rotation approximation by extending the Hartle-Thorne expansion to seventh order in the dimensionless spin parameter. This extension enables a more precise modeling of the spacetime geometry and associated observables, particularly in the context of fast-rotating pulsars. The central questions we address are:
\textit{(i)} For a given desired accuracy in an observable, what order in the slow-rotation approximation is required to reliably model a neutron star of fixed rotational frequency? and more importantly
\textit{(ii)} What is the magnitude of the systematic error introduced by truncating the perturbative expansion at a given order?

To answer these questions, we construct the metric and fluid perturbations order by order in spin, compute the resulting multipole moments and observable quantities, and quantify their dependence on the rotation rate of the star. By pushing the expansion to seventh order, we are able to track the behavior of the approximation well beyond the second-, third-, and fourth-order truncations studied previously~\cite{Benhar:2005gi,Yagi:2014bxa}. This level of precision is crucial for evaluating the reliability of the perturbative framework in modern astrophysical contexts, particularly in light of the increasing accuracy of observational data. This work constitutes the first in a two-part series, focused on constructing the perturbative equations and deriving expressions for extracting the spacetime’s multipole moments within the slow-rotation approach.

\subsection{Executive Summary}

In this first paper of a two-part series, we construct the equations of stellar structure within the Hartle-Thorne approximation, extended to seventh order in the angular spin-frequency of the star $\Omega$. This formalism models isolated, unmagnetized neutron stars that rotate slowly relative to their mass-shedding frequency, $\Omega_{\textrm{sh}}$. We build on the original Hartle-Thorne framework by incorporating higher-order perturbative corrections to a spherically-symmetric, non-rotating configuration, expressed as expansions in powers of the spin frequency up to the seventh power.
Working in a suitable coordinate system, we decouple the Einstein field equations via harmonic decomposition and classify the resulting system into two sectors based on their parity properties. These equations, commonly referred to as the stellar structure equations, reduce to linear, radial ordinary differential equations for the metric perturbation modes at each fixed order in spin frequency.
The even-parity sector governs the mass multipole moments of the neutron star, $M_{\ell}$, and arises at \emph{even} orders in the spin-frequency expansion, while the odd-parity sector encodes the mass-current multipole moments, $S_{\ell}$, and appears at \emph{odd} orders in $\Omega$.

\begin{figure}[h]
  \centering
  \includegraphics[width=0.42\textwidth]{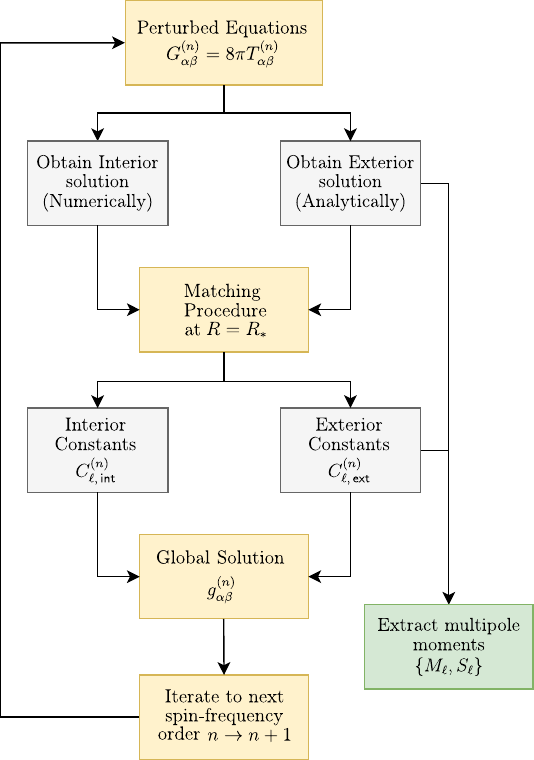}
  \caption{Flowchart illustrating the iterative procedure for extracting the star’s multipole moments within the Hartle-Thorne approximation at each spin-frequency order $n$.}
  \label{fig:flowchart}
\end{figure}

The full iterative procedure is summarized schematically in Fig.~\ref{fig:flowchart}, which outlines the step-by-step construction of the stellar structure equations and the extraction of multipole moments at each order in spin. At a given perturbative order $n$, we construct the corresponding Einstein equations through harmonic modes, ensuring that the spacetime symmetries—stationarity, axisymmetry, and reflection symmetry about the equatorial plane—are preserved. The resulting field equations for the metric perturbation functions at order $n$ are linear, coupled ordinary differential equations in the radial coordinate, with source terms that depend on solutions obtained at previous orders.

These equations must be solved separately in the stellar interior and the vacuum exterior. In the interior, they must be integrated numerically from the center of the star up to the surface, and we here carry out a local analysis around the core to find the correct boundary conditions. For the exterior, we derive exact analytical solutions in closed form at each perturbative order. Each region contributes two \emph{types} of integration constants per set of differential equations for each perturbation function, as the system can be reformulated as a set of two, coupled, first-order, ordinary differential equations for each harmonic mode (except for the \(\ell = 0\) mode; see Section~VI for details). Imposing regularity at the center and asymptotic flatness at spatial infinity eliminates one constant per region. The remaining constants must be determined through a matching procedure at the stellar surface, located at $R=R_{*}$ where the interior and exterior solutions are smoothly connected.
 
The exterior integration constant at order \(n\) determines the \emph{multipole moments} of the star, which must be extracted by analyzing the asymptotic behavior of the exterior solution at spatial infinity. Owing to the reflection symmetry of the spacetime about the equatorial plane, only the \emph{even} \(\ell\) \emph{mass multipole moments} \(M_\ell\) and the \emph{odd} \(\ell\) \emph{mass-current multipole moments} \(S_\ell\) are nonvanishing. Performing the slow-rotation expansion to order \(n\) allows us to determine all multipole moments \(\{M_\ell, S_\ell\}\) for \(\ell \leq n\), with each successive order contributing the leading-order spin correction to a new, higher-order multipole, as well as additional spin-induced corrections to lower-order moments. The interior integration constant, on the other hand, enters the global interior solution at order \(n\) and contributes to the \emph{source terms} at order \(n+1\), thereby feeding into the next step of the iteration. The recursive framework put forth here provides a controlled and systematic method for extracting the full multipolar structure of a slowly-rotating neutron star to seventh order in spin.

In Table \ref{table:multipoles}, we summarize all the multipole moments and their spin-induced corrections that can be extracted by extending the slow-rotation approximation to seventh order in spin. The parameter \(\epsilon\) denotes the expansion order in spin frequency (for details, see Sec.~\ref{sec:slow-rot}). As shown in the table, the \emph{mass multipole moments} receive corrections at \emph{even} powers of \(\epsilon\), while the \emph{mass-current multipole moments} receive corrections at \emph{odd} powers.
The first column lists each multipole moment, with the 0th column indicating its leading-order contribution in the spin expansion. The subsequent columns show the first (1st), second (2nd), and third (3rd) order spin contributions, which appear at progressively higher orders in \(\epsilon\).
The leading contribution to the mass monopole \(M_0\) corresponds to the non-rotating background mass (the TOV mass), with spin corrections entering at \(\mathcal{O}(\epsilon^2)\), \(\mathcal{O}(\epsilon^4)\), and \(\mathcal{O}(\epsilon^6)\), respectively. Similarly, the spin angular momentum \(S_1\) appears at linear order, with additional spin corrections at \(\mathcal{O}(\epsilon^3)\), \(\mathcal{O}(\epsilon^5)\), and \(\mathcal{O}(\epsilon^7)\).
Following the conventions in the literature, we express spin corrections \emph{relative} to the leading-order appearance of each multipole. In this sense, both \(M_0\) and \(S_1\) receive second-, fourth-, and sixth-order spin corrections relative to their leading terms. The next set of moments—\(M_2\) and \(S_3\)—receive second- and fourth-order relative spin corrections. The higher multipoles \(M_4\) and \(S_5\) have only one additional relative second-order spin correction, while the highest-order moments shown, \(M_6\) and \(S_7\) , appear only at leading order in the spin approximation. The shaded regions in Table \ref{table:multipoles} highlight the new contributions obtained through the extended perturbative expansion developed in this work. While some of the perturbation equations at \(\mathcal{O}(\epsilon^{4})\) have been previously explored in \cite{Yagi:2014bxa}, such study did not compute the corresponding correction to the monopole moment \(M_0\), and effectively nothing was known about fifth, sixth and seventh order in rotation until this work. Moreover, certain inconsistencies were present in the literature when formulating the equations—particularly regarding the coordinate transformation to the Hartle-Thorne frame-which we have addressed and resolved in this paper. Additionally, we uncovered a novel dependency of higher-order multipole moments and their spin corrections on higher derivatives of the speed of sound squared, $c_{s}^{2}$.

\begin{table}[]
\renewcommand{\arraystretch}{1.5}
\begin{tabular}{|cccccc|}
\hline
\textbf{Multipole}                                   & \textbf{Symbol}              & \textbf{0th}                             & \textbf{1st }                   & \textbf{2nd}                   & \textbf{3rd } \\ \hline
\multicolumn{1}{|c|}{\textit{Monopole}}           & \multicolumn{1}{c|}{$M_{0}$} & \multicolumn{1}{c|}{$\mathcal{O}(1)$}        & \multicolumn{1}{c|}{$\mathcal{O}(\epsilon^{2})$} & \multicolumn{1}{c|}{\cellcolor{yellow!15}$\mathcal{O}(\epsilon^{4})$} & \cellcolor{yellow!15}$\mathcal{O}(\epsilon^{6})$ \\[0.5ex] \hline
\multicolumn{1}{|c|}{\textit{Dipole}}             & \multicolumn{1}{c|}{$S_{1}$} & \multicolumn{1}{c|}{$\mathcal{O}(\epsilon)$}     & \multicolumn{1}{c|}{$\mathcal{O}(\epsilon^{3})$} & \multicolumn{1}{c|}{\cellcolor{yellow!15}$\mathcal{O}(\epsilon^{5})$} & \cellcolor{yellow!15}$\mathcal{O}(\epsilon^{7})$ \\[0.5ex] \hline
\multicolumn{1}{|c|}{\textit{Quadrupole}}         & \multicolumn{1}{c|}{$M_{2}$} & \multicolumn{1}{c|}{$\mathcal{O}(\epsilon^{2})$} & \multicolumn{1}{c|}{$\mathcal{O}(\epsilon^{4})$} & \multicolumn{1}{c|}{\cellcolor{yellow!15}$\mathcal{O}(\epsilon^{6})$} & $\cdot$ \\[0.5ex] \hline
\multicolumn{1}{|c|}{\textit{Octupole}}           & \multicolumn{1}{c|}{$S_{3}$} & \multicolumn{1}{c|}{$\mathcal{O}(\epsilon^{3})$} & \multicolumn{1}{c|}{\cellcolor{yellow!15}$\mathcal{O}(\epsilon^{5})$} & \multicolumn{1}{c|}{\cellcolor{yellow!15}$\mathcal{O}(\epsilon^{7})$} & $\cdot$ \\[0.5ex] \hline
\multicolumn{1}{|c|}{\textit{Hexadecapole}}       & \multicolumn{1}{c|}{$M_{4}$} & \multicolumn{1}{c|}{$\mathcal{O}(\epsilon^{4})$} & \multicolumn{1}{c|}{\cellcolor{yellow!15}$\mathcal{O}(\epsilon^{6})$} & \multicolumn{1}{c|}{$\cdot$}             & $\cdot$ \\[0.5ex] \hline
\multicolumn{1}{|c|}{\textit{Dotriacontapole}}    & \multicolumn{1}{c|}{$S_{5}$} & \multicolumn{1}{c|}{\cellcolor{yellow!15}$\mathcal{O}(\epsilon^{5})$} & \multicolumn{1}{c|}{\cellcolor{yellow!15}$\mathcal{O}(\epsilon^{7})$} & \multicolumn{1}{c|}{ $\cdot$ }               & $\cdot$ \\[0.5ex] \hline
\multicolumn{1}{|c|}{\textit{Hexacontatetrapole}} & \multicolumn{1}{c|}{$M_{6}$} & \multicolumn{1}{c|}{\cellcolor{yellow!15}$\mathcal{O}(\epsilon^{6})$} & \multicolumn{1}{c|}{$\cdot$}                 & \multicolumn{1}{c|}{$\cdot$}                 & $\cdot$ \\[0.5ex] \hline
\multicolumn{1}{|c|}{\textit{Hectoicosaoctapole}} & \multicolumn{1}{c|}{$S_{7}$} & \multicolumn{1}{c|}{\cellcolor{yellow!15}$\mathcal{O}(\epsilon^{7})$} & \multicolumn{1}{c|}{$\cdot$}                 & \multicolumn{1}{c|}{$\cdot$}                 & $\cdot$ \\[0.5ex] \hline
\end{tabular}
\caption{Spin-frequency orders of approximation for the mass $M_{\ell}$ and current $S_{\ell}$ multipole moments within the slow-rotation expansion. The first and second columns list the names and symbols that represent the multipole moments we study, with the type (mass or current) suppressed. The third column shows the (0th) leading-order contribution of each multipole, such as the TOV mass $M_* = M_0$ at ${\cal{O}}(\epsilon^0)$.  
The subsequent columns list the first (1st), second (2nd), and third (3rd) order corrections, which may appear at different orders in $\epsilon$. The shaded cells highlight new contributions obtained through the perturbative analysis carried out in this work.  }
\label{table:multipoles}
\end{table}

The full expressions for the multipole moments are provided in Appendix \ref{apx:multipoles} and are written in terms of the exterior integration constants determined at previous orders in the spin-frequency expansion. We extracted and validated these expressions using two independent methods: (i) Thorne’s expansion in Asymptotically Cartesian Mass-Centered (ACMC) coordinates, and (ii) Ryan’s procedure, which relies on Fodor’s algorithm within the Geroch–Hansen invariant formalism. In particular, we extended Ryan’s method to compute the mass-current multipole moment \(S_7\), which had not been previously derived using this approach.

\subsection{Organization and Convention}

The remainder of this paper is organized as follows. In Sec.~\ref{sec:slow-rot}, we review the slow-rotation expansion, and present the symmetries and assumptions of the model. We construct an extension of the spacetime metric up to seventh order in spin frequency and describe the harmonic decomposition procedure used to decouple the Einstein equations, setting the Hartle-Thorne frame as the basis for the perturbative analysis. In Sec.~\ref{sec:structure}, we derive the stellar structure equations, separating them into even- and odd-parity sectors, and outline the extraction of the corresponding modes from the harmonic decomposition. Section \ref{sec:local} presents an asymptotic analysis of the stellar structure equations near the origin, which we use to derive boundary conditions for numerical integration from the center to the surface of the star. In Sec.~\ref{sec:exterior}, we obtain exact analytical solutions for the exterior metric for both parity sectors. Section \ref{sec:multipoles} details the matching of interior and exterior solutions to construct global solutions, and describes two independent methods for extracting the multipole moments. Finally, in Sec.~\ref{sec:discussion}, we discuss the physical implications of achieving this level of modeling accuracy, including the appearance of higher-order derivatives of $c_{s}^{2}$ in the source terms. We conclude with a summary of our results and their relevance for future multimessenger observations. The numerical evolution of the equations derived in this work, as well as an analysis of observables is left for the second paper in this two-paper series. 

Throughout this work, we adopt geometric units with \( G = 1 = c \), and use the metric signature \((- + + +)\). The total energy density of the star is denoted by \( \varepsilon \), while \( \epsilon \) refers to the expansion parameter in the slow-rotation approximation. Quantities labeled with ``\textsf{HT}'' are defined in the Hartle-Thorne frame. The radial coordinate \( R \) refers to the Hartle-Thorne radial coordinate, while \( R_* \) denotes the Tolman–Oppenheimer–Volkoff (TOV) stellar radius in the non-rotating configuration; similarly, \( M_* \) denotes the TOV stellar mass in the non-rotating configuration. The mass and mass-current scalar multipole moments are defined following the Geroch–Hansen convention and are denoted by \( M_\ell \) and \( S_\ell \), respectively.

We also adopt the following notational conventions:
\begin{itemize}
  \setlength{\itemsep}{2pt}
  \setlength{\parskip}{0pt}
  \setlength{\parsep}{0pt}
    \item A superscript in parentheses, \( (n) \), indicates a perturbative quantity of order \( n \) in the slow-rotation expansion, where \( n \) corresponds to the power of the star’s angular spin frequency. Since \( n \) can be even or odd, we reserve \( n = s \) for even-order (even-parity) perturbations and \( n = k \) for odd-order (odd-parity) perturbations. Superscripts \( (s) \) and \( (k) \) are used accordingly to distinguish between the two sectors.
    \item The subindices \( \ell \) and \( m \) label the spherical-harmonic decomposition, where \( \ell = 0, 1, 2, \ldots \) and \( m = -\ell, \ldots, \ell \).
    \item For each metric perturbation function at spin-frequency order \( n \) and mode \( \ell \), we use the superscript \( \textsf{H} \) for the homogeneous and \( \textsf{P} \) for the particular solution of the corresponding differential equation.
\end{itemize}

\section{Slow-rotation expansion}
\label{sec:slow-rot}

We focus on unmagnetized neutron stars that are isolated and rotate uniformly at a slow rate. In equilibrium, these compact objects can be mathematically described using perturbation theory in GR, where the small perturbation parameter is related to the star's spin-frequency. 
This framework, known as the Hartle-Thorne (\textsf{HT}) method, involves expanding the metric and the stress-energy momentum tensor in powers of a slow-rotation perturbation parameter $\epsilon$, and then computing the Einstein equations perturbatively.
In this section, we review the method's assumptions and extend the approximation to $\mathcal{O}(\epsilon^{7})$. To decouple the Einstein equations, we perform a spherical-harmonic decomposition in a fixed gauge, choosing an appropriate coordinate system to ensure that the matter perturbations due to rotation remain small.
This procedure will provide the necessary elements to derive the equations of stellar structure in the subsequent section.

\subsection{Hartle-Thorne Method}

Consider an isolated, static neutron star in equilibrium. As the star spins, the stars' matter rearranges itself to minimize its energy, resulting in an oblate spheroidal shape (a Maclaurin spheroid, to be precise) \cite{chandrasekhar1987ellipsoidal}. When the star rotates slowly, small deviations from sphericity can be treated as minor disturbances. According to GR, such disturbances may induce slight changes in the spacetime geometry in response to the mass-energy configuration of the rotating star. This motivates the expansion of the metric in a slow-rotation parameter around the non-rotating configuration. This idea, originally formulated by Hartle and Thorne in the late 1960s \cite{Hartle:1967he, Hartle:1968si}, is based on the following assumptions:
\begin{itemize}
 \setlength{\itemsep}{2pt}
  \setlength{\parskip}{0pt}
  \setlength{\parsep}{0pt}
    \item[(I)] \textit{Matter description-equation of state}: We assume matter follows a barotropic equation of state, $p=p(\varepsilon)$ where $p$ is the pressure and $\varepsilon$ is the total mass-energy density. This means, in particular, that we neglect the influence of temperature or magnetic fields in the equation of state. These assumptions are reasonable for neutrons stars for the following reasons. The Fermi temperature of neutron stars is much greater than the star's temperature, so corrections caused by thermal agitation are small. Moreover, the contribution of the magnetic field energy to the energy density $\varepsilon$ of neutron stars is small (except for magnetars or proto-neutron stars, which we do not consider in this paper).  
   \item[(II)] \textit{Matter description-stress-energy tensor:} We assume that the stress-energy tensor can be described by that of a perfect fluid. This means we assume the star is in equilibrium, and so viscous or heat-conductivity effects are not present. We also neglect magnetic fields, whose topology will usually break isotropy. Such assumptions are appropriate for old and isolated neutron stars, i.e.~those that are not currently accreting material from a companion or in the late stages of merging with another body. 
    \item[(III)] \textit{Slow rotation:} The star's spin must be low enough that the fractional changes in pressure and energy density due to rotation are small enough to be treated perturbatively. 
    Then, we can define a dimensionless spin-frequency parameter $\epsilon$ as
    \begin{equation}
        \epsilon \equiv  \dfrac{\Omega}{\Omega_{\textrm{sh}}} \ll 1 \ .
    \end{equation}
    Here, \(\Omega \) is the neutron star's spin angular frequency as measured by an observer at infinity, and \(\Omega_{\textrm{sh}}\) is the mass-shedding spin angular frequency, beyond which the star may be disrupted. For most equations of state in the literature, the star's spin frequency is much lower than the mass-shedding limit. However, the \textsf{HT} method breaks down when the spin frequency approaches the mass-shedding limit, i.e., when \(\epsilon \sim 1\). As shown in \cite{Berti:2004ny}, even when modeling the fastest pulsars, the \textsf{HT} method remains a good approximation. Therefore, the slow-rotation approximation is still valid for studying rotating neutron stars.
    \item[(IV)] \textit{Uniform rotation:} As in the Newtonian case, in GR, the rotating configurations that minimize the system's mass-energy must rotate uniformly \cite{hartle1967, boshkayev2012gravitational}. While differential rotation, particularly in newly-formed neutron stars, has been studied \cite{Hartle1970diff, Chirenti:2013xm}, it has been shown that stars with differential rotation generally evolve toward a state of uniform rotational equilibrium \cite{Duez:2006qe}. Therefore, assuming uniform rotation is a valid approximation for most cases, and we can take \(\Omega \equiv \textrm{constant}\), where \(\Omega\) is the star's angular spin frequency.
    \item[(V)] \textit{Stationary and axial symmetry:} The rotating-configuration is stationary and axially symmetric with respect to an arbitrary axis, which can always be assumed to be aligned with the star's rotation axis. Thus, the metric components must be independent of the time coordinate $t$ and the azimuthal angle $\phi$, i.e., $g_{\alpha \beta} = g_{\alpha \beta}(r, \theta)$, where $t^\alpha=[1,0,0,0]$ and $\phi^\alpha=[0,0,0,1]$ are Killing vectors in this coordinate system. Furthermore, the fluid motion is circular and perpendicular to the axis of rotation, driven only by the star's rotation. This implies the configuration is invariant under the simultaneous inversion $t\rightarrow -t$ and $\phi \rightarrow - \phi$. However, note that such symmetry does not hold for meridional currents, as such simultaneous inversion changes the direction of the flow \cite{Friedman:2013xza}.
    \item[(VI)] \textit{Reflection symmetry:} The matter distribution has reflection symmetry across a plane perpendicular to the axis of rotation. This means that the configuration should be invariant under the transformation $\theta \rightarrow \pi - \theta$, where $\theta$ denotes the polar angle.
\end{itemize}

The rotating configuration is described by expanding the metric in powers of $\epsilon$ around the non-rotating, spherically-symmetric background. Based on the symmetry assumptions discussed earlier, it can be shown that there is a spherical coordinate system where the non-zero components representing the geometry of a rotating object are $g_{tt}$, $g_{rr}$, $g_{\theta\theta}$, $g_{\phi\phi}$, and $g_{t\phi}$ \cite{hartle1967}. We can describe this system using Boyer-Lindquist-type coordinates $(t, r, \theta, \phi)$. The metric in these coordinates can then be expanded in powers of $\epsilon$, up to $\mathcal{O}(\epsilon^{7})$, as corrections to the background metric,
\begin{widetext}
    \begin{equation}
     \begin{split}
        ds^{2} &= -e^{\nu} \left[1 \, + \, 2 \epsilon^{2}  h^{(2)} \, + \,  2\epsilon^{4}  h^{(4)} \, + \, 2\epsilon^{6} h^{(6)} \right] dt^{2} 
         \, + \, e^{\lambda} \left[1 \, + \, \dfrac{2 \epsilon^{2} m^{(2)} \, + \, 2 \epsilon^{4} m^{(4)} \, + \, 2 \epsilon^{6} m^{(6)}}{r-2M} \right] dr^{2} \\[1ex]
        & \, + \,  r^{2} \left[ 1  \, + \,  2 \epsilon^{2} k^{(2)} \, + \, 2 \epsilon^{4} k^{(4)} \, + \, 2 \epsilon^{6} k^{(6)} \right] 
          \Big( d\theta^{2} \, + \, \sin^{2}\theta \bigl\{ d \phi \, - \, \left[ \epsilon \omega^{(1)} \, + \,  \epsilon^{3} \omega^{(3)} \, + \,  \epsilon^{5} \omega^{(5)} \, + \, \epsilon^{7} \omega^{(7)} \right] dt \bigr\}^{2} \Big)
     \end{split}
     \label{eq:metric}
    \end{equation}
\end{widetext}
where the background metric functions  $\nu$, $\lambda$ and $M$ depend only on the radial coordinate $r$, with $\lambda$ and $M$ related by
\begin{equation}
    e^{\lambda(r)} \equiv \left[ 1 -  \dfrac{2M(r)}{r} \right]^{-1} \ .
\end{equation}
The metric perturbation functions are denoted by 
$h$, $m$, $k$ and $\omega$, each with a superscript 
$(n)$ that indicates their dependence on the power $n$ of the spin-frequency parameter $\epsilon$. Due to the symmetry assumptions, these perturbations can only depend on $r$ and $\theta$ and must be invariant under reflection symmetry across the equator. 
Additionally, in Eq.~\eqref{eq:metric}, the perturbations exhibit even powers of $\epsilon$ in the $(t,t)$, $(r,r)$, $(\theta, \theta)$, and $(\phi,\phi)$ components, and odd-powers of $\epsilon$ in the $(t,\phi)$ component. This is consistent with the requirement that time reversal $t \rightarrow -t$ must be equivalent to the transformation  $\Omega \rightarrow - \Omega$. Perturbations along other component directions or different power parities are excluded by symmetry arguments or can be removed under a coordinate transformation\footnote{See reference \cite{hartle1967} for more details. }. 
If we aim to restore staticity by ensuring that the metric remains invariant under the transformation $t \rightarrow -t$, 
all the $\omega$ functions must be set to zero. Consequently, these functions are associated with the rotation of spacetime itself, a phenomenon often referred to as the dragging of inertial frames.

The spacetime geometry is being sourced by the mass-energy distribution of the rotating star. For simplicity, we assume the matter-stress energy tensor $T^{\mu \nu}$ to be a perfect fluid, so any out-of-equilibrium effects inside the star will be neglected\footnote{ Anisotropy effects have been explored in \cite{Cadogan:2024ywc, Yagi:2015hda, Beltracchi:2024dfb}.}. 
Thus, $T_{\mu \nu}$ is given by 
\begin{equation}
\label{eq:perfect_fluid}
    T_{\mu \nu} = (\varepsilon + p) u_{\mu} u_{\nu} + pg_{\mu \nu} 
\end{equation}
where $u^{\mu}$ is the fluid's four-velocity. Due to rotation, the matter distribution of the star will also develop perturbations. Therefore, the pressure $p$, the total mass-energy density $\varepsilon$ and the four-velocity $u^{\mu}$ can be written as a one-parameter family of perturbed quantities,
\begin{align}
\label{eq:ppert}
  p(r,\theta) &= p^{(0)}(r) + \epsilon p^{(1)}(r,\theta) + \frac{1}{2} \epsilon^{2} p^{(2)} (r,\theta) + \cdots \, , \\
  \label{eq:epert}
 \varepsilon(r,\theta) &= \varepsilon^{(0)}(r) + \epsilon \varepsilon^{(1)}(r,\theta) + \frac{1}{2} \epsilon^{2} \varepsilon^{(2)} (r,\theta) + \cdots \, , \\
 \label{eq:upert}
   u_{\mu}(r,\theta) &= u_{\mu}^{(0)}(r) + \epsilon u_{\mu}^{(1)}(r,\theta) + \frac{1}{2} \epsilon^{2} u_{\mu}^{(2)} (r,\theta) + \cdots 
\end{align}
where the functions $\varepsilon^{(0)}$ and $p^{(0)}$, and the vector field $u_{\mu}^{(0)}$, depend only on the radial coordinate $r$ since these refer to the spherically-symmetric background configuration.
The fluid's four-velocity is parameterized as
\begin{equation}
\label{eq:4-velocity}
    u^{\mu} = (u^{t},0,0,\epsilon \, \Omega \, u^{t}) 
\end{equation}
where the component $u^{t}$ can be obtained from the normalization condition $u^{\mu}u_{\mu}=-1$. Note that Eq.~\eqref{eq:4-velocity} is in agreement with the assumption of circularity and thus $u^{r}$ and $u^{\theta}$ must vanish.

The slow-rotation approximation is characterized by the perturbed metric given in Eq.~\eqref{eq:metric}, together with a perturbed perfect fluid stress-energy tensor that accounts for the changes in pressure and energy density induced by rotation. The next step in the Hartle-Thorne framework is to construct the Einstein equations systematically, proceeding order by order in the spin-frequency parameter, $\epsilon$. Two additional challenges arise in this process: (i) the Einstein equations are coupled in the radial and polar angle coordinates, $r$ and $\theta$, and (ii) an appropriate reference frame must be chosen to ensure that the fractional changes  in pressure and energy density between the background and perturbed configurations remain small. The following two subsections address these issues before proceeding with the computation of the Einstein equations.

\subsection{Harmonic decomposition}

Since the rotating configuration is stationary and axisymmetric, the metric perturbation functions depend on both the radial $r$ and  polar angle $\theta$ coordinates only. To decouple the Einstein equations, we decompose the metric perturbations into spherical harmonics. The process is analogous to decoupling Schrödinger's equation in quantum mechanics in spherical coordinates. However, due to the tensorial nature of the Einstein equations, a standard scalar harmonic decomposition of the metric components is insufficient, as not all metric components transform as scalars under rotations. We then introduce a scalar, vector and tensor harmonic decomposition following the same notation as in \cite{Martel:2005ir} for black hole perturbation theory. Lowercase Latin component indices  run over the coordinates $x^{a}=(t,r)$, while uppercase Latin indices run over the angular coordinates $x^{A}=(\theta, \phi)$. In general, the metric component perturbations $p_{\alpha \beta}$  can then be expanded at each spin frequency order $(n)$ as 
\begin{align}
\label{eq:scalar}
p^{(n)}_{ab} &= \sum_{\ell m} h^{(n)\ell m}_{ab}(x^{c}) \; Y^{\ell m} \ , \\[1ex]
\label{eq:vector}
p^{(n)}_{aA} &= \sum_{\ell m} j^{(n)\ell m}_{a}(x^{c}) \; Y^{\ell m}_{A} + \sum_{\ell m} h^{(n)\ell m} _{a}(x^{c}) \; X^{\ell m}_{A} \ , \\[1ex]
\label{eq:tensor}
p^{(n)}_{AB} &= r^{2}\sum_{\ell m} \left[  K^{(n)\ell m} (x^{c}) \; \Omega_{AB} \; Y^{\ell m} + G^{(n)\ell m}(x^{c}) \; Y^{\ell m}_{AB}\right] \nonumber  \\[1ex] 
&+ \sum_{\ell m} h^{(n)\ell m}_{2}(x^{c}) \; X^{\ell m}_{AB} \ .
\end{align}
Here, $Y^{\ell m}=Y^{\ell m}(x^{A})$ represents the usual scalar spherical harmonics. From Eq.~\eqref{eq:scalar}, we see that $p_{ab}$  transforms as a scalar under rotations on the unit two-sphere $\mathcal{S}^{2}$. The quantities $Y^{\ell m}_{A} = Y^{\ell m}_{A}(x^{A})$ and $X^{\ell m}_{A} = X^{\ell m}_{A}(x^{A})$ represent the vector spherical harmonics of even and odd parities, respectively. They form a complete basis for $p_{aA}$ which behaves as a vector on $\mathcal{S}^{2}$. Finally, $Y^{\ell m}_{AB} = Y^{\ell m}_{AB}(x^{A})$ and $X^{\ell m}_{AB} = X^{\ell m}_{AB}(x^{A})$ are the trace-free tensor spherical harmonics of even and odd parity, respectively, and they constitute a basis for the trace-free part of $p_{AB}$. The first term under the summation symbol in Eq.~\eqref{eq:tensor} denotes the trace piece, which transforms as a scalar, where $\Omega_{AB}=\mathrm{diag}(1,\sin^{2}\theta)$ is the metric on $\mathcal{S}^{2}$. 

Perturbations can be classified into two sectors based on the parity properties of the scalar, vector, and tensor spherical harmonics. These sectors do not couple to each other and transform differently. Under a parity transformation, where $\theta \rightarrow \pi - \theta$ and $\phi \rightarrow \phi + \pi$ are applied simultaneously, the spherical harmonics $Y^{\ell m}$, $Y^{\ell m}_{A}dx^{A}$, $\Omega_{AB}Y^{\ell m}dx^{A}dx^{B}$, and $Y_{AB}^{\ell m}dx^{A}dx^{B}$ all transform with a factor of $(-1)^{\ell}$, indicating they have \textit{even-parity}. In contrast, the spherical harmonics $X^{\ell m}_{A}dx^{A}$ and $X^{\ell m}_{AB}dx^{A}dx^{B}$ transform with a factor of $(-1)^{\ell+1}$ under the same parity transformation, and are therefore said to have \textit{odd-parity}. The metric components of the perturbations expanded in terms of the even-parity spherical harmonics represent the even-parity sector, while those expanded in terms of the odd-parity spherical harmonics represent the odd-parity sector. 

In GR, there is gauge freedom when describing perturbations due to the intrinsic diffeomorphism invariance of the theory. That is, we can always perform an infinitesimal diffemorphism on a given metric tensor, which will generate a different perturbed metric but represent the same physical system. Two equivalent perturbations are therefore related by 
\begin{equation} p^{(n)}_{\alpha \beta , \, \textsf{new}} - p^{(n)}_{\alpha \beta , \, \textsf{old}} = 2 \nabla_{(\alpha}\Xi^{(n)}_{\beta)} , 
\label{eq:diff}
\end{equation} where $\nabla$ denotes the covariant derivative associated with the background spacetime, and the parentheses around the indices in Eq.~\eqref{eq:diff} indicate symmetrization. This freedom allows us to choose a convenient generator $\Xi_{\alpha}$ for the gauge transformation, thereby enabling the vanishing of the following functions,
\begin{align}
\label{eq:gauge}
     \hspace{0.2cm} j_{a}^{(n)\ell m}(r) = 0 \ , 
     \hspace{0.2cm} G^{(n)\ell m}(r) = 0 \ ,
     \hspace{0.2cm} h^{(n)\ell m}_{2}(r) = 0 \ .
\end{align}
By ``fixing the gauge'' in this way, the perturbed Einstein equations can be greatly simplified. Henceforth, we work in this gauge, which is called the \textit{Regge-Wheeler} gauge \cite{Regge:1957td, Martel:2005ir}. 

Following the symmetry assumptions for the metric in Boyer-Lindquist-type coodinates given in Eq.~\eqref{eq:metric}, the general harmonic decomposition expansion in Eqs.~\eqref{eq:scalar}-\eqref{eq:tensor} can be simplified and split into odd and even sectors as
\begin{align}
p^{(s)}_{\alpha \beta,\textsf{even}} &= \sum_{\substack{\ell = 0 \\[0.2ex] ( \ell \textsf{ even} ) }}^{s}
\begin{pmatrix}
 h^{(s)\ell 0}_{tt}&     0&   0&   0      \\
 0& h^{(s)\ell 0}_{rr}&  0&   0             \\
 0&   0& \tilde{K}^{(s)\ell0}&        0             \\
 0&   0&    0&   \tilde{K}^{(s)\ell0}\sin^{2}\theta \\
\end{pmatrix}  
\label{eq:peven}
Y^{\ell 0} \, , \\[1ex] 
p^{(k)}_{\alpha \beta, \mathsf{odd}} &=
\sum_{\substack{\ell = 1 \\[0.2ex] ( \ell \textsf{ odd} ) }}^{k}
\begin{pmatrix}
 0&  \, 0& \, 0& h^{(k) \ell 0}_{t}  \\
 0&  \, 0& \, 0&  0  \,  \\
 0&  \, 0& \, 0&  0  \,  \\
 h^{(k) \ell 0}_{t}&  \, 0& \, 0&  0  \,  \\
\end{pmatrix}X^{\ell 0}_{\phi} \, 
\label{eq:podd}
\end{align}
where $\tilde{K}^{(s)\ell 0}\equiv r^{2} K^{(s)\ell 0} $ and the functions $h^{(s)\ell 0}_{tt}$, $h^{(s)\ell 0}_{rr}$, $\tilde{K}^{(s)\ell 0}$ and $h^{(k)\ell 0}_{t}$ depend only on the radial coordinate $r$. The superscript in parentheses, $(s)$, can only take even values, while $(k)$ can take only odd values\footnote{Since we are performing the calculation to $\mathcal{O}(\epsilon^{7})$ in the spin frequency, the values that
$k$ can take are 1, 3, 5, and 7, while the values 
$s$ can take are 2, 4, and 6.}. These indices label the spin-frequency order of even- and odd-parity perturbations. 

The metric perturbation components in Eqs.~\eqref{eq:peven}-\eqref{eq:podd} can be obtained from the harmonic expansion in Eqs.~\eqref{eq:scalar}--\eqref{eq:tensor} in Boyer-Lindquist-type coordinates as follows. First, by fixing the gauge using Eq.~\eqref{eq:gauge}, only the expansion of the metric components in terms of $X^{\ell m}_{A}$
is included in Eq.~\eqref{eq:vector}, while only the metric components in terms of the trace piece $\Omega_{AB}Y^{\ell m}$ are considered in Eq.~\eqref{eq:tensor}. Second, under the assumption of azimuthal symmetry and stationarity, we set $m=0$ in all spherical harmonics and the metric perturbation functions should not depend on $t$. Third, the metric must be invariant under reflection across the equator by making $\theta \rightarrow \pi - \theta$. Therefore, since under reflection $Y^{\ell 0}$ transforms with a factor of $(-1)^{\ell}$, while $X^{\ell 0}_{\phi}d\phi$ transforms with a factor of $(-1)^{\ell+1}$, then $\ell$ can take only even values in Eq.~\eqref{eq:peven} and odd values in Eq.~\eqref{eq:podd}, such that the metric remains invariant under equatorial reflection. Finally, note that the upper limits of both summations in Eqs.~\eqref{eq:peven} and \eqref{eq:podd} coincide with the same spin-frequency order, i.e., $s$ and $k$, respectively. This is explained by requiring that the solution of the Einstein equations be regular at the center of the neutron star and asymptotically flat at spatial infinity \cite{Hartle:1967he}. It is sufficient to prove that if $\ell=1$ is the only nonzero mode solution of the Einstein equations at $\mathcal{O}(\epsilon)$ that satisfies such two conditions, then the maximum value of $\ell$ for any nonlinear spin-frequency perturbation must be equal to the same spin-frequency order. For instance, the perturbed Einstein equations at $\mathcal{O}(\epsilon^{2})$ will have driving terms that are quadratic in the linear perturbation functions. Therefore, at most, the harmonic expansion of the perturbations at $\mathcal{O}(\epsilon^{2})$ must have $\ell=2$. Similar arguments can be given for the higher order spin-frequency perturbations.
The uniqueness proof for $\ell=1$ at $\mathcal{O}(\epsilon)$ is shown in detail in Appendix \ref{apx:unique}.

A further simplification can be made in Eqs.~\eqref{eq:peven} and \eqref{eq:podd}
by using the explicit expressions for the vector and scalar spherical harmonics in terms of Legendre polynomials. The vector spherical harmonic of odd parity are defined as
\begin{equation}
X_{A}^{\ell m} := -\varepsilon_{A}^{\ B}D_{B}Y^{\ell m}    
\end{equation}
where $\varepsilon_{AB}$ is the antisymmetric Levi-Civita  tensor on $\mathcal{S}^{2}$ with $\varepsilon_{AB}=\sin\theta$, and $D_{A}$ is the covariant derivative on $\mathcal{S}^{2}$, which is compatible with the metric $\Omega_{AB}$. Then, due to azimuthal symmetry, the quantities $Y^{\ell 0}$ and $X^{\ell 0}_{\phi}$ can be expressed in terms of Legendre polynomials $P_{\ell}$ as
\begin{align}
   Y^{\ell 0} &= \alpha P_{\ell}(\cos\theta) \, , \\[1ex]
   X^{\ell 0}_{\phi} &= \alpha \sin \theta \dfrac{d P_{\ell}(\cos\theta)}{d \theta} \, ,
\end{align}
where $\alpha$ is a normalization constant, given by $\alpha \equiv \sqrt{(2\ell +1)/4\pi}$.

From the line element in Eq.~\eqref{eq:metric}, we note that the metric perturbation function \( h^{(s)} \) appears only in the \( g_{tt} \) component, while \( m^{(s)} \) contributes only in the \( g_{rr} \) component. The function \( k^{(s)} \) appears in the \( g_{tt} \), \( g_{rr} \), \( g_{\theta \theta} \), and \( g_{t\phi} \) components. These metric perturbation functions correspond to the even sector. In contrast, the function \( \omega^{(k)} \) enters the \( g_{tt} \) and \( g_{t\phi} \) components, representing the odd sector. Let us then perform a spherical-harmonic decomposition of the metric perturbation functions, extending the original expansion presented by Hartle \cite{Hartle:1967he}. The even-parity sector functions are descomposed as
\begin{align}
\label{eq:hexp}
h^{(s)}(r,\theta) &= \sum_{\substack{\ell = 0 \,  \\[0.2ex] ( \ell \textsf{ even} ) }}^{s} h^{(s)}_{\ell}(r) P_{\ell}(\cos \theta) \, , \\[1ex]
\label{eq:kexp}
k^{(s)}(r,\theta) &=  \sum_{\substack{\ell = 2 \,  \\[0.2ex] ( \ell \textsf{ even} ) }}^{s} k^{(s)}_{\ell}(r) P_{\ell}(\cos \theta) \, , \\[1ex]
\label{eq:mexp}
m^{(s)}(r,\theta) &=  \sum_{\substack{\ell = 0 \\[0.2ex] ( \ell \textsf{ even} ) }}^{s} m^{(s)}_{\ell}(r) P_{\ell}(\cos \theta) \, .
\end{align}
while the odd-sector functions are expanded as,
\begin{equation}
\label{eq:wexp}
\omega^{(k)}(r,\theta) =  \sum_{\substack{\ell = 1 \\[0.2ex] ( \ell \, \textsf{odd} ) }}^{k} \omega^{(k)}_{\ell}(r) \dfrac{dP_{\ell}(\cos \theta)}{d\cos\theta}  \, .
\end{equation}
Notice that the expansion in ~Eq.\ \eqref{eq:kexp} starts from $\ell = 2$. This is because the $\ell=0$ term only contributes with a term that only depends on the radial coordinate, shifting the spacetime areal radius. Such a term can always be removed through a coordinate transformation, allowing us to set at each even spin-frequency order,
\begin{equation}
    k^{(s)}_{0}(r) = 0 \ .
\end{equation}

By using Eqs.~\eqref{eq:hexp}--\eqref{eq:wexp} in the metric of Eq.~\eqref{eq:metric}, and factoring out the angular part, the metric perturbations can be recast into the form presented in Eqs.~\eqref{eq:peven} and \eqref{eq:podd}. Then, the radial functions for the even- and odd-parity sectors\footnote{Note that these functions are time independent since the spacetime is stationary.},
\begin{align}
   \left\{ j^{(s)\ell 0}_{r}, \ h^{(s)\ell 0}_{tt}, \ h^{(s)\ell 0}_{rr}, \ K^{(s)\ell 0} \right\} \  \ , \  \  \left\{ h^{(k)\ell 0}_{t}  \right\} 
\end{align}
can be identified in terms of the metric perturbation functions $h^{(s)}$, $ k^{(s)}$, $m^{(s)}$ and $\omega^{(k)}$. Further details of the calculation are provided in a Mathematica script available in the repository associated with this work \cite{conde2025hartlethorne}.

\subsection{The Hartle-Thorne frame}

Eventhough we have chosen the Regge-Wheeler gauge for our perturbation, a residual gauge freedom exists in terms of a radial rescaling, infinitesimal diffeomorphism. 
As pointed out by Hartle \cite{Hartle:1967he}, a suitable choice of coordinates must be taken into account if the fractional changes in pressure and energy density are required to be small. 
When the star is rotating, isodensity and isobaric contours are no longer located at constant $r$. 
To overcome this issue, we map our radial coordinate $R$ for a non-rotating star that is spherically symmetric (i.e. $R={\rm{const}}$ translates to a spherical shell of constant baryon number density $n_B$) to the radial coordinate $r(R,\Theta)$ of the deformed star that is along a surface of constant $n_B$ equivalent to that of $R$.
Since the equation of state gives us a direct relation between the pressure $p(n_B)$ and energy density $\varepsilon(n_B)$ for a given $n_B$ (when  the star is at $T=0$ and in $\beta$-equilibrium), this implies that\footnote{This also follows from the Gibbs-Duhem relation, $\varepsilon + p = n_{B} \mu_{B}$.}
\begin{align}
\label{eq:pHT}
    p[r(R,\Theta), \theta] &= p(R) = p^{(0)}(R) \ , \\[1ex]
    \varepsilon[r(R,\Theta), \theta] &= \varepsilon(R) = \varepsilon^{(0)}(R) \ ,
\label{eq:eHT}
\end{align}
where $p^{(0)}$ and $\varepsilon^{(0)}$ are the pressure and energy density in the non-rotating configuration, respectively, and $\Theta$ is the same polar angle as $\theta$. Let us refer to this new frame with coordinates $(t, R,\Theta, \phi)$ as the \textit{Hartle-Thorne} (\textsf{HT}) frame. 

The mapping between the contours of both configurations is defined through a function $\xi$, obtained from the following coordinate transformation,
\begin{align}
 r(R,\Theta) = R + \xi(R,\Theta) \ \ ; \ \  \theta(R, \Theta) = \Theta \ .
 \label{eq:transf}
\end{align}
Then, $\xi(R,\Theta)$ can be expanded in slow-rotation in terms of the dimensionless spin-frequency parameter as\footnote{The expansion is performed in even powers of $\epsilon$ to preserve the metric symmetries. Note that at $\mathcal{O}(\epsilon^{0})$ there is no radial displacement $\xi$ since the star is spherically symmetric.
}
\begin{align}
 \xi(R,\Theta) = \epsilon^{2}\xi^{(2)}(R,\Theta) + \epsilon^{4}\xi^{(4)}(R,\Theta) + \epsilon^{6}\xi^{(6)}(R,\Theta) \ .
 \label{eq:transfII}
\end{align} 
The function $\xi$ is interpreted as a radial displacement away from sphericity due to the deformation of the star caused by its rotation.
Since $\xi(R,\Theta)$ behaves as a scalar under rotations, it can be decomposed into scalar spherical harmonics as
\begin{align}
\xi^{(s)}(R,\Theta) =  \sum_{\substack{\ell = 0 \\[0.2ex] ( \ell \textsf{ even} ) }}^{s} \xi^{(s)}_{\ell}(R) P_{\ell}(\cos \Theta) \, .
\label{eq:xiexp}
\end{align}

By construction of the \textsf{HT} frame, there are no pressure or energy density spin-frequency perturbations in the \textsf{HT} frame, i.e.,
\begin{equation}
    p_{\textsf{HT}}^{(n)}(R, \Theta) = 0 \ \  ; \ \ \varepsilon^{(n)}_{\textsf{HT}}(R,\Theta) = 0 \ .
    \label{eq:peHT}
\end{equation}
This is because in the \textsf{HT} frame the pressure and energy density contours are located at shells of constant $R$ as seen in Eqs.~\eqref{eq:pHT}--\eqref{eq:eHT}. In fact, this is the whole reason why an \textsf{HT} frame is sought.
In this frame, the function $\xi$ plays the role of pressure and energy density perturbations and is only well-defined inside the star, while outside it is assumed to take a constant value. 

We choose \textsf{HT} coordinates $(t, R, \Theta, \phi)$ as the frame for computing the perturbed Einstein equations. The decoupled perturbed metric in the \textsf{HT} frame can be obtained by applying the coordinate transformation in Eq.~\eqref{eq:transf}, followed by the spherical-harmonic decomposition outlined in Eqs.\ \eqref{eq:wexp}--\eqref{eq:mexp}. This process involves Taylor expanding all perturbed metric functions about the radial coordinate $R$ and retaining terms up to $\mathcal{O}(\epsilon^{7})$. An alternative approach, as used in \cite{Yagi:2013mbt}, is to first perform the radial coordinate transformation perturbatively and then apply the harmonic decomposition, using the same functional form as in Eqs.\ \eqref{eq:wexp}--\eqref{eq:mexp}, with the substitutions $r \rightarrow R$ and $\theta \rightarrow \Theta$. Both methods rely on a Taylor expansion, which is well-justified by the slow-rotation expansion, and they yield the same result. We provide the transformation of the perturbed metric up to $\mathcal{O}(\epsilon^{7})$ in 
Appendix \ref{apx:HTmetric}.

The stress-energy tensor must be computed from Eq.~\eqref{eq:perfect_fluid} in \textsf{HT} coordinates. 
Since the coordinate transformation in Eq.~\eqref{eq:transf} is performed only along the radial direction, the form of $u^{\alpha}_{\textsf{HT}}$ is the same as given in Eq.~\eqref{eq:4-velocity}. Additionally, the component $u^{t}_{\textsf{HT}}$ can be found from the normalization condition $u^{\textsf{HT}}_{\alpha}u^{\alpha}_{\textsf{HT}}=-1$ from which one obtains
\begin{equation}
    u^{t}_{\textsf{HT}} = \left( \dfrac{-1}{g^{\textsf{HT}}_{tt} \, + \, 2 \epsilon \Omega g^{\textsf{HT}}_{t \phi} \, + \, \epsilon^{2} \Omega^{2} g^{\textsf{HT}}_{\phi \phi}} \right)^{1/2} \ .
    \label{eq:ut}
\end{equation}
Now we can substitute the $ u^{t}_{\textsf{HT}}$ component in Eq.~\eqref{eq:ut} into  Eq.~\eqref{eq:4-velocity} to obtain the $4$-velocity vector. 
Then, using this vector, the $\textsf{HT}$ constraints for pressure and energy density in Eq.~\eqref{eq:peHT}, and the $T^{\mu\nu}$ in Eq.~\eqref{eq:perfect_fluid}, we obtain the stress-matter energy tensor $T^{\textsf{HT}}_{\mu\nu}$ for a perfect fluid in the \textsf{HT} frame. 

Once the metric and the stress-energy tensor are defined in the \textsf{HT} frame, the Einstein equations can be expressed as a power series in $\epsilon$,
\begin{equation} \sum_{n=0}^{7} \dfrac{\epsilon^{n}}{n!} G_{\alpha}^{(n)\beta} = 8 \pi \left[ \, \sum_{n=0}^{7} \dfrac{\epsilon^{n}}{n!} T_{\alpha}^{(n)\beta} \, \right] . \label{eq:EinsteinEQ} 
\end{equation} 
By comparing terms of equal power in $\epsilon$ on both sides of Eq.~\eqref{eq:EinsteinEQ}, we obtain the perturbed Einstein equations at each spin-frequency order $n$, \begin{equation} E_{\alpha}^{(n)\beta} \equiv G_{\alpha}^{(n)\beta} - 8 \pi T_{\alpha}^{(n)\beta} = 0 \ .
\label{eq:perturbedEinstein}
\end{equation} 
In the next section we present these equations in a systematic form at each spin-frequency order up to $\mathcal{O}(\epsilon^{7})$ for each parity sector.

\section{Equations of Stellar Structure}
\label{sec:structure}

The gravitational field within a slowly-rotating star is described by the perturbed Einstein equations, as shown in Eq.~\eqref{eq:perturbedEinstein}. These equations, often referred to as the equations of stellar structure, characterize the star's equilibrium properties. One approach to deriving them involves calculating the Einstein tensor $G_{\alpha}{}^{\beta}$ and the stress-energy tensor $T_{\alpha}{}^{\beta}$ up to the desired order, while neglecting higher-order terms in $\epsilon$. However, although this method is straightforward, we found it computationally expensive. To address this, we reformulate the perturbation equations in terms of the covariant derivatives of the background solution. This approach offers two key advantages: (i) the expressions are gauge-invariant and applicable in any reference frame, and (ii) it avoids the need to compute higher-order terms beyond $\mathcal{O}(\epsilon^{7})$, thus reducing computational demands.

We used the \texttt{xAct} suite of free tensor packages to compute the perturbation equations \cite{xAct}. The process begins by setting up the general perturbation equations with \texttt{xPert} and expanding them to the seventh order in the spin-frequency parameter $\epsilon$ \cite{Brizuela:2008ra}. Next, we used \texttt{xCoba} to define the \textsf{HT} chart, which allows us to compute the Einstein equations in \textsf{HT} coordinates at each order. This method involves using the equations from the previous order to compute the next, a technique known as \textit{order reduction}. This approach is particularly useful for solving the equations numerically because, at higher orders, the equations may include higher derivatives of metric perturbation functions from previous orders. Without order reduction, these derivatives would have to be computed numerically, which can introduce numerical noise. By using order reduction, we avoid this issue and ensure that the equations to be solved numerically are set up correctly. As a consequence, due to this process, the equations at higher orders can grow very quickly, and the expressions can become quite large.

In this section, we begin by reviewing the background equations in the non-rotating limit at $\mathcal{O}(\epsilon^{0})$, i.e., the TOV equations, for consistency.
We then construct the radial equations of stellar structure that govern the equilibrium properties of a slowly-rotating neutron star up to $\mathcal{O}(\epsilon^{7})$ in the \textsf{HT} frame. These equations are derived separately for each parity sector and for different modes, as they remain uncoupled for a fixed spin-frequency order $n$. We detail the process for extracting each  $\ell$ mode within each parity sector and present the general form of these equations. 
Explicit expressions for the structure equations, provided up to third order, are included below. For higher-order expressions, the equations are available in three formats--CForm, C++ (CPP), and Mathematica InputForm--as \texttt{.txt} files. These files are included as supplementary material to this paper, while both the \texttt{.txt} files and the Mathematica scripts in \texttt{.nb} format used to generate them are available in the repository associated with this work \cite{conde2025hartlethorne}.

\subsection{TOV Equations: \texorpdfstring{$\mathcal{O}(\epsilon^{0})$}{}}

The background metric is obtained from Eq.~\eqref{eq:metric} or the metric form given in Appendix \ref{apx:HTmetric} by setting\footnote{Note that the metric transformation to the \textsf{HT} frame begin at $\mathcal{O}(\epsilon^{2})$, as shown in Eqs.~\eqref{eq:transf} and \eqref{eq:transfII}. Thus, we can obtain the metric in \textsf{HT} coordinates up to $\mathcal{O}(\epsilon)$ by making the identifications $r \rightarrow R$ and $\theta \rightarrow \Theta$ in Eq.~\eqref{eq:metric}.} 
$\epsilon=0$. The  line element at $\mathcal{O}(\epsilon^{0})$ then reads
\begin{equation}
ds^{2} = e^{\nu} dt^{2} + e^{\lambda} dR^{2} + R^{2} \left( d\Theta^{2} + \sin^{2}\Theta d\phi^{2}  \right)  \ . 
\end{equation}
Computing the $(t,t)$ and the $(R,R)$ components of the Einstein equations yields the following equations, respectively
\begin{align}
\label{eq:enclosed}
 \frac{dM}{dR} &= 4 \pi R^{2} \varepsilon \, ,  \\
 \frac{d\nu}{dR} &= 2 \frac{M + 4 \pi R^{3} p }{R(R-2M)} \, .
 \label{eq:nu}
\end{align}
In addition, combining the radial component of the equation of motion, $\nabla^{\mu}T_{\mu R}=0$ with Eq.~\eqref{eq:nu}, one obtains the Tolman-Oppenheimer-Volkov (TOV) equation,
\begin{equation}
\frac{dp}{dR} = -(\varepsilon + p) \frac{M + 4\pi R^{3} p }{R(R-2M)}   \ .
\label{eq:TOV}
\end{equation}
Equations \eqref{eq:enclosed}-\eqref{eq:TOV} are often referred to as the TOV equations. Together with a specified barotropic equation of state $p=p(\varepsilon)$, they form a closed system of ordinary differential equations for the functions $M$, $\nu$, $p$ and $\varepsilon$. The solution to these equations gives us the TOV radius of the star (i.e.~the radial coordinate at which the pressure vanishes $p(R=R_*) = 0$) and the TOV mass of the star (i.e.~the value of the enclosed mass function at the stellar radius $M(R=R_*) = M_*$).

\subsection{Even parity sector}
\label{sec:even_parity_equations}

The even-parity sector refers to metric perturbations that can be expressed in terms of even-parity spherical harmonics. In the \textsf{HT} frame, the metric perturbation components for this sector include
the $(t, t)$, $(R, R)$, $(\Theta,\Theta)$, $(\phi, \phi)$ and $(R, \Theta)$ components. For practical purposes, it is often more convenient to work with the mixed tensor components of the Einstein equations, $E_{\alpha}^{(s) \beta}$,
as these simplify the equations and make them more manageable. Then, at each even spin-frequency order $s$, we calculate the following components
\begin{equation}
 E^{(s)t}_{t} \ ,  \  E^{(s)R}_{R} \ ,  \ E^{(s)\Theta}_{\Theta} \ , \ E^{(s)\phi}_{\phi} \ , \ \text{and} \  E^{(s)\Theta}_{R} \ .
\end{equation}

By using the explicit form of the Legendre polynomials in the harmonic decomposition, each corresponding metric component can be expressed as a power series in even powers of $\cos\Theta$  as 
\begin{equation}
\label{eq:EevenPower}
    E^{(s)\beta}_{\alpha}(R,\Theta) = \sum_{q=0,2,...}^{s} \mathcal{E}^{(s)\beta}_{\alpha}(R) \cos^{q}\Theta \, ,
\end{equation}
where $\mathcal{E}^{(s)\beta}_{\alpha}$ are functions that depend only on the radial coordinate $R$ and are formed from combinations of the radial part of the metric perturbation expansions. Since the power $q$ in Eq.~\eqref{eq:EevenPower} is even, the factors of $\cos^{q}\Theta$ can be expanded in Legendre polynomials $P_{\ell}(\cos\Theta)$, where $\ell$ is an even integer. This is accomplished using the following expression \cite{abramowitz1968}
\begin{equation}
\label{eq:powers_cos}
 \cos^{q}\Theta = \sum_{\ell =q, q-2, ...}\dfrac{(2\ell + 1)q! \, P_{\ell}(\cos\Theta)}{2^{\frac{s-\ell}{2}}  \left[ \frac{1}{2}  (q-\ell)\right]!(\ell + q + 1)!} \, .
\end{equation}
Thus, using Eq.~\eqref{eq:powers_cos} in Eq.~\eqref{eq:EevenPower}, the components $E_{\alpha}^{(s)\beta}$ can be written in terms of the Legendre polynomials basis. This allows us to extract every even $\ell$ mode of the perturbed Einstein equations by performing the following operation
\begin{equation}
\label{eq:extracteven}
    \int_{0}^{\pi} P_{\ell}(\cos\Theta) E^{(s) \beta}_{\alpha} (R,\Theta)\sin\Theta d\Theta = 0
\end{equation}
and using the orthogonality relation between Legendre polynomials, given by
\begin{equation}
    \int_{0}^{\pi} P_{\ell}(\cos\Theta)P_{\ell'}(\cos\Theta) \sin\Theta d\Theta = \dfrac{2}{2\ell + 1}\delta_{\ell \ell'} \ .
\end{equation}

The general structure of the perturbed Einstein equations for the even-parity sector can be divided into two distinct sets. The first set, corresponding to the $\ell = 0$ mode, describes the corrections to the mass monopole, the total baryon number, and the binding energy of the source distribution. The second set, which includes up to $\ell = 2, 4, 6$ modes, describes the higher mass-multipole moments of the source distribution and their corrections\footnote{For a full discussion of the extraction of such quantities at $\mathcal{O}(\epsilon^{2})$, please refer to reference \cite{Hartle:1967he}.}. Each set is presented and described in more detail below.

\subsubsection*{Equations of structure: $\ell=0$}

The $\ell = 0$ monopole mode of the even-parity sector is described by a set of radial, first-order, coupled, and inhomogeneous ordinary differential equations for the metric perturbation functions $m^{(s)}_{0}$, $h^{(s)}_{0}$, and $\xi^{(s)}_{0}$ at each even spin-frequency order $s$. Extracting the $\ell=0$ piece using Eq.~\eqref{eq:extracteven} for  $E^{(s) t}_{t}$ and $E^{(s) R}_{R}$ gives differential equations for $m^{(s)}_{0}$ and $h^{(s)}_{0}$, respectively. On the other hand, inserting the equation for $h^{(s)}_{0}$ into the radial component of the Bianchi identities, given by the expression
\begin{equation}
\label{eq:Bianchi}
    \left[ \nabla^{\alpha} T^{R}{}_{\alpha} \right]^{(s)} = 0 \, ,
\end{equation}
gives an equation for $\xi^{(s)}_{0}$. The superscript index $(s)$ notation in Eq.~\eqref{eq:Bianchi} means that the Bianchi identities are computed up to the specified spin-frequency order $s$, where $s$ could take the values $s=2,4$ and $6$. We follow the same approach described at the beginning of this section: we first perform the perturbations of the Bianchi identities in a general frame using \texttt{xPert}, and then use \texttt{xCoba} to express the results in the \textsf{HT} chart.

The extraction of the $\ell=0$ piece from Eq.~\eqref{eq:Bianchi} is obtained following the same procedure previously described for $E^{(s)\beta}_{\alpha}$. We express Eq.~\eqref{eq:Bianchi} in terms of even powers of $\cos\Theta$ and use Eq.~\eqref{eq:powers_cos} to collect terms in the Legendre polynomials basis $P_{\ell}$ with $\ell$ being even. Then, similarly as done in Eq.~\eqref{eq:extracteven}, we perform the operation
\allowdisplaybreaks[4]
\begin{equation}
\label{eq:modeBianchi}
\int_{0}^{\pi} P_{\ell}(\cos\Theta) \left[ \nabla^{\alpha} T^{\beta}{}_{\alpha} \right]^{(s)}(R,\Theta) \sin\Theta d\Theta = 0 \, ,
\end{equation}
setting $\beta=R$ to extract the $R$ component. 
The system of equations for the functions 
$m^{(s)}_{0}$, $h^{(s)}_{0}$ and $\xi^{(s)}_{0}$ with $s=2,4,6$ and $\ell=0$ are given by
\begin{widetext}
\begin{align}
\label{eq:m0}
\dfrac{dm^{(s)}_{0}}{dR} &=  \, - 4 \pi R^{2} \dfrac{d\varepsilon}{dR}\xi^{(s)}_{0} \, + \, S_{m^{(s)}_{0}} \ , \\[1ex]
\label{eq:h0}
\dfrac{dh^{(s)}_{0}}{dR} &=  \, \dfrac{1 + 8 \pi p R^{2}}{R^{2}}e^{2 \lambda} m^{(s)}_{0} \, + \,  \dfrac{4 \pi R (M + 4 \pi p R^{3})(p +\varepsilon)}{R^{2}}e^{2\lambda} \xi^{(s)}_{0} \, + \,  S_{h^{(s)}_{0}} \, , \\[1ex] 
\dfrac{d\xi^{(s)}_{0}}{dR}  &=  \,  \dfrac{2[M(R+8\pi p R^{3}) - M^{2} - 2 \pi R^{4} (p + \varepsilon + 8 \pi p R^{2} \varepsilon )]}{R^{2} (M+ 4\pi p R^{3} )}e^{\lambda} \xi^{(s)}_{0}  \, - \,   \dfrac{1+8 \pi p R^{2}}{M + 4 \pi p R^{3} }e^{\lambda} m^{(s)}_{0} \,  + \, S_{\xi^{(s)}_{0}} \, .
\label{eq:xi0}
\end{align}
\end{widetext}
For a fixed spin-frequency order $s$, these equations can be solved from the center of the star outward to its surface using an equation of state $p=p(\varepsilon)$, along with the metric perturbation solutions from the equations at previous orders.
The following table shows the perturbation functions needed for solving the $\ell=0$ equations for the even-parity sector.
\begin{table}[htb]
\renewcommand{\arraystretch}{2} 
\setlength{\tabcolsep}{5pt}      
\begin{tabular}{c|c|c|c|}
\cline{2-4}
\multicolumn{1}{l|}{}     & $s=2$                                         & $s=4$                                         & $s=6$                                         \\ \hline
\multicolumn{1}{|c|}{$\ell=0$} & \begin{tabular}[c]{@{}c@{}}$m^{(2)}_{0}, \, h^{(2)}_{0}, \, \xi^{(2)}_{0}$\end{tabular} & \begin{tabular}[c]{@{}c@{}}$m^{(4)}_{0}, \, h^{(4)}_{0}, \, \xi^{(4)}_{0}$\end{tabular} & \begin{tabular}[c]{@{}c@{}}$m^{(6)}_{0}, \, h^{(6)}_{0}, \, \xi^{(6)}_{0}$\end{tabular} \\ \hline
\end{tabular}
\caption{Mode perturbation functions for $\ell=0$ and spin-frequency orders $s=2,4,6$. These functions are associated to the mass monopole corrections of the star. (see Sec.~\ref{sec:exterior} and \ref{sec:multipoles}).}
\label{table:evenfun0}
\end{table}

The symbols introduced in Eqs.~\eqref{eq:m0}--\eqref{eq:xi0}, $S_{m^{(s)}_{0}}$, $S_{h^{(s)}_{0}}$ and $S_{\xi^{(s)}_{0}}$, are source functions that depend only on the radial coordinate $R$ and involve non-linear combinations of perturbation functions from spin-frequency orders lower than $s$. Once the spin-frequency order $s$ is fixed, such combinations are constructed from terms expressed as products of lower-order metric functions, resulting in terms of order $s$. For example, the source terms at $s=2$ are,
 \begin{align}   
 \nonumber
 \label{eq:Sm20}
S_{m^{(2)}_{0}} &= \frac{1}{12} R^3 e^{-\nu} \bigg\{32 \pi  R \left( \varpi^{(1)}_{1} \right)^{2} (p +\varepsilon) \\ 
&+ \left( R-2 M \right)   \left(\bar{\alpha}^{(1)}_{1} \right)^{2}\bigg\} \, ,\\
 \label{eq:Sh20}
S_{h^{(2)}_{0}} &= -\frac{1}{12} R^3 e^{-\nu} \left(\bar{\alpha}^{(1)}_{1} \right)^{2}\,  , \\ \nonumber
\label{eq:Sxi20}
S_{\xi^{(2)}_{0}} &= \dfrac{R^2 e^{-\nu} }{12 \left(M +4 \pi  R^3 p\right)} \bigg\{ R^{2} (R-2 M) \left( \bar{\alpha}^{(1)}_{1} \right)^{2} -8 \left( \varpi^{(1)}_{1} \right)^{2}  \\  
& \times \left(3 M + 4 \pi  R^3 p - R \right) 
+ 8 R (R-2 M) \varpi^{(1)}_{1} \bar{\alpha}^{(1)}_{1} \bigg\} \, , 
\end{align}  
where the quantities $\varpi^{(1)}_{1}$ and $\bar{\alpha}^{(1)}_{1}$ are related to $\omega^{(1)}_{1}$ via $\varpi^{(1)}_{1} \equiv \Omega - \omega^{(1)}_{1}$ and $\bar{\alpha}^{(1)}_{1} =d\omega^{(1)}_{1}/dR$. For more details on working with the function $\varpi_{1}^{(1)}$, please refer to Appendix~\ref{apx:unique}.

\subsubsection*{Equations of structure: $\ell=2,4$ and $6$}

For $\ell = 2,4$ and $6$ harmonics, a set of radial, first-order and coupled, inhomogeneous, ordinary differential equations can be constructed for the metric perturbation functions $k^{(s)}_{\ell}$ and $h^{(s)}_{\ell}$. These equations, along with two algebraic equations for $m^{(s)}_{\ell}$ and $\xi^{(s)}_{\ell}$, form the entire set of equations describing the shape of the star through the mass multipole moments of the source distribution.
The leading-order terms of the quadrupole, hexadecapole, and hexacontatetrapole moments can be extracted at the spin-frequency orders $s=2,4$ and $6$ for the modes $\ell=2,4$, and $6$, respectively. Additionally, the first- and second-order corrections to the quadrupole moment for the $\ell=2$ mode enter at $s=4$ and $s=6$ respectively, while the first-order correction to the hexadecapole moment for the $\ell=4$ mode enters at $s=6$. For more details, please refer to Sec.~\ref{sec:multipoles}.

The equation for $k^{(s)}_{\ell}$ is derived from the $(R,\Theta)$ component of the perturbed Einstein equations, $E^{(s) \Theta}_{R}=0$, while 
$h^{(s)}_{\ell}$  is obtained from the $(R,R)$ component,
$E^{(s) R}_{R}=0$. To isolate a specific even 
$\ell$ mode, we apply the operation defined in Eq.~\eqref{eq:extracteven}. Once a fixed $\ell$ mode is extracted, it does not couple to other modes of the same spin-frequency order, forming independent sets of differential equations. We present the general form of these equations in compact notation as follows:
\begin{widetext}
    \begin{align}
    \label{eq:keven}
             \dfrac{dk^{(s)}_{\ell}}{dR} \, &= \,  -  \dfrac{dh^{(s)}_{\ell}}{dR} \, + \, \dfrac{R-3M-4\pi p R^{3}}{R^{2}}e^{\lambda}h^{(s)}_{\ell} \, + \, \dfrac{R-M+4p R^{3}}{R^{3}}e^{2\lambda}m^{(s)}_{ \ell} \, + \, S_{k^{(s)}_{\ell}} \ , \\[1ex]  \nonumber
              \dfrac{dh^{(s)}_{\ell}}{dR} \, &= \,  -  \frac{R-M+ 4\pi p R^{3}}{R} e^{\lambda} \dfrac{dk^{(s)}_{\ell}}{dR} \, + \, \left[ \frac{\ell(\ell+1)}{2R} - 4\pi(\varepsilon + p)R \right]e^{\lambda} h^{(s)}_{\ell}  \, + \, \frac{(\ell+2)(\ell-1)}{2R} e^{\lambda} k^{(s)}_{\ell}  \\[1ex] \ 
              &
              \, + \, \frac{1+8\pi p R^{2}}{R^{2}} e^{2\lambda} m^{(s)}_{\ell} \, + \,  S_{h^{(s)}_{\ell}} \ .
     \label{eq:heven}
    \end{align}
\end{widetext}

The structure of the differential operators acting on the functions $k^{(s)}_{\ell}$ and $h^{(s)}_{\ell}$ in Eqs.~\eqref{eq:keven} and \eqref{eq:heven} has the same form as the corresponding equations for $s=2$ and $\ell=2$ presented in~\cite{Yagi:2013awa}. 

The algebraic equation for $m^{(s)}_{\ell}$ and $\xi^{(s)}_{\ell}$ are obtained as follows. The former is obtained from the $(\Theta, \Theta)$ and the $(\phi,\phi)$ components of the perturbed Einstein equations. By taking the difference $E^{(s)\Theta}_{\Theta}-E^{(s)\phi}_{\phi}= 0$
we can isolate the function $m^{(s)}_{\ell}$. Extracting a particular $\ell$ mode follows the same approach as described in Eq.~\eqref{eq:extracteven}. The general form is given by the following expression:
\begin{equation}
\label{eq:meven}
     m^{(s)}_{\ell} \, = \,   - (R - 2M) h^{(s)}_{\ell} + S_{m^{(s)}_{\ell}} \ .
\end{equation}
The other algebraic equation, the one for the functions $\xi^{(s)}_{\ell}$, can be obtained from the $\Theta$ component of the Bianchi identities,
\begin{equation}
    \left[ \nabla^{\mu} T_{\mu}{}^{\Theta}\right]^{(s)} = 0 \ .
\end{equation}
Setting $\beta=\Theta$ in Eq.~\eqref{eq:modeBianchi}, we can extract any even $\ell$ mode. The general structure of the algebraic equation for the functions $\xi^{(s)}_{\ell}$ can be written as
\begin{equation}
\label{eq:xieven}
    \xi^{(s)}_{\ell} \, =  \,  - \dfrac{R(R-2M)}{M + 4 \pi p R^{3}}h^{(s)}_{\ell} + S_{\xi^{(s)}_{\ell}} \ .
\end{equation}

As introduced in Eqs.~\eqref{eq:m0}--\eqref{eq:xi0}, we define the quantities $S_{k^{(s)}_{\ell}}$, $S_{h^{(s)}_{\ell}}$, $S_{m^{(s)}_{\ell}}$ and $S_{\xi^{(s)}_{\ell}}$
in Eqs.~\eqref{eq:keven}, \eqref{eq:heven}, \eqref{eq:meven}, and \eqref{eq:xieven} as functions of the $\textsf{HT}$ coordinate $R$. These functions depend on non-linear combinations of terms from lower spin-frequency orders, contributing terms of fixed order $s$. For instance, for $s=2$ these functions are
\begin{align}
S_{k^{(2)}_{2}} &= 0 \, , \\
S_{h^{(2)}_{2}} &= \dfrac{R^{3}}{12e^{\nu}} \left( \bar{\alpha}^{(1)}_{1} \right)^{2} - \dfrac{4\pi (\varepsilon + p)R^{4}}{3R e^{\nu-\lambda}} \left( \varpi^{(1)}_{1} \right)^{2}  \, , \\ \nonumber
S_{m^{(2)}_{2}} &= \dfrac{1}{6} R^{4} e^{-(\nu + \lambda)} \bigg[ Re^{-\lambda} \left( \bar{\alpha}^{(1)}_{1} \right)^{2} \\
&+ 16 \pi R \left( \varpi^{(1)}_{1} \right)^{2} (\varepsilon + p) \bigg] \, , \\
S_{\xi^{(2)}_{2}} &=  - \dfrac{R^{4} e^{-(\lambda+\nu)}}{3(M+4\pi p R^{3})} \left( \varpi^{(1)}_{1} \right)^{2} \, .
\end{align}
These source functions at a given spin order $s$ can then be calculated explicitly once the $s-1$ order solution has been found.

Table \ref{table:evenfun} summarizes all mode perturbation functions that can be obtained at a fixed spin-frequency order $s$ and mode $\ell$ with $\ell \neq 0$ by solving Eqs.~\eqref{eq:keven} and \eqref{eq:heven}. 
Note, however, that the system of equations in~\eqref{eq:keven} and \eqref{eq:heven} is not in a convenient form for numerical implementation, as the derivatives of both \(k^{(s)}_{\ell}\) and \(h^{(s)}_{\ell}\) appear on the right-hand side of both equations. By substituting the expression for \(m^{(s)}_{\ell}\) from Eq.~\eqref{eq:meven} into Eqs.~\eqref{eq:keven} and \eqref{eq:heven} and solving for the derivatives of \(k^{(s)}_{\ell}\) and \(h^{(s)}_{\ell}\), we can transform the system into two first-order, linear differential equations for \(k^{(s)}_{\ell}\) and \(h^{(s)}_{\ell}\). We do not present the explicit expressions for these equations here, as they are lengthy and not particularly illuminating. However, they are available in the repository associated with this work \cite{conde2025hartlethorne}, both as Mathematica scripts and as \texttt{.txt} files in CForm, C++ (CPP), and Mathematica InputForm formats. The \texttt{.txt} files are also included as supplementary material to this paper.
\begin{table}[htb]
\renewcommand{\arraystretch}{2} 
\setlength{\tabcolsep}{10pt}      
\begin{tabular}{c|c|c|c|}
\cline{2-4}
\multicolumn{1}{l|}{}     & $\ell=2$                                           & $\ell=4$                                           & $\ell=6$                                           \\ \hline
\multicolumn{1}{|c|}{$s=2$} & \begin{tabular}[c]{@{}c@{}}$k^{(2)}_{2}, \, h^{(2)}_{2}$\\[-1ex] $m^{(2)}_{2}$, \,$\xi^{(2)}_{2}$\end{tabular} & \begin{tabular}[c]{@{}c@{}}---\end{tabular} & \begin{tabular}[c]{@{}c@{}}---\end{tabular} \\ \hline
\multicolumn{1}{|c|}{$s=4$} & \begin{tabular}[c]{@{}c@{}}$k^{(4)}_{2}, \, h^{(4)}_{2}$\\[-1ex] $m^{(4)}_{2}$, \,$\xi^{(4)}_{2}$\end{tabular}                                             & \begin{tabular}[c]{@{}c@{}}$k^{(4)}_{4}, \, h^{(4)}_{4}$\\[-1ex] $m^{(4)}_{4}$, \,$\xi^{(4)}_{4}$\end{tabular} & --- \\ \hline
\multicolumn{1}{|c|}{$s=6$} & \begin{tabular}[c]{@{}c@{}}$k^{(6)}_{2}, \, h^{(6)}_{2}$\\[-1ex] $m^{(6)}_{2}$, \,$\xi^{(6)}_{2}$\end{tabular}                                             & \begin{tabular}[c]{@{}c@{}}$k^{(6)}_{4}, \, h^{(6)}_{4}$\\[-1ex] $m^{(6)}_{4}$, \,$\xi^{(6)}_{4}$\end{tabular}                                             & \begin{tabular}[c]{@{}c@{}}$k^{(6)}_{6}, \, h^{(6)}_{6}$\\[-1ex] $m^{(6)}_{6}$, \,$\xi^{(6)}_{6}$\end{tabular} \\ \hline
\end{tabular}
\caption{Mode perturbation functions describing the equations of structure for spin-frequency orders $s=2,4$ and $6$. These functions are associated to the mass multipole moments of the star and their corrections (see Sec.~\ref{sec:exterior} and \ref{sec:multipoles}).}
\label{table:evenfun}
\end{table}

\subsection{Odd parity sector}
\label{sec:odd_parity_equations}

The odd-parity sector refers to the metric perturbations that can be expressed using odd-parity spherical harmonics. In the \textsf{HT} frame, the only non-zero component of the metric perturbation that is of odd parity is the $(t, \phi)$ component, which contributes only at odd spin-frequency orders $k$. As a result, the structure equations for this sector can be derived from the $(t, \phi)$ component of the perturbed Einstein equations, $E^{(k)\phi}_{t}$. These equations are then expressed in terms of vector spherical harmonics, $X^{\ell0}{\phi}$, or equivalently, $P'_{\ell}(\cos\Theta)$, where the prime indicates differentiation with respect to $\cos\Theta$. 
From these equations, we can derive a set of second-order differential equations for the metric perturbation functions $\omega^{(k)}_{\ell}$ corresponding to each odd $\ell$ mode.

The strategy for extracting each mode is similar to the approach used for the even-sector equations. First, we explicitly apply the harmonic decomposition and collect terms proportional to even powers of $\cos\Theta$, as shown in Eq.~\eqref{eq:EevenPower}. We then rewrite these even powers as linear combinations of $P'_{\ell}(\cos\Theta)$ by differentiating Eq.~\eqref{eq:powers_cos} with respect to $\cos\Theta$, i.e., 
\begin{equation}
 k\cos^{k-1}\Theta = \sum_{\ell =k, k-2, ...}\dfrac{(2\ell + 1)k! \, P_{\ell}'(\cos\Theta)}{2^{\frac{k-\ell}{2}}  \left[ \frac{1}{2}  (k-\ell)\right]!(\ell + k + 1)!!} 
\end{equation}
where $k$ takes only odd integer values. In this way, the $(t,\phi)$ component of the perturbed Einstein equations is written in the odd-parity vector spherical harmonic basis, as this component transforms as a vector under rotations.

To extract a particular odd $\ell$ mode we can use the vector spherical harmonic orthogonality relation
\begin{equation}
\label{eq:orthogonalX}
    \int \bar{X}^{A}_{\ell m} X_{A}^{\ell m} d \Omega = \ell (\ell + 1) \delta_{\ell \ell'} \delta_{m m'}
\end{equation}
where the overhead bar denotes complex conjugation, and $d\Omega := \sin\Theta d\Theta d\phi$ is the element of solid angle\footnote{This line element should not be confused with the metric tensor $\Omega_{AB}$ on $\mathcal{S}^{2}$.}. 
Equation \eqref{eq:orthogonalX} can be further simplifed using $X^{A}_{\ell m} = \Omega^{AB}X^{\ell m}_{B}$ and applying azimuthal symmetry, i.e., $m=0$. The expression in Eq.~\eqref{eq:orthogonalX} simplifies to
\begin{equation}
    \int_{0}^{\pi} \dfrac{dP_{\ell}(\cos\Theta)}{d\Theta} \dfrac{dP_{\ell'}(\cos\Theta)}{d\Theta} \sin\Theta d\Theta = \dfrac{2 \ell(\ell+1)}{2\ell + 1}\delta_{\ell \ell'} \, .
\end{equation}
Therefore, in order to obtain any odd $\ell$ mode equation from the $(t,\phi)$ component of the perturbed Einstein equations, we perform the following operation 
\begin{equation}
    \int_{0}^{\pi} \dfrac{dP_{\ell}(\cos\Theta)}{d\Theta} \sin\Theta E^{(k)\phi}_{t}(R,\Theta) d\Theta = 0 \, .
\end{equation}
This procedure yields a set of radial, second-order, inhomogeneous differential equations for the functions $\omega^{(k)}_{\ell}$ which can be expressed in a compact form as
\begin{align}
\label{eq:wodd}
\nonumber
&\frac{d^{2}\omega^{(k)}_{\ell}}{dR^{2}} \, + \, 4 \frac{1-\pi R^{2}(\varepsilon + p)e^{\lambda}}{R} \frac{d\omega^{(k)}_{\ell}}{dR} \\
&\, - \,  \left[ \dfrac{(\ell+2)(\ell-1)}{R^{2}} + 16\pi (\varepsilon + p) \right] e^{\lambda} \omega^{(k)}_{\ell} = S_{\omega^{(k)}_{\ell}}
\end{align} 
where $\ell$ and $k$ are odd integers with $\ell \leq k$. 

The source functions $S_{\omega^{(k)}_{\ell}}$ are radial functions that depend on non-linear combinations of perturbation functions from lower spin-frequency orders $n < k$. 
In particular, at spin-frequency orders $k=1$ and $k=3$, the source terms are
\begin{align}
\label{eq:S11}
S_{\omega^{(1)}_{1}} &= 16 \pi   \Omega  (\varepsilon + p)e^{\lambda} \, , \\ 
S_{\omega^{(3)}_{1}} &= D^{(3)}_{0} - \dfrac{1}{5} D^{(3)}_{2} \, , \\
S_{\omega^{(3)}_{3}} &= \dfrac{1}{5}D^{(3)}_{2} \, , 
\end{align}
where we have defined
\begin{align}
D^{(3)}_{0} &= - \bar{\alpha}^{(1)}_{1} j \dfrac{d}{dR} \left( \dfrac{m^{(2)}_{0}}{R-2M} + h^{(2)}_{0} \right) \\
&+ \dfrac{4}{R} j \varpi^{(1)}_{1} \bigg[ \dfrac{2m^{(2)}_{0}}{R-2M} + \left( \dfrac{1}{dp/d\varepsilon} + 1 \right) \delta p^{(2)}_{0} \\ 
& +\dfrac{2}{3} e^{-\nu} R^{2} \left( \varpi^{(1)}_{1} \right)^{2} \bigg] \\ \, 
D^{(3)}_{2} &= \bar{\alpha}^{(1)}_{1} j \dfrac{d}{dR}\left( 4 k^{(2)}_{2} - \dfrac{m^{(2)}_{2}}{R-2M} - h^{(2)}_{2} \right) \\
&+ \dfrac{4}{R} \dfrac{dj}{dR} \varpi^{(1)}_{1}  \bigg[ \dfrac{2m^{(2)}_{0}}{R-2M} + \left( \dfrac{1}{dp/d\varepsilon} + 1 \right) \delta p^{(2)}_{2} \\ 
& +\dfrac{2}{3} e^{-\nu} R^{2} \left( \varpi^{(1)}_{1} \right)^{2} \bigg]
\end{align}
with $j = e^{-(\nu + \lambda)/2}$ and
\begin{align}
\delta p^{(2)}_{0} = - \dfrac{\xi^{(2)}_{0}}{\varepsilon + p} \dfrac{dp}{d\varepsilon} \dfrac{d\varepsilon}{dR} \ \ , \ \ \delta p^{(2)}_{2} = - \dfrac{\xi^{(2)}_{2}}{\varepsilon + p} \dfrac{dp}{d\varepsilon} \dfrac{d\varepsilon}{dR} \, .
\end{align}
The source term at linear order in spin and for $\ell = 1$ is given in Eq.~\eqref{eq:S11}. 

Equation~\eqref{eq:wodd} is not in an ideal form for numerical implementation yet. First, as before, we transform this equation to the quantity $\varpi^{(1)}_{1} \equiv \Omega - \omega^{(1)}_{1}$, as this transformation makes Eq.~\eqref{eq:wodd} a homogeneous ordinary differential equation for the function $\varpi^{(1)}_{1}$ (see also App.~\ref{apx:unique}). Moreover, while the source terms at $k=3$ are written in a compact form, they are not suitable for numerical implementation. To make them applicable in practice, they must be rewritten using \textit{order reduction} from the previous spin-frequency orders, i.e., $k=1$ and $s=2$, to avoid taking derivatives of the numerical solutions from earlier orders. With these modifications at hand, one can solve Eq.~\eqref{eq:wodd} for all the possible functions of the odd-parity sector up to $\mathcal{O}(\epsilon^{7})$, as shown in Table~\ref{table:oddfun}.
\begin{table}[htb]
\renewcommand{\arraystretch}{2} 
\setlength{\tabcolsep}{5.5pt}      
\begin{tabular}{c|c|c|c|c|}
\cline{2-5}
\multicolumn{1}{l|}{}     & $\ell=1$                                           & $\ell=3$                                           & $\ell=5$                                           & $\ell=7$                                           \\ \hline
\multicolumn{1}{|c|}{$k=1$} & \begin{tabular}[c]{@{}c@{}}$\omega^{(1)}_{1}, \, \alpha^{(1)}_{1}$\end{tabular} & \begin{tabular}[c]{@{}c@{}}---\end{tabular} & \begin{tabular}[c]{@{}c@{}}---\end{tabular} & \begin{tabular}[c]{@{}c@{}}---\end{tabular} \\ \hline
\multicolumn{1}{|c|}{$k=3$} & \begin{tabular}[c]{@{}c@{}}$\omega^{(3)}_{1}, \, \alpha^{(3)}_{1}$\end{tabular}                                             & \begin{tabular}[c]{@{}c@{}}$\omega^{(3)}_{1}, \, \alpha^{(3)}_{1}$\end{tabular} & --- & --- \\ \hline
\multicolumn{1}{|c|}{$k=5$} & \begin{tabular}[c]{@{}c@{}}$\omega^{(5)}_{1}, \, \alpha^{(5)}_{1}$\end{tabular}                                             & \begin{tabular}[c]{@{}c@{}}$\omega^{(5)}_{3}, \, \alpha^{(5)}_{3}$\end{tabular}                                             & \begin{tabular}[c]{@{}c@{}}$\omega^{(5)}_{5}, \, \alpha^{(5)}_{5}$\end{tabular} & --- \\ \hline
\multicolumn{1}{|c|}{$k=7$} & \begin{tabular}[c]{@{}c@{}}$\omega^{(7)}_{1}, \, \alpha^{(7)}_{1}$\end{tabular} & \begin{tabular}[c]{@{}c@{}}$\omega^{(7)}_{3}, \, \alpha^{(7)}_{3}$\end{tabular} & \begin{tabular}[c]{@{}c@{}}$\omega^{(7)}_{5}, \, \alpha^{(7)}_{5}$\end{tabular} & \begin{tabular}[c]{@{}c@{}}$\omega^{(7)}_{7}, \, \alpha^{(7)}_{7}$\end{tabular} \\ \hline
\end{tabular}
\caption{Mode perturbation functions describing the equations of structure for spin-frequency orders $k=1,3,5$ and $7$, where $\alpha^{(k)}_{\ell} \equiv d \omega^{(k)}_{\ell} /dR$. These functions are associated to the mass-current multipole moments of the star and their corrections (see Sec.~\ref{sec:exterior} and \ref{sec:multipoles}).
}
\label{table:oddfun}
\end{table}

Unlike for the even-parity sector, the $4$-vector $\nabla_{\alpha}T^{\alpha \beta}$ vanishes for the odd-parity sector and no additional constraints can be obtained. To see this, we split the Bianchi identities following \cite{Martel:2005ir} as
\begin{align}
\label{eq:bianchiI}
    \tilde{\nabla}_{b}T^{ab} + D_{B}T^{aB} + \dfrac{2}{r}r_{b}T^{ab} -r \, r^{a}\Omega_{AB} T^{AB} = 0 \\[1ex]
\label{eq:bianchiII}    
    \tilde{\nabla}_{a}T^{aA} + D_{B}T^{AB} + \dfrac{4}{r} \, r_{a}T^{aA} = 0
\end{align}
where $\tilde{\nabla}$ is the covariant derivative on the 2-dimensional manifold $\mathcal{M}^{2}$ that corresponds to the ``$t-R$'' plane, and $r^{a}:=\partial_{a}r$ is a vector pointing orthogonal to the $r = {\rm{const}}$ hypersurface. Since the only nonzero component of the stress-energy tensor for the odd-parity sector is the $(t,\phi)$ component, Eqs.~\eqref{eq:bianchiI} and \eqref{eq:bianchiII} reduce to 
\begin{align}
 \label{eq:bI}
     D_{\phi}T^{t\phi} = 0 \, , \\[1ex]
\label{eq:bII}    
    \tilde{\nabla}_{t}T^{t\phi} = 0  \, . 
\end{align}
Using the fact that the source is stationary and axisymmetric, the left-hand sides of Eqs.~\eqref{eq:bI} and \eqref{eq:bII} vanish, as the connections $\Gamma^{B}_{B\phi}=0$ on $\mathcal{S}^{2}$ and $\Gamma^{t}_{tt}=0$ on $\mathcal{M}^{2}$.

The full set of stellar structure equations for both parity sectors forms a closed system for all metric perturbation functions. As shown in Tables \ref{table:evenfun0}, \ref{table:evenfun}, and \ref{table:oddfun}, there are 33 equations for the even sector and 20 for the odd sector up to the spin order we consider here. Together with the 4 TOV equations \cite{Tolman:1939jz,Oppenheimer:1939ne},
this results in a system of 57 equations for 57 unknown metric functions. Of these, only 12 are algebraic equations for the metric functions $m^{(s)}_{\ell}$ and $\xi^{(2)}_{\ell}$ with $\ell \neq 0$. The remaining 45 equations are ordinary differential equations that must be solved within the star’s interior to determine the gravitational field and extract physical quantities as explained in Sec.~\ref{sec:multipoles}. The next section addresses the boundary conditions necessary to solve the entire system of equations.

\section{Local Analysis of equations of stellar structure}
\label{sec:local}

In the previous section, we constructed the complete set of perturbation equations at each spin-frequency order for all unknown mode functions in the metric. To obtain a numerical solution for a single neutron star up to $\mathcal{O}(\epsilon^{7})$, these equations must be solved sequentially—starting with the background (TOV) equations and then solving the perturbed Einstein equations order by order. This process requires specifying boundary conditions at the star’s center, $R = 0$, and integrating outward to the surface at $R = R_{*}$. However, all differential equations governing the stellar geometry, including both the background and perturbed systems, diverge at the origin. To address this, we perform a local analysis around $R = 0$ to derive regular boundary conditions. This section presents the necessary central boundary conditions for solving each differential equation.

We begin by performing a local analysis of the background equations, i.e., at order \(\mathcal{O}(\epsilon^{0})\). Let \(x\) denote any of the background quantities: the enclosed mass \(M(R)\), the metric function \(\nu(R)\), the pressure \(p(R)\), or the total energy density \(\varepsilon(R)\). Assuming regularity of these functions at the origin, each can be expanded as a power series in the radial coordinate \(R\):
\begin{equation}
x(R) = x_{c} + \sum_{j=1}^{\infty} x_{j} R^{j},    
\label{eq:asymX}
\end{equation}
where \(x_{c} = x(R=0)\) denotes the central value of the quantity \(x\). We refer to \(x_{c}\) and the coefficients \(x_{j}\) as the \textit{local asymptotic constants of the background functions}.
Substituting the expansion in Eq.~\eqref{eq:asymX} into the TOV equations—namely, Eqs.~\eqref{eq:enclosed}--\eqref{eq:TOV}—along with the equation of state \(p = p(\varepsilon)\), and matching terms order-by-order in powers of \(R\), yields a set of algebraic relations among the asymptotic constants \(x_{j}\). These can be expressed in terms of the central values \(x_{c}\), which are chosen initially to solve the TOV equations.

In practice, the most convenient way to carry out this procedure is to first rewrite each differential equation so that both sides are free of fractions involving metric functions. For example, consider the TOV equation, given in Eq.~\eqref{eq:TOV}. Its right-hand side contains a fraction, with the metric function \( M \) in the denominator. By multiplying both sides by \( R(R - 2M) \)—effectively clearing the denominator—we avoid the need to expand \( M \) within the denominator of a fractional expression; although this manipulation is obvious at the background level, we find that the systematization of this procedure helps when going to higher order. Once the equation is free of such terms, we substitute the power series expansions of the relevant metric functions and match coefficients of like powers of \( R \) on both sides. This yields a system of algebraic equations that can be solved to determine the asymptotic constants $x_{j}$ and characterize the behavior near the origin.

At zeroth order in spin-frequency, the local solutions for the background metric functions up to $\mathcal{O}(R^{3})$ are
\begin{align}
%
%
\label{eq:Msol}
M(R) &= \frac{4\pi}{3}\varepsilon_{c} R^{3} + \mathcal{O}(R^{5}), \\
\label{eq:psol}
p(R) &= p_c  -  \frac{2\pi}{3}    \left(p_c+\varepsilon _c\right) \left(3 p_c+\varepsilon _c\right) R^{2} +\mathcal{O}(R^{4}), \\
\label{eq:esol}
\varepsilon(R) &= \varepsilon _c  -  \frac{2\pi}{3 p'_{c}} \left(p_c+\varepsilon _c\right)\left(3p_c+\varepsilon _c\right) R^{2} + \mathcal{O}(R^{4}), \\
\label{eq:nusol}
\nu(R) &= \nu_c  +  \frac{4\pi}{3} \left(3 p_c+\varepsilon _c\right)R^{2} + \mathcal{O}(R^{4}) . 
\end{align}
Here, the central asymptotic constants $\varepsilon_{c}$, $p_{c}$ and $\nu_{c}$, represents the total energy density, pressure and the metric function $\nu$ evaluated at the center of the star, respectively, and $p'_{c} \equiv ( dp/d\varepsilon )_{\varepsilon = \varepsilon_{c}} $. In particular, note that $M_{c}$, $M_{1}$, and $M_{2}$ vanish at the center, and the asymptotic constants $p_{2}$, $\varepsilon_{2}$ and $\nu_{2}$ are written in terms of the central asymptotic constants $\varepsilon_{c}$ and $p_{c}$. 

As alluded to before, the procedure of removing fractions is not strictly necessary, but it is highly convenient. Even if the original equations contained fractions with metric functions in the denominator, one could still substitute the power series expansions directly into these expressions. In such cases, the entire differential equation can be re-expanded in a Taylor series about $R=0$. However, since the equations at higher spin-frequency orders often contain large and complicated source terms, performing a full Taylor expansion in the presence of such fractions is generically very computationally expensive. Therefore, clearing denominators beforehand significantly simplifies both the algebraic structure of the equations and their implementation in symbolic computations.

The asymptotic solution to the perturbed equations can be constructed systematically by using the solution at each lower order as input for the next. Let \( y^{(n)}_{\ell} \) denote any metric perturbation function at order \( n \) and mode $\ell$, such as \( \omega^{(k)}_{\ell} \), \( h^{(s)}_{\ell} \), \( k^{(s)}_{\ell} \), \( m^{(s)}_{\ell} \), or \( \xi^{(s)}_{\ell} \). Each of these functions can be expanded as a power series in the radial coordinate \( R \) about the center of the star as
\begin{equation}
y^{(n)}_{\ell}(R) = y^{(n)}_{\ell,c} + \sum_{j=1}^{\infty} y^{(n)}_{\ell, j} R^{j},
\label{eq:yexp}
\end{equation}
where \( y^{(n)}_{\ell,c} = y^{(n)}_{\ell}(R=0) \) denotes the central value of the quantity $y^{(n)}_{\ell}$. We refer to $ y^{(n)}_{\ell,c}$ and  the coefficients $ y^{(n)}_{\ell,j}$ as \textit{local asymptotic constants of the perturbed equations}. These constants are similar to the $x_c$ and $x_j$ of the background equations, but instead they refer to the equations at higher order in spin frequency.

At first order in the angular spin-frequency, the asymptotic solution for the function \( \varpi^{(1)}_{1} \) is obtained by substituting the zeroth-order asymptotic solution from Eqs.~\eqref{eq:Msol}--\eqref{eq:nusol} into Eq.~\eqref{eq:Eqvarpi1}, along with the power series expansion for \( \varpi^{(1)}_{1} \), given in Eq.~\eqref{eq:yexp}. Following the procedure outlined earlier, we eliminate any fractional terms on both sides of the equation and collect terms of equal powers in \( R \). This yields an algebraic system for the coefficients \( \varpi^{(1)}_{1,j} \). 
The solution to this system expresses each coefficient in terms of the central asymptotic constant \( \varpi^{(1)}_{1,c} \), as well as the central asymptotic constants of the zeroth-order background quantities, i.e., $\varepsilon_{c} $, $p_{c}$ and $\nu_{c}$. The local solution for the metric function $\varpi^{(1)}_{1}$ up to $\mathcal{O}(R^{2})$ around the origin is then given by
\begin{equation}
%
%
\label{eq:w11sol}
\varpi^{(1)}_{1}(R) = \varpi^{(1)}_{1,c} +  \frac{8\pi}{5} \varpi^{(1)}_{1,c} \left(p_c+\varepsilon _c\right) R^{2} + \mathcal{O}(R^{4}), \\    
\end{equation}
and the solution for $\bar{\alpha}^{(1)}_{1}=d \varpi^{(1)}_{1}/dR$ is obtained by taking the derivative of  Eq.~\eqref{eq:w11sol} with respect to $R$. 

Using the background and first-order local solutions, we proceed with the asymptotic analysis at second order in the spin frequency for the $\ell=0$ and $\ell=2$ modes. For the $\ell=0$ mode, we expand the functions $m^{(2)}_{0}$, $h^{(2)}_{0}$, and $\xi^{(2)}_{0}$ as described in Eq.~\eqref{eq:yexp}, and substitute these expansions into the system of equations governing the $\ell=0$ mode, given in Eqs.~\eqref{eq:m0}--\eqref{eq:xi0}, along with the background and first-order asymptotic solutions. Following the procedure outlined previously, we obtain the following solution, valid up to order $\mathcal{O}(R^{5})$:
\begin{align}
%
\label{eq:m20asympt}
m^{(2)}_{0}(R) &= \frac{4 \pi   \left( \varpi^{(1)}_{1,c}\right)^{2}  \left(1 + 2 p'_{c}\right)}{15 e^{\nu_c} p'_{c} \left(p_c+\varepsilon _c\right)^{-1}} R^5 + \mathcal{O}(R^{7}),   \\
h^{(2)}_{0}(R) &= h^{(2)}_{0,c} + \frac{\pi   \left( \varpi^{(1)}_{1,c}\right)^{2}  \left(1 + 7 p'_{c}\right) }{15 e^{\nu_c} p'_{c} \left(p_c+\varepsilon _c\right)^{-1}}R^{4}  + \mathcal{O}(R^{6}), \\
\xi^{(2)}_{0}(R) &= \frac{ e^{-\nu_c}  \left( \varpi^{(1)}_{1,c}\right)^{2}}{4 \pi \left( 3 p_c+ \varepsilon_c \right)}R + \mathcal{O}(R^{3}) \, .
\end{align}

For the $\ell=2$ mode, we obtain the asymptotic solution using Eqs.~\eqref{eq:keven}, \eqref{eq:heven}, \eqref{eq:meven}, and \eqref{eq:xieven}. We begin by constructing a second-order differential equation for the metric function $h^{(2)}_{2}$ by combining Eqs.~\eqref{eq:keven} and \eqref{eq:heven}. We then insert the asymptotic solutions from the previous orders, along with a power-series expansion for $h^{(2)}_{2}$ of the form given in Eq.~\eqref{eq:yexp}. Following the same procedure used at previous orders we determine the asymptotic coefficients $h^{(2)}_{2,j}$ in terms of the central asymptotic constants obtained from lower orders. Notably, the asymptotic constant $h^{(2)}_{2,2}$, which multiplies the $R^2$ term, cannot be expressed in terms of lower-order central asymptotic constants and must, therefore, be treated as a free parameter. We denote this constant as $h^{(2)}_{2,2} \equiv A^{(2)}_{2,\textsf{int}}$, where `$A$' denotes an asymptotic free parameter, and the subscript ‘$\textsf{int}$’ refers to the stellar interior.

Once the local solution for $h^{(2)}_{2}$ is obtained, we use the differential equation for $k^{(2)}_{2}$ (given in Eq.~\eqref{eq:keven}) to solve for its local behavior in terms of the free constant $A^{(2)}_{2,\textsf{int}}$. Similarly, the local solutions for $m^{(2)}_{2}$ and $\xi^{(2)}_{2}$ can be determined using Eqs.~\eqref{eq:meven} and \eqref{eq:xieven}, respectively. The final expressions for the $\ell=2$ mode, valid up to order $\mathcal{O}(R^{3})$, are given by
\begin{align}
k^{(2)}_{2}(R) &=  - A^{(2)}_{2, \textsf{int}} R^{2}  + \mathcal{O}(R^{4}), \\
h^{(2)}_{2}(R) &= A^{(2)}_{2, \textsf{int}} R^{2}  + \mathcal{O}(R^{4}), \\
m^{(2)}_{2}(R) &= - A^{(2)}_{2, \textsf{int}} R^{3}  + \mathcal{O}(R^{5}), \\
\xi^{(2)}_{2}(R) &= -\frac{e^{-\nu_c}  \left( \varpi^{(1)}_{1,c}\right)^{2} + 3 A^{(2)}_{2, \textsf{int}} }{4 \pi \left( 3 p_c+ \varepsilon_c \right)} R  + \mathcal{O}(R^{3}) .
\end{align}

The local solution to third order in the spin-frequency is obtained by using the solutions of the previous orders into the equations of structure at $\mathcal{O}(\epsilon^{3})$ from Eq.~\eqref{eq:wodd}. The expressions for the modes $\ell=1$ and $\ell=3$ are given by
\begin{align}
%
%
\label{eq:w31asymp}
\omega^{(3)}_{1}(R) &= \omega^{(3)}_{1,c} + \frac{8\pi}{5} \omega^{(3)}_{1,c} \left(p_c+\varepsilon _c\right)R^{2} +  \mathcal{O}(R^{4}) , \\ 
\omega^{(3)}_{3}(R) &=A^{(3)}_{3,\textsf{int}} R^{2} + \mathcal{O}(R^{4}) ,
\label{eq:w33asymp}
\end{align}
where the solutions for $\alpha^{(3)}_{1}$ and $\alpha^{(3)}_{3}$ are obtained by taking the derivative of Eqs.~\eqref{eq:w31asymp} and \eqref{eq:w33asymp} with respect to $R$.
For the $\ell=3$ mode, the asymptotic constant proportional to $R^{2}$, namely $\omega^{(3)}_{3,2}$, appears as a new free parameter that cannot be expressed in terms of the central asymptotic constants from previous orders. We denote this free constant by $\omega^{(3)}_{3,2} \equiv A^{(3)}_{3,\textsf{int}}$.

By continuing the same procedure at higher orders, we obtain all local solutions up to $\mathcal{O}(\epsilon^{7})$ for all $\ell$ modes. It is important to note that some asymptotic solutions for the metric functions begin at higher powers of $R$, a feature that arises from the structure of the differential equations governing each metric function, as well as from the solutions at lower orders.
Specifying values of the asymptotic constants yields a unique neutron-star solution inside the star. In this sense, the asymptotic constants form a parametric family of neutron-star solutions. 
Therefore, the parameteric family of neutron star solutions, valid to $\mathcal{O}(\epsilon^{7})$ in the small-spin expansion, is characterized by the following asymptotic constants:
\begin{align}
\nonumber
\big\{ &\varepsilon_{c} \, , \, \nu_{c} \, ,  \, \varpi^{(1)}_{1,c} \, , \,  h^{(2)}_{0,c} \, , \, A^{(2)}_{2,\textsf{int}} \, , \, \omega^{(3)}_{1,c} \, , \,  A^{(3)}_{3,\textsf{int}} \ , \ 
h^{(4)}_{0,c} \ , \ \\ \nonumber
&A^{(4)}_{2,\textsf{int}} \, , \,  A^{(4)}_{4,\textsf{int}} \, , \, \omega^{(5)}_{1,c} \, , \, A^{(5)}_{3,\textsf{int}} \, , \, A^{(5)}_{5,\textsf{int}} \, , \,  h^{(6)}_{0,c} \, , \, A^{(6)}_{2,\textsf{int}} \, , \, \\ 
&A^{(6)}_{4,\textsf{int}} \, , \, A^{(6)}_{6,\textsf{int}} \, , \, \omega^{(7)}_{1,c} \ , \ A^{(7)}_{3,\textsf{int}} \ , \ A^{(7)}_{5,\textsf{int}} \ , \ A^{(7)}_{7,\textsf{int}} \big\} \ .
\end{align}
These constants affect various physical properties of the star. For example, $\varepsilon_c$ effectively controls the neutron star mass and $ \varpi^{(1)}_{1,c}$ the star's moment of inertia. Generically, the asymptotic constants \( A^{(s)}_{\ell,\textsf{int}} \), where recall that $s$ is an even integer, affect the \( \ell \)-th mass multipole moment when \( s = \ell \), and its spin-induced corrections when \( s \neq \ell \). Similarly, \( A^{(k)}_{\ell,\textsf{int}} \), where recall $k$ is an odd integer, affect the \( \ell \)-th mass-current multipole moment when \( k = \ell \), and its spin-induced corrections when \( k \neq \ell \). The central asymptotic constants \( \varpi^{(k)}_{1,c} \) for \( k \neq 1 \), and \( h^{(s)}_{\ell, c} \), contribute to the spin-induced corrections to the moment of inertia and the mass of the star, respectively.

As discussed in Sec.~\ref{sec:multipoles}, it is convenient to express the interior solution as the sum of a homogeneous and a particular solution. 
As expected, the particular solutions solve the inhomogeneous differential equations, while the homogeneous ones solve the inhomogeneous differential equations in the limit of all source terms set to zero.  
Both the inhomogeneous and homogeneous systems must be solved numerically by imposing boundary conditions at the origin. The local analysis presented thus far has focused exclusively on the inhomogeneous systems. We have verified that the local solutions to the corresponding homogeneous systems can be obtained by setting all asymptotic constants from previous orders—up to a given order $n$—to zero, since these constants arise from the source terms only. This result has been independently confirmed through a separate asymptotic analysis of the homogeneous system, with consistent results.
All asymptotic solutions up to $\mathcal{O}(\epsilon^{7})$ are provided in the Supplementary Material of this paper as \texttt{.txt} files in CForm, C++, and Mathematica InputForm, including terms at higher powers of $R$. A Mathematica script containing the full asymptotic analysis is also available in the repository associated with this work \cite{conde2025hartlethorne}.

\section{Exterior Solutions}
\label{sec:exterior}

The gravitational field inside a star is described by the full set of perturbed equations of stellar structure, with boundary conditions at the center determined through a local analysis. To understand the gravitational field throughout the entire space, we must find the perturbed field equations outside the star. This process is similar to the interior case; however, in the external region, no matter is present, as it is vacuum. Thus, by setting $p = 0$ and $\varepsilon = 0$ in the interior equations, we can derive the equations for the region outside the star at each spin-frequency order. In principle, one could perform an asymptotic analysis near spatial infinity; however, we have found exact analytical solutions at each order up to seventh order in the spin-frequency.

In this section, we outline the general structure of the exterior solutions for each parity sector. We present the solutions to the homogeneous system of exterior equations up to $\mathcal{O}(\epsilon^{7})$, and provide the particular solutions up to $\mathcal{O}(\epsilon^{3})$ here. The remaining particular solutions are provided in the Supplemental Material accompanying this paper, available as \texttt{.txt} files in CForm, C++, and Mathematica Input Form. Additionally, a Mathematica script containing the symbolic calculations used to derive the exterior solutions is included in the repository \cite{conde2025hartlethorne}.

\subsection{Even parity sector}

In the exterior region, the $\ell = 0$ modes at each even spin-frequency perturbation order are governed by a set of two, first-order, ordinary differential equations for the metric perturbation functions $h^{(s)}_{0, \mathsf{ext}}(R)$ and $m^{(s)}_{0, \mathsf{ext}}(R)$. These equations are derived by setting $p$ and $\varepsilon$ to zero in Eqs.~\eqref{eq:m0}-\eqref{eq:h0} and using the exterior solutions from previous spin-frequency orders. 
Since there are two first-order equations, the solution will involve two integration constants. However, to ensure that the spacetime is asymptotically flat at spatial infinity, one of these constants can be determined, leaving only one free constant. The procedure is as follows: for a fixed value of $s$, and by substituting the exterior solutions from earlier spin-frequency orders, the ordinary differential equation for $m^{(s)}_{0, \mathsf{ext}}(R)$ decouples from the equation for $h^{(s)}_{0, \mathsf{ext}}(R)$. 
Thus, we first solve for $m^{(s)}_{0, \mathsf{ext}}(R)$ and use this solution in the second ordinary differential equation to find $h^{(s)}_{0, \mathsf{ext}}(R)$. Afterward, we impose the condition of asymptotic flatness on the solution for $h^{(s)}_{0, \mathsf{ext}}(R)$ to determine the value of one integration constant. Finally, this value is substituted back into the solution for $m^{(s)}_{0, \mathsf{ext}}(R)$.

The general structure of the exact exterior solutions at even orders for $\ell = 0$ is given by the following two expressions:
\begin{align}
\label{eq:h0ext}
h^{(s)}_{0, \mathsf{ext}}(R) &=  h^{(s) \, \textsf{P}}_{0, \mathsf{ext}}(R) - \dfrac{C^{ \, (s)}_{0,\mathsf{ext}}}{R-2M_{*}}  \, , \\[1ex]
\label{eq:m0ext}
m^{(s)}_{0, \mathsf{ext}}(R) &=  m^{(s) \, \textsf{P}}_{0, \mathsf{ext}}(R) + C^{ \, (s)}_{0,\mathsf{ext}} \,,
\end{align}
where the superscript \textsf{P} stands for ``particular''.
The quantities $ h^{(s) \, \textsf{P}}_{0, \mathsf{ext}}(R) $ and $ m^{(s) \, \textsf{P}}_{0, \mathsf{ext}}(R) $
represent the (particular) solution to the inhomogeneous ordinary differential equation system in the exterior, while the second terms in Eqs.~\eqref{eq:h0ext} and \eqref{eq:m0ext} are the homogeneous solution. The functions $  h^{(s) \, \textsf{P}}_{0, \mathsf{ext}}(R) $ and $  m^{(s) \, \textsf{P}}_{0, \mathsf{ext}}(R) $ are exact analytical functions of the radial coordinate $R$, and depend on the exterior integration constants from lower spin-frequency orders. We denote the exterior integration constants for the mode $\ell = 0$ and spin-frequency order $s$ as $C^{(s)}_{0, \mathsf{ext}}$, which are related to the corrections of the mass monopole of the source distribution. For instance, at $\mathcal{O}(\epsilon^{2})$, Eqs.~\eqref{eq:h0ext} and \eqref{eq:m0ext} are \cite{Hartle:1967he} 
\begin{align}
\label{eq:h20ext}
h^{(2)}_{0,\mathsf{ext}}(R) &= \frac{S^2}{R^{3} \left(R-2 M_{* }\right)}-\frac{C^{ \, (2)}_{0,\mathsf{ext}} }{R-2 M_{* }} \, , \\
m^{(2)}_{0,\mathsf{ext}}(R) &= -\dfrac{S^{2}}{R^{3}} + C^{ \, (2)}_{0,\mathsf{ext}} \, .
\label{eq:m20ext}
\end{align}
We provide the particular solutions \( h^{(s) \, \textsf{P}}_{0, \mathsf{ext}}(R) \) and \( m^{(s) \, \textsf{P}}_{0, \mathsf{ext}}(R) \) for orders \( s = 2, 4, \) and \( 6 \) in the Supplemental Material. Due to their length, these expressions are included as \texttt{.txt} files, formatted as previously described.

On the other hand, for the modes $\ell = 2, 4, 6$, we have two, first-order, ordinary differential equations for the metric perturbation functions $k^{(s)}_{\ell, \mathsf{ext}}(R)$ and $h^{(s)}_{\ell, \mathsf{ext}}(R)$, which are obtained from Eqs.~\eqref{eq:keven} and \eqref{eq:heven} by setting $p$ and $\varepsilon$ to zero and substituting the exterior solutions from previous orders. The general solution is then the sum of the particular solution plus a constant times the homogeneous solution,
\begin{align}
\label{eq:solh}
h^{(s)}_{\ell, \mathsf{ext}}(R) &= h^{(s) \,  \mathsf{P}}_{\ell, \mathsf{ext}}(R) + C^{ \, (s)}_{\ell,  \, \mathsf{ext}} h^{(s) \,  \mathsf{H}}_{\ell, \mathsf{ext}}(R) \, , \\[1ex]
k^{(s)}_{\ell, \mathsf{ext}}(R) &= k^{(s) \,  \mathsf{P}}_{\ell, \mathsf{ext}}(R) + C^{ \, (s)}_{\ell, \, \mathsf{ext}} k^{(s) \,  \mathsf{H}}_{\ell, \mathsf{ext}}(R) \,,
\label{eq:solk}
\end{align}
where the superscript \textsf{H} stands for ``homogeneous''.
The homogeneous system of equations for \( h^{(s) \, \textsf{H}}_{\ell, \mathsf{ext}}(R) \) and \( k^{(s) \, \textsf{H}}_{\ell, \mathsf{ext}}(R) \)
is described by 
\begin{align}
\label{eq:heqH}
&\dfrac{dh^{(s) \, \textsf{H}}_{\ell, \mathsf{ext}}}{dR} + \dfrac{(\ell + 2)(\ell - 1)}{2 M_{*}} k^{(s) \, \textsf{H}}_{\ell, \mathsf{ext}} \\ \nonumber
&+ \dfrac{\tfrac{1}{2}(\ell + 2)(\ell - 1) R^{2} - (\ell^{2} + \ell - 4) M_{*} R - 2 M_{*}^{2}}{M_{*} R(R - 2 M_{*})} h^{(s) \, \textsf{H}}_{\ell, \mathsf{ext}} = 0, \\ \nonumber
\label{eq:keqH}
&\dfrac{dk^{(s) \, \textsf{H}}_{\ell, \mathsf{ext}}}{dR} - \dfrac{(\ell + 2)(\ell - 1)}{2 M_{*}} k^{(s) \, \textsf{H}}_{\ell, \mathsf{ext}} \\
&\quad - \dfrac{\tfrac{1}{2} (\ell + 2)(\ell - 1) R + 2 M_{*}}{M_{*} R} h^{(s) \, \textsf{H}}_{\ell, \mathsf{ext}} = 0.
\end{align}   
These two equations can be combined to derive a second-order ordinary differential equation for \( h^{(s) \, \textsf{H}}_{\ell, \mathsf{ext}} \). Defining \( \zeta \equiv ({R}/{M_{*}}) - 1 \), 
the second-order equation can be rewritten as
\begin{align}
\nonumber
&(1 - \zeta^{2}) \dfrac{d^{2} h^{(s) \, \textsf{H}}_{\ell, \mathsf{ext}}}{d \zeta^{2}} - 2 \zeta \dfrac{d h^{(s) \, \textsf{H}}_{\ell, \mathsf{ext}}}{d \zeta} \\
&\quad + \left[ \ell(\ell + 1) - \dfrac{4}{1 - \zeta^{2}} \right] h^{(s) \, \textsf{H}}_{\ell, \mathsf{ext}} = 0.
\label{eq:legendre}
\end{align}
The general solution to this equation is given by
\[
h^{(s) \, \textsf{H}}_{\ell, \mathsf{ext}}(\zeta) = A P^{2}_{\ell}(\zeta) + B Q^{2}_{\ell}(\zeta),
\]
where \( P^{m}_{\ell}(\zeta) \) are the associated Legendre polynomials and \( Q^{m}_{\ell}(\zeta) \) are the associated Legendre functions of the second kind. We set \( A = 0 \) to ensure the correct asymptotic behavior at infinity. This leads to the solution
\begin{equation}
\label{eq:hHsol}
h^{(s) \, \textsf{H}}_{\ell, \mathsf{ext}}(R) = Q^{2}_{\ell}(\zeta),
\end{equation}
up to a proportionality constant. Substituting Eq.~\eqref{eq:hHsol} into Eq.~\eqref{eq:heqH}, and solving for \( k^{(s) \, \textsf{H}}_{\ell, \mathsf{ext}} \), we obtain
\begin{equation}
k^{(s) \, \textsf{H}}_{\ell, \mathsf{ext}}(\zeta) = \dfrac{2 M_{*}}{\sqrt{R(R - 2 M_{*})}} Q^{1}_{\ell}(\zeta) - Q^{2}_{\ell}(\zeta).
\label{eq:hextsolH}
\end{equation}
where the following recursion relation was used,
\begin{align}
(\zeta^{2} - 1) \dfrac{d Q^{2}_{\ell}}{d \zeta} = - (\ell + 2)(\ell - 1) \sqrt{\zeta^{2} - 1} Q^{1}_{\ell} - 2 \zeta Q^{2}_{\ell}.
\label{eq:kextsolH}
\end{align}
Since Eq.~\eqref{eq:legendre} is linear and homogeneous, any constant can be multiplied by the solutions in Eqs.~\eqref{eq:hextsolH} and \eqref{eq:kextsolH}, and the result will still be a valid solution. We denote this constant by \( C^{\, (s)}_{\ell, \, \textsf{ext}} \), as shown in Eqs.~\eqref{eq:solh} and \eqref{eq:solk}. Additionally, the functional form of the solutions in Eqs.~\eqref{eq:hextsolH} and \eqref{eq:kextsolH} mirrors the solution for \( s = 2 \) and \( \ell = 2 \) from Hartle's work \cite{Hartle:1967he}, with the only difference being that \( \ell \) is generalized to take different even values, i.e., $\ell=2,4$ and $6$. 

The associated Legendre functions of the second kind, \( Q^{2}_{\ell} \) and \( Q^{1}_{\ell} \), for \( \ell = 2, 4 \) and $6$ are
\begin{align}
\label{eq:Q22}
Q^{2}_{2}(\zeta) &= \dfrac{3}{2}\left( \zeta^{2} - 1 \right) \log\left[ \dfrac{\zeta + 1}{\zeta - 1} \right] - \dfrac{3 \zeta^{3} - 5\zeta }{\zeta^{2} - 1} \, ,  \\ \nonumber
Q^{2}_{4}(\zeta) &= \frac{15}{4} \left(7 \zeta ^2-1\right) \left(\zeta ^2-1\right) \log \left[ \frac{\zeta +1}{\zeta -1}\right]  \\ 
&-\frac{105 \zeta ^5-190 \zeta ^3+81 \zeta }{2 \left(\zeta ^2-1\right)} \, ,  \\ \nonumber
Q^{2}_{6}(\zeta) &= \frac{105}{16} \left(33 \zeta ^4-18 \zeta ^2+1\right) \left(\zeta ^2-1\right) \log \left[\frac{\zeta +1}{\zeta -1}\right] \\
&-\frac{3465 \zeta ^7-7665 \zeta ^5+5103 \zeta ^3-919 \zeta }{8 \left(\zeta ^2-1\right)} \, , \\
Q^{1}_{2}(\zeta) &= \sqrt{\zeta ^2-1} \left(\frac{3 \zeta ^2-2}{\zeta ^2-1}-\frac{3}{2} \zeta  \log \left[\frac{\zeta +1}{\zeta -1}\right]\right) \, , \\ \nonumber
Q^{1}_{4}(\zeta) &= \frac{1}{6} \sqrt{\zeta ^2-1} \bigg[\frac{105 \zeta ^4-115 \zeta ^2+16}{\zeta ^2-1} \\  
&-\frac{15}{2} \zeta  \left(7 \zeta ^2-3\right) \log \left(\frac{\zeta +1}{\zeta -1}\right)\bigg] \, , \\ \nonumber
Q^{1}_{6}(\zeta) &= \frac{1}{40} \sqrt{\zeta ^2-1} \bigg[ \frac{3465 \zeta ^6-5460 \zeta ^4+2163 \zeta ^2-128}{\zeta ^2-1} \\ 
&-\frac{105}{2} \zeta  \left(33 \zeta ^4-30 \zeta ^2+5\right) \log \left(\frac{\zeta +1}{\zeta -1}\right) \bigg] \, .
\label{eq:Q16}
\end{align}
Note that the definitions of \( Q^{2}_{\ell} \) and \( Q^{1}_{\ell} \) in Eqs.~\eqref{eq:Q22}--\eqref{eq:Q16} use slightly different sign conventions than the standard form, as also defined in \cite{Hartle:1967he}. However, these representations are consistent when considering their parity transformations and assuming \( \zeta > 1 \), which follows from the condition \( R > 2 M_{*} \); the latter condition holds because in the exterior $R > R_*$ by definition and $M_*/R_* < 1$ for a neutron star.

To obtain the particular solutions, the inhomogeneous system of equations must be solved iteratively at each order in spin. As the spin-frequency order increases, the metric perturbation functions from previous orders serve as source terms for the equations at the next order, leading to differential equations with distinct functional forms. The general structure of the particular solutions depends on the integration constants, \( C^{\, (s)}_{\ell, \, \textsf{ext}} \), from the previous orders and involves logarithmic functions of the radial coordinate \( R \) and the TOV mass \( M_{*} \), after imposing asymptotic flatness. Once the solutions for $h^{\, (s)}_{\ell}$ and $k^{\, (s)}_{\ell}$ are found, the solution for $m^{\, (s)}_{\ell, \, \textsf{ext}}$ can be obtained through the algebraic expression in Eq.~\eqref{eq:meven}. Note from such algebraic expression that the first term is proportional to $h^{\, (s)}_{\ell, \textsf{ext}}$, and thus, there is a piece of the solution for $m^{\, (s)}_{\ell, \, \textsf{ext}}$ given by $-(R- 2M_{*}) C^{ \, (s)}_{\ell, \, \textsf{ext}} Q^{2}_{\ell}(\zeta)$.

For instance, at order $s=2$, the solutions are given by
\begin{align}
h^{\, (2) \, \textsf{P}}_{2, \textsf{ext}}(R) &= \dfrac{1}{M_{*} R^{3}} \left( 1 + \dfrac{M_{*}}{R} \right)S^{2} \, , \\
k^{\, (2) \, \textsf{P}}_{2, \textsf{ext}}(R) &= - \dfrac{1}{M_{*} R^{3}} \left( 1 + \dfrac{2M_{*}}{R} \right) S^{2} \, .
\end{align}
while the entire solution for $m^{(2)}_{2, \, \textsf{ext}}$ is
\begin{align}
\nonumber
m^{(2)}_{2, \, \textsf{ext}}(R) &= - \dfrac{1}{M_{*}R^{2}} \left( 1 - 7 \dfrac{M_{*}}{R} + 10 \dfrac{M_{*}^{2}}{R^{2}} \right) S^{2} \\
&-(R- 2M_{*}) C^{\, (2)}_{2, \, \textsf{ext}}  Q^{2}_{2}\left( \dfrac{R}{M_{*}} - 1 \right) \, ,
\end{align}
where \( S \) is an integration constant and represents the leading contribution of the star's spin angular momentum, which appears at \( \mathcal{O}(\epsilon) \), as explained in the next section. The full expressions for the solutions corresponding to other even spin-frequency orders and modes are provided in \texttt{.txt} files, using the formats described earlier, and can be found in the Supplemental Material of this paper.

\subsection{Odd parity sector}
For the odd-parity sector, with spin-frequency orders $k = 1, 3, 5,$ and $7$, the exterior equations are similarly obtained by setting $p$ and $\varepsilon$ to zero in Eq.~\eqref{eq:wodd} and using the exterior solutions from the previous spin-frequency orders. Since the odd-parity sector is described by a single, second-order, ordinary differential equation, there are two integration constants. One of them can be obtained by imposing asymptotic flatness at spatial infinity. The general solution can be written as the sum of the particular solution and a constant times the homogeneous solution,
\begin{align}
\label{eq:wPH}
\omega^{(k)}_{\ell, \, \mathsf{ext}} &= \omega^{(k), \mathsf{P}}_{\ell, \, \mathsf{ext}}(R) + C^{(k)}_{\ell, \, \mathsf{ext}} \omega^{(k), \mathsf{H}}_{\ell, \, \mathsf{ext}}(R) \, , \\[1ex]
\alpha^{(k)}_{\ell, \, \mathsf{ext}} &= \alpha^{(k), \mathsf{P}}_{\ell, \, \mathsf{ext}}(R) + C^{(k)}_{\ell, \, \mathsf{ext}} \alpha^{(k), \mathsf{H}}_{\ell, \, \mathsf{ext}}(R) \, ,
\end{align}
where $\alpha^{(k)}_{\ell, \, \mathsf{ext}} \equiv  d \omega^{(k)}_{\ell, \, \mathsf{ext}}/dR $. The homogeneous equation for $\ell=1,3,5$ and $7$ can be written in a compact form as
\begin{align}
\label{eq:woddExt}
\frac{d^{2}\omega^{(k) \, \textsf{H}}_{\ell, \, \textsf{ext}}}{dR^{2}}  +  \dfrac{4}{R}  \frac{d\omega^{(k) \, \textsf{H}}_{\ell, \, \textsf{ext}}}{dR}  -  \left[ \dfrac{(\ell+2)(\ell-1)}{R(R- 2M_{*})} \right] \omega^{(k) \, \textsf{H}}_{\ell, \, \textsf{ext}} = 0 \, .
\end{align} 
The solutions at each $\ell$ mode that are asymptotically flat at spatial infinity (up to a proportionality constant) are 
\begin{align}
\label{eq:w11solExt}
\omega^{(k) \, \textsf{H} }_{1\, , \textsf{ext} }(R) &= -\dfrac{1}{3 R^{3}} \, , \\[1ex] \nonumber
\omega^{(k) \, \, \textsf{H}}_{3, \textsf{ext}}(R) &= \dfrac{105 R}{64 M_{*}^7} \left(3 R-4 M_{*} \right) f(R) \log f(R)  \\ \nonumber
&+ \frac{7}{32 R^3 M_{*}^6}  \bigg(30 R^2 M_{*}^2+10 R M_{*}^3+4 M_{*}^4 \\
&+45 R^4 -105 R^3 M_{*}\bigg) \, , \\ \nonumber
\omega^{(k) \, \, \textsf{H}}_{5, \, \textsf{ext}}(R) &= -\dfrac{105 R}{128 M_{*}^{11}} \bigg[15 R^3 -54 R^2 M_{* }+60 R M_{* }^2 \\ \nonumber
&-20 M_{* }^3 \bigg] f(R) \log f(R) - \dfrac{1}{64 R^3 M_{* }^{10}} \bigg[ \\ \nonumber
&1575 R^6 -7245 R^5 M_{*}+10920 R^4 M_{*}^2 + 8 M_{*}^6 \\
&-5670 R^3 M_{*}^3+420 R^2 M_{*}^4+56 R M_{*}^5\bigg] \, , \\  \nonumber
\omega^{(k) \, \, \textsf{H}}_{7, \, \textsf{ext}}(R) &= -\dfrac{3003R}{16384 M_{*}^{15}} \bigg[1001 R^5 -5720 R^4 M_{* } \\ \nonumber
&+12320 R^3 M_{* }^2-12320 R^2 M_{* }^3 \\ \nonumber &+5600 R M_{* }^4 
-896 M_{* }^5 \bigg] f(R) \log f(R) \\ \nonumber
&-\dfrac{143}{122880 R^3 M_{*}^{14}} \bigg[ -2117115 R^7 M_{*} \\ \nonumber
&+5472390 R^6 M_{*}^2-6770610 R^5 M_{*}^3 \\ \nonumber
&+4006548 R^4 M_{*}^4-941136 R^3 M_{*}^5 \\ \nonumber
&+30240 R^2 M_{*}^6 +2160 R M_{*}^7+160 M_{*}^8 \\
&+315315 R^8 \bigg] \, 
\end{align}
where $f(R) \equiv 1- {2M_{*}}/{R}$. As shown in App.~\ref{apx:unique}, the ordinary differential equation for $\varpi^{(1)}_{1, \, \textsf{ext}}$ in Eq.~\eqref{eq:Eqvarpi}—when $p$ and $\varepsilon$ are set to zero—has the same form as Eq.~\eqref{eq:woddExt}, indicating that it is shift-invariant. This occurs because the term proportional to $\varpi^{(1)}_{1, \, \textsf{ext}}$ vanishes when $\ell=1$. Consequently, both $\varpi^{(1)}_{1, \, \textsf{ext}}$ and $\omega^{(1)}_{1, \, \textsf{ext}}$ share the same general solution form, which is
$
\varpi^{(1)}_{1, \, \textsf{ext}}(R) = A - {B}/({3R^{3}}).   
$
Since $\varpi^{(1)}_{1, \, \textsf{ext}}(R)$ is defined as $\varpi^{(1)}_{1, \, \textsf{ext}}(R) = \Omega - \omega^{(1)}_{1, \, \textsf{ext}}(R)$, and knowing that $\omega^{(1)}_{1, \, \textsf{ext}}(R)$ must vanish at spatial infinity by imposing asymptotic flatness, we conclude that $A = \Omega$. A convenient choice for $B$ is $B = 6S$, where $S$ is the spin angular momentum of the star, as discussed in Sec.~\ref{sec:multipoles}. Therefore,
\begin{align}
\varpi^{(1)}_{1, \, \textsf{ext}}(R) &= \Omega -\dfrac{2S}{R^{3}}  \, .
\label{eq:wbext}
\end{align}
From Eq.~\eqref{eq:wbext}, it follows that $\omega^{(1)}_{1, \, \textsf{ext}} = {2S}/{R^{3}}$. Additionally, using Eqs.~\eqref{eq:wPH} and \eqref{eq:w11solExt} with $k=1$ and $\ell=1$,  we have that $C^{(1)}_{1, \, \textsf{ext}} = -2S$.

On the other hand, similarly as explained for the even-parity section, the inhomogeneous system of equations contains source terms that depend on the exterior solutions of the previous order, and thus, the general structure of the ordinary differential equations changes at each spin-frequency order. Below, we list the particular solutions up to $\mathcal{O}(\epsilon^{3})$,
\begin{align}
\omega^{(1) \, \textsf{P}}_{1, \, \textsf{ext}}(R) &= 0 \, , \\ \nonumber
 \omega^{(3) \, \textsf{P}}_{1, \, \textsf{ext}}(R) &= -\frac{33 S C^{\, (2)}_{2, \mathsf{ext}}}{40 M_{* }^3}-\frac{4 S^3}{5 R^6 M_{* }}-\frac{12 S^3}{5 R^7} \, , \\ \nonumber
 &+ \dfrac{S C^{\, (2)}_{2, \textsf{ext}}}{40 R^{4}M^{3}_{*}} \bigg[ 
 33 R^{4} - 120 R^{4} \ln f(R) -240 R^3 M_{* } \\ \nonumber
 &+ 288 R^3 M_{* } \ln f(R) + 336 R^2 M_{* }^2+256 R M_{* }^3 \\
 &- 192 R M_{* }^3 \ln f(R)-96 M_{* }^4 \bigg] \, , \\ \nonumber
 \omega^{(3) \, \textsf{P}}_{3, \, \textsf{ext}}(R) &= \dfrac{S^{3}}{240 R^7 M_{* }^8} \bigg\{ M_{* } \big[ 2 M_{* } \big(175 R^5 M_{* }+70 R^4 M_{* }^2 \\ \nonumber
 &- 80 R^2 M_{* }^4+16 R M_{* }^5+288 M_{* }^6+525 R^6\big) \\ \nonumber 
 &-3675 R^7\big] +1575 R^8\bigg\} + \frac{S}{15 R^4 M_{* }^4} \big(315 R^5  \\  \nonumber
 &-735 R^4 M_{* }+210 R^3 M_{* }^2+34 R^2 M_{* }^3 \\ 
 &+52 R M_{* }^4+36 M_{* }^5 \big) C^{ \, (2)}_{2, \textsf{ext}}  \, .
\end{align}
The particular solutions for the remaining spin-frequency orders and modes are provided in the Supplemental Material as \texttt{.txt} files, using the format described earlier.
After imposing asymptotic flatness at spatial infinity, these solutions depend on the integration constants $C^{(n)}_{\ell , \, \textsf{ext}}$ from previous orders, with some terms involving logarithmic functions of the radial coordinate $R$.

\section{Global solution and Multipole Moments}
\label{sec:multipoles}

In Sec.\ \ref{sec:structure}, we constructed the equations of stellar structure along with the asymptotic boundary conditions at the center of the star, as explained in Sec.\ \ref{sec:local}. Given an equation of state, these equations must be solved numerically inside the star up to its surface. Meanwhile, the exterior problem was discussed in the previous section, and despite the length of the source terms involving the exterior equations, we have found exact analytical solutions. 
What remains to be done is to connect the exterior and interior solutions such that we have global solutions. 

In this section, we derive the global solutions for each spin-frequency order and extract the star's multipole moments, including their spin corrections. To ensure the continuity of the global solution at the star's surface, we match the interior and exterior metric solutions at the star's boundary.
Once the global solution is obtained, we extract the star's mass and mass-current multipole moments using two distinct methods. The first method relies on Thorne's expansion in \textit{Asymptotically-Cartesian Mass-Centered} (ACMC) coordinates. The second method, known as \textit{Ryan's method}, offers a more formal and gauge-invariant approach for extracting multipoles; this method is based on the Geroch-Hansen multipole moments formalism in GR.
We have verified that the analytical expressions for the multipole moments that we obtain are identical, irrespective of the method we employ.

\subsection{Matching procedure}
\label{subsec:matching}

In order to ensure continuity of the global solution at the star's surface, we match the interior and exterior metric perturbation solutions at the surface boundary, \( R = R_{*} \). This matching procedure allows us to determine two types of integration constants at each spin-frequency order---one from the interior region and another from the exterior region. The integration constants from the interior region are used to construct the global solution at a previous spin-frequency order, and these are then iterated to the next order. In contrast, the exterior constants are directly related to the star's multipole moments.

\subsubsection{Matching conditions at $\mathcal{O}(\epsilon)$}

At linear order in the spin-frequency approximation, there is only one metric function: $\varpi^{(1)}_{1}$ and its derivative $ \bar{\alpha}^{(1)}_{1} \equiv d \varpi^{(1)}_{1}/dR$. We impose continuity between the interior and exterior solutions at $\mathcal{O}(\epsilon)$ for these functions at $R=R_{*}$ (or equivalently, we require that $\varpi^{(1)}_{1}$ be $C^1$ smooth at the surface), obtaining the following matching conditions,
\begin{align}
\label{eq:wbmatch1}
\varpi^{(1)}_{1, \textsf{int}}(R_{*}) &= \varpi^{(1)}_{1, \textsf{ext}}(R_{*})   \, , \\[1ex]
\bar{\alpha}^{(1)}_{1, \textsf{int}}(R_{*}) &= \left.  \dfrac{d \varpi^{(1)}_{1, \textsf{ext}}}{dR} \right |_{R_{*}} \, .
\label{eq:wbmatch2}
\end{align}
Substituting Eq.~\eqref{eq:wbext} into Eqs.~\eqref{eq:wbmatch1}--\eqref{eq:wbmatch2}, we obtain two equations for the two integration constants, \( S \) and \( \Omega \), which can be solved for and expressed in terms of the interior solutions as 
\begin{equation}
S = \dfrac{1}{6}R_{*}^{4} \bar{\alpha}^{(1)}_{1, \, \textsf{int}}(R_{*}) \ \ \  ; \ \ \ \Omega = \varpi^{(1)}_{1, \, \textsf{int}}(R_{*}) + \dfrac{2S}{R^{3}_{*}} \, .
\label{eq:constO1}
\end{equation}
Note that the integration constants in Eqs.~\eqref{eq:constO1} depend on the interior solution \( \varpi^{(1)}_{1, \, \textsf{int}} \), which must be obtained numerically from the center of the star to its surface. In general, the solution depends on the choice of \( \varpi^{(1)}_{1, \, c} \), which appears in the boundary condition in Eq.~\eqref{eq:w11sol}. Since the TOV equations must also be solved for a specified central energy density \( \varepsilon_{c} \) (or equivalently \( p_{c} \)) before solving the first-order equations, the quantities \( (\varepsilon_{c}, \varpi^{(1)}_{1, c}) \) form a two-parameter family of equilibrium solutions up to \( \mathcal{O}(\epsilon) \), corresponding to a single value of \( S \) and \( \Omega \).
However, since Eq.~\eqref{eq:Eqvarpi1} is homogeneous, a rescaled function \( \varpi^{(1)}_{1, \, \textsf{int}} \rightarrow C \varpi^{(1)}_{1, \, \textsf{int}} \), where \( C \) is a constant, will also be a solution. To obtain a star with any desired angular spin-frequency \( \Omega^{\textsf{new}} \), we rescale the solution as \cite{Hartle:1968si} \( \varpi^{(1), \, \textsf{new}}_{1, \, \textsf{int}} = \varpi^{(1), \, \textsf{old}}_{1, \, \textsf{int}} (\Omega^{\textsf{new}} / \Omega) \), where \( \Omega \) is the original spin frequency obtained using an arbitrary value of \( \varpi^{(1)}_{1, \, c} \) . After rescaling, the new value of \( S^{\textsf{new}} \) can be obtained from Eq.~\eqref{eq:constO1}.

\subsubsection{Matching conditions at $\mathcal{O}(\epsilon^{s})$ with even-$s$ and $\ell=0$} 

As mentioned in Sec.~\ref{sec:even_parity_equations}, the metric perturbations describing the mass monopole corrections of the star ($\ell=0$ and $s$ being even) are described by the functions $m^{(s)}_{0}$, $h^{(s)}_{0}$, and $\xi^{(s)}_{0}$. We impose continuity of these functions at the boundary $R=R_{*}$ and thus, the matching conditions are given by,
\begin{align}
\label{eq:matchm0}
m^{(s)}_{0, \, \textsf{int}}(R_{*}) &= m^{(s) \, \textsf{P}}_{0, \mathsf{ext}}(R_{*}) + C^{ \, (s)}_{0,\mathsf{ext}} \, , \\[1ex] \label{eq:mathch0}
h^{(s) \, \textsf{P}}_{0, \, \textsf{int}}(R_{*}) + h^{(s)}_{0, \, c} &=  h^{(s) \, \textsf{P}}_{0, \mathsf{ext}}(R_{*}) - \dfrac{C^{ \, (s)}_{0,\mathsf{ext}}}{R_{*}-2M_{*}} \, , \\[1ex]
\xi^{(s)}_{0, \, \textsf{int}}(R_{*}) &= C^{(s)}_{\xi_{0}} = \textsf{const.} 
\label{eq:xis0}
\end{align}
From Eq.~\eqref{eq:mathch0}, we observe that the left-hand side can be split into two distinct terms for the interior solution \( h^{(s)}_{0, \textsf{int}} \). The first term represents a particular solution, while the second term is a constant at exactly the center of the star. Since the first-order, ordinary differential equation in Eq.~\eqref{eq:h0} for \( h^{(s)}_{0, \textsf{int}} \) does not depend on \( h^{(s)}_{0} \) on the right-hand side, the equation is shift-invariant\footnote{Compare this property with the TOV equation for the background metric function 
$\nu$ in Eq.~\eqref{eq:nu}, which exhibits a similar behavior.}. Therefore, any solution of the form \( h^{(s)}_{0, \textsf{int}} + A^{(s)}_{0, \, \textsf{int}} \), where \( A^{(s)}_{0, \, \textsf{int}} \) is a constant, is also a valid solution. For the particular solution, we set \( h^{(s) \, \textsf{P}}_{0, \, \textsf{int}} \) to vanish at \( R = 0 \), so that \( h^{(s)}_{0, \, \textsf{int}}(0) = h^{(s) \, \textsf{P}}_{0, \, \textsf{int}}(0) + A^{(s)}_{0, \, \textsf{int}} = h^{(s)}_{0, \, c} \), which implies that \( A^{(s)}_{0, \, \textsf{int}} = h^{(s)}_{0, \, c} \). A similar argument can be made for Eq.~\eqref{eq:matchm0}, as the first-order ordinary differential equation for \( m^{(s)}_{0, \, \textsf{int}} \) is also shift-invariant. However, since \( m^{(s)}_{0, \, \textsf{int}}(0) = 0 \), the shift constant must be zero\footnote{Refer to the asymptotic expansion about $R=0$ in Eq.~\eqref{eq:m20asympt} for $m^{(2)}_{0}$ or the expansions for $m^{(4)}_{0}$ and $m^{(6)}_{0}$ in the Supplemental Material. The first term in all expansions goes as $\sim R^{5}$. }. 

Using Eqs.~\eqref{eq:mathch0}--\eqref{eq:mathch0}, one can solve for the integration constants \( C^{(s)}_{0, \, \textsf{ext}} \) and \( h^{(s)}_{0, \, c} \) to obtain
\begin{align}
\label{eq:Cs0ext}
C^{(s)}_{0 \, , \textsf{ext}} &= m^{(s)}_{0, \, \textsf{int}}(R_{*}) - m^{(s) \, \textsf{P}}_{0, \mathsf{ext}}(R_{*}) \\[1ex]
h^{(s)}_{0, \, c} &= h^{(s) \, \textsf{P}}_{0, \mathsf{ext}}(R_{*}) -  h^{(s) \, \textsf{P}}_{0, \mathsf{int}}(R_{*}) - \dfrac{m^{(s)}_{0, \, \textsf{int}}(R_{*}) - m^{(s) \, \textsf{P}}_{0, \mathsf{ext}}(R_{*})}{R_{*}-2M_{*}} 
\label{eq:hs0c}
\end{align}
In general, Eqs.~\eqref{eq:xis0}, \eqref{eq:Cs0ext}, and \eqref{eq:hs0c} are the integration constants of the perturbation equations for $\ell=0$ at order $\mathcal{O}(\epsilon^{s})$. It is worth mentioning that, as shown in \cite{Hartle:1968si}, at the spin-frequency order $s=2$, there is another approach for solving for the integration constant $h^{(2)}_{0, \, c}$. That strategy is to use the $R$ component of the Bianchi identities, $ (\nabla_{\alpha}T^{\alpha}{}_{R})^{(2)}_{\ell=0}=0$ to $\mathcal{O}(\epsilon^{2})$ and solve the first-order, ordinary differential equation for $h^{(2)}_{0, \, \textsf{int}}(R)$. The solution is,
\begin{equation}
\label{eq:h20int}
h^{(2)}_{0, \, \textsf{int}} = h^{(2)}_{0, \, c} - \dfrac{M + 4\pi p R^{3}}{R(R-2M)} \xi^{(2)}_{0} + \dfrac{1}{3}e^{-\nu } R^{2} \left( \varpi^{(1)}_{1} \right)^{2}
\end{equation}
On the other hand, the exterior solution $h^{(2)}_{0, \, \textsf{ext}}$ is given by Eq.~\eqref{eq:h20ext}. Thus, by matching the interior and exterior solutions, $h^{(2)}_{0, \, \textsf{int}}(R_{*}) = h^{(2)}_{0, \, \textsf{ext}}(R_{*}) $ we can solve for $h^{(2)}_{0, \, c}$,
\begin{align}
h^{(2)}_{0, \, c } &= \frac{S^2}{R_{*}^{3} \left(R_{*}-2 M_{* }\right)}-\frac{C^{ \, (2)}_{0,\mathsf{ext}} }{R_{*}-2 M_{* }} \\ \nonumber
&+  \dfrac{M_{*} \xi^{(2)}_{0}(R_{*})}{R_{*}(R_{*}-2M_{*})}  - \dfrac{1}{3} \dfrac{R^{3}_{*}}{R_{*}-2M_{*}} \left( \Omega - \dfrac{2S}{R_{*}^{3}} \right)^{2} 
\end{align}
where we have used the exterior solutions for the functions $\nu$ and $\varpi^{(1)}_{1}$ when matching at the boundary. The integration constant $C^{(2)}_{0, \, \textsf{ext}}$ is obtained from the matching $m^{(2)}_{0, \textsf{int}}(R_{*}) = m^{(2)}_{0, \, \textsf{ext}}(R_{*})$ and using Eq.~\eqref{eq:m20ext} for $m^{(2)}_{0, \, \textsf{ext}}$, i.e., 
\begin{equation}
C^{(2)}_{0, \, \textsf{ext}} = m^{(2)}_{0, \, \textsf{int}}(R_{*}) + \dfrac{S^{2}}{R_{*}^{3}} \, .
\end{equation}
The advantage of this approach is that there is no need to solve the ordinary differential equation for $h^{(2)}_{0, \textsf{int}}$ from the Einstein equations in Eq.~\eqref{eq:h0}, since it could be obtained through Eq.~\eqref{eq:h20int}.

In general, the values of the integration constants $C^{(s)}_{0, \, \textsf{ext}}$ and $h^{(s)}_{0, \, c}$ depend on the star's angular spin frequency, $\Omega$. Following a procedure similar to that at $\mathcal{O}(\epsilon)$, the functions $m^{(s)}_{0}$, $h^{(s)}_{0}$, and $\xi^{(s)}_{0}$ can be rescaled to correspond to a desired angular spin frequency $\Omega^{\textsf{new}}$ by multiplying them by $(\Omega^{\textsf{new}}/\Omega)^{s}$. Consequently, the integration constants are also rescaled by the same factor to correspond to $\Omega^{\textsf{new}}$.

\subsubsection{Matching conditions at $\mathcal{O}(\epsilon^{n})$ with $2 < n \leq 7$ and $\ell \neq 0$} 

The strategy for matching the metric perturbation functions at spin-frequency orders $2< n \leq 7$ and modes $\ell \neq 0$ is the same. The ordinary differential equations from the odd-parity sector are described by the functions $\{\omega^{(k)}_{\ell}, \alpha^{(k)}_{\ell}\}$, while for the even-parity sector we have $\{h^{(s)}_{\ell}, k^{(s)}_{\ell}\}$ plus two algebraic functions for $m^{(s)}_{\ell}$ and $\xi^{(s)}_{\ell}$. Let us define the pair of metric perturbation functions $\{ f^{(n)}_{\ell}, g^{(n)}_{\ell} \}$ to be either $\{\omega^{(k)}_{\ell}, \alpha^{(k)}_{\ell}\}$ or $\{h^{(s)}_{\ell}, k^{(s)}_{\ell}\}$ with $\ell \neq 0$. As explained in the previous sections, the general solutions in both the interior and exterior regions for $f^{(n)}_{\ell}$ and $g^{(n)}_{\ell}$ are,
\begin{align}
f^{(n)}_{\ell, \, \textsf{int}}(R) &= f^{(n) \, \textsf{P}}_{\ell, \, \textsf{int}}(R) + C^{(n)}_{\ell, \, \textsf{int}} f^{(n) \, \textsf{H}}_{\ell, \, \textsf{int}}(R) \\[0.5ex]
f^{(n)}_{\ell, \, \textsf{ext}}(R) &= f^{(n) \, \textsf{P}}_{\ell, \, \textsf{ext}}(R) + C^{(n)}_{\ell, \, \textsf{ext}} f^{(n) \, \textsf{H}}_{\ell, \, \textsf{ext}} (R) \\[0.5ex]
g^{(n)}_{\ell, \, \textsf{int}}(R) &= g^{(n) \, \textsf{P}}_{\ell, \, \textsf{int}}(R) + C^{(n)}_{\ell, \, \textsf{int}} g^{(n) \, \textsf{H}}_{\ell, \, \textsf{int}}(R) \\[0.5ex]
g^{(n)}_{\ell, \, \textsf{ext}}(R) &= g^{(n) \, \textsf{P}}_{\ell, \, \textsf{ext}}(R) + C^{(n)}_{\ell, \, \textsf{ext}} g^{(n) \, \textsf{H}}_{\ell, \, \textsf{ext}}(R)
\end{align}
Imposing continuity at the boundary of the star located at $R=R_{*}$ implies,
\begin{align}
\label{eq:fmatch}
f^{(n)}_{\ell, \, \textsf{int}}(R_{*}) = f^{(n)}_{\ell, \, \textsf{ext}}(R_{*}) \, , \\[1ex]
g^{(n)}_{\ell, \, \textsf{int}}(R_{*}) = g^{(n)}_{\ell, \, \textsf{ext}}(R_{*}) \, .
\label{eq:gmatch}
\end{align}
Equations \eqref{eq:fmatch} and \eqref{eq:gmatch} constitute an algebraic system for the integration constants \( C^{(n)}_{\ell, \, \textsf{int}} \) and \( C^{(n)}_{\ell, \, \textsf{ext}} \), which can be solved to obtain
\allowdisplaybreaks[4]
\begin{widetext}
\begin{align}
\label{eq:Cnlint}
    C^{\, (n)}_{\ell, \, \mathsf{int}} =& \dfrac{ \left[ f^{\, (n) \, \mathsf{P}}_{\ell, \, \mathsf{ext}}(R_{*}) - f^{\, (n) \, \mathsf{P}}_{\ell, \, \mathsf{int}}(R_{*})\right] g^{\, (n) \, \mathsf{H}}_{\ell, \, \mathsf{ext}}(R_{*}) +  f^{\, (n) \, \mathsf{H}}_{\ell, \, \mathsf{ext}}(R_{*}) \left[ g^{\, (n) \, \mathsf{P}}_{\ell, \, \mathsf{int}}(R_{*}) - g^{\, (n) \, \mathsf{P}}_{\ell, \, \mathsf{ext}}(R_{*})  \right] }{ f^{\, (n) \, \mathsf{H}}_{\ell, \, \mathsf{int}}(R_{*}) g^{\, (n) \, \mathsf{H}}_{\ell, \, \mathsf{ext}}(R_{*}) -  f^{\, (n) \, \mathsf{H}}_{\ell, \, \mathsf{ext}}(R_{*}) g^{\, (n) \, \mathsf{H}}_{\ell, \, \mathsf{int}}(R_{*}) } \, ,
     \\[1ex]
    C^{\, (n)}_{\ell, \, \mathsf{ext}} =&  \dfrac{ \left[ f^{\, (n) \, \mathsf{P}}_{\ell, \, \mathsf{ext}}(R_{*}) - f^{\, (n) \, \mathsf{P}}_{\ell, \, \mathsf{int}}(R_{*})\right] g^{\, (n) \, \mathsf{H}}_{\ell, \, \mathsf{int}}(R_{*}) +  f^{\, (n) \, \mathsf{H}}_{\ell, \, \mathsf{int}}(R_{*}) \left[ g^{\, (n) \, \mathsf{P}}_{\ell, \, \mathsf{int}}(R_{*}) - g^{\, (n) \, \mathsf{P}}_{\ell, \, \mathsf{ext}}(R_{*})  \right] }{ f^{\, (n) \, \mathsf{H}}_{\ell, \, \mathsf{int}}(R_{*}) g^{\, (n) \, \mathsf{H}}_{\ell, \, \mathsf{ext}}(R_{*}) -  f^{\, (n) \, \mathsf{H}}_{\ell, \, \mathsf{ext}}(R_{*}) g^{\, (n) \, \mathsf{H}}_{\ell, \, \mathsf{int}}(R_{*}) } \, .
    \label{eq:Cnlext}
\end{align}    
\end{widetext}

The integration constants \( C^{(n)}_{\ell, \, \textsf{int}} \) in Eq.~\eqref{eq:Cnlint} are used to construct the interior solutions for a fixed spin-frequency order \( n \), ensuring appropriate boundary matching on the surface. These solutions are then iterated in the subsequent spin-frequency-order approximation. In contrast, the exterior integration constants \( C^{(n)}_{\ell, \, \textsf{ext}} \) in Eq.~\eqref{eq:Cnlext} not only help construct the matched exterior solution at the boundary, but are also related to the mass and mass-current multipole moments of the star, as explained in the next subsection.

Similar to the other integration constants discussed earlier, the constants \( C^{(n)}_{\ell, \, \textsf{int}} \) and \( C^{(n)}_{\ell, \, \textsf{ext}} \) depend in general on the star's angular spin frequency \( \Omega \), which is determined once \( \varpi_{c} \) is specified. From Eqs.~\eqref{eq:Cnlint} and \eqref{eq:Cnlext}, the only terms that depend on \( \Omega \) are the particular solutions both inside and outside the star. The homogeneous solutions, however, are unaffected by any angular spin-frequency rescaling because the corresponding ordinary differential equations do not involve any lower-order metric perturbation functions, depending solely on the background metric functions. For the interior region, this can be shown by setting the source terms to zero in the ordinary differential equations for \( k^{(s)}_{\ell} \), \( h^{(s)}_{\ell} \), and \( \omega^{(k)}_{\ell} \) in Eqs.~\eqref{eq:keven}, \eqref{eq:heven}, and \eqref{eq:wodd}, respectively. Similarly, for the exterior region, the homogeneous ordinary differential equations governing these perturbation functions are given in Eqs.~\eqref{eq:heqH}, \eqref{eq:keqH}, and \eqref{eq:woddExt}. Therefore, the integration constants \( C^{(n)}_{\ell, \, \textsf{int}} \) and \( C^{(n)}_{\ell, \, \textsf{ext}} \) can be determined for any initial value of \( \varpi_{c} \) (i.e., when solving the interior problem) and then rescaled by multiplying by $(\Omega^{\textsf{new}}/\Omega)^{n}$ to correspond to a new and chosen angular spin frequency \( \Omega^{\textsf{new}} \).

\subsubsection{Matching conditions for unbounded functions (e.g. quark stars)} 

The previously discussed matching procedure relied on imposing continuity at \( R = R_{*} \). However, a more careful approach is necessary when unbounded terms appear in the ordinary differential equations that describe the metric perturbations \cite{Reina:2014fga, Reina:2017mbi}. For example, in the case of quark stars, the energy density does not smoothly approach zero at the star’s surface. Instead, it remains finite at the surface when approaching from the interior, but abruptly drops to zero just outside. As a result, any term involving derivatives of the energy density \( \varepsilon \) becomes unbounded, including the first term on the right-hand side of the ordinary differential equation for \( m^{(s)}_{\ell} \) in Eq.~\eqref{eq:m0}. To address this issue, the strategy is to integrate the ordinary differential equations over the range \( [R_{*} - \delta, R_{*} + \delta] \) and then take the limit as \( \delta \to 0 \). This approach allows us extract the discontinuity at the boundary \cite{ Yagi:2014bxa, Yagi:2013mbt, Adam:2020aza}. However, for neutron stars, the energy density decreases smoothly to zero at the star's surface and therefore its derivative vanishes. We will focus on neutron stars in the present work. 

\subsection{Thorne's expansion}

The multipole moments of the source distribution for a stationary, isolated object can be determined by analyzing the asymptotic behavior of the spacetime at spatial infinity. Thorne's approach involves selecting a specific coordinate system (known as \textit{Asymptotically-Cartesian Mass-Centered} (ACMC)) in which the coefficients of the expanded metric—expressed in inverse powers of an appropriate radial coordinate—correspond to the multipole moments \cite{Thorne:1980ru}. 
In this system, the coordinates are asymptotically Cartesian at large distances from the source, with the origin located at the center of mass of the distribution. Since multipole moments are read out from the expansion in this specific coordinate frame, this formulation is not gauge-invariant. However, it has been shown that Thorne's expansion and the gauge-invariant formulation of multipole moments by Geroch and Hansen (see next subsubsection) are equivalent \cite{gursel1983multipole, Mayerson:2022ekj}. 

The metric components in ACMC coordinates, expanded in a suitable radial coordinate $r$, are given by \cite{Cardoso:2016ryw}
\begin{align}
\label{eq:g00ACMC}
g_{00} &= - 1 + \dfrac{2M}{r} + \sum_{\ell \geq 2} \dfrac{1}{r^{\ell +1}} 
\left[   \dfrac{2}{\ell!} M^{\langle a_{1} \cdots a_{\ell} \rangle} n^{\langle a_{1} \cdots a_{\ell} \rangle} + \mathcal{S}_{\ell-1}  \right] \, , \\ \nonumber
g_{0j} &= - 2 \sum^{\infty}_{\ell \geq 1} \dfrac{1}{r^{\ell + 1}} \Bigg[ \dfrac{1}{\ell!}\epsilon^{jka_{\ell} } S^{\langle k a_{1} \cdots a_{\ell-1} \rangle} n^{\langle a_{1} \cdots a_{\ell} \rangle} + \mathcal{S}_{j, \ell-1} \\ 
& + \left( \ell-\textrm{harmonics with parity }(-1)^{\ell} \right) \Bigg] \, .
\label{eq:g0jACMC}
\end{align}
The symbol $\langle \cdot \rangle$ denotes the Symmetric Trace-Free (STF) operation applied to the enclosed indices. The notation $n^{a_1 \cdots a_\ell} \equiv n^{a_1} \cdots n^{a_\ell}$ represents products of the unit radial vector components, where $n^j = x^j / r$. The Levi-Civita symbol is denoted by $\epsilon^{ijk}$, and the quantities $M^{\langle a_1 \cdots a_\ell \rangle}$ and $S^{\langle a_1 \cdots a_\ell \rangle}$ represent the mass and mass-current multipole tensor moments, respectively. The quantities ${\cal{S}}_{\ell-1}$ and ${\cal{S}}_{j,\ell-1}$ depend on fewer powers of these units vectors for a fixed $\ell$, as we explain below. Moreover, following the approach in \cite{Cardoso:2016ryw}, we adopt the standard normalization factors for the multipole moments\footnote{Note that Thorne uses a different normalization convention in \cite{Thorne:1980ru}.} in Eqs.~\eqref{eq:g00ACMC} and \eqref{eq:g0jACMC}, as established by Geroch and Hansen \cite{Geroch:1970cd, Hansen:1974zz}.

Thorne introduced a criterion to ensure that ACMC coordinates remain asymptotically flat at spatial infinity. One key condition is that, at each order in the asymptotic expansion, there must be no spherical harmonics\footnote{Recall that any STF tensor can be written in terms of spherical harmonics.} with angular degree higher than $\ell$. This means that the quantities $\mathcal{S}_{\ell}$ and $\mathcal{S}_{j, \ell}$ can depend on products of $n^{i}$, such as $n^{a_{1} \cdots a_{\ell}}, n^{a_{1} \cdots a_{\ell-1}}$, and so on, but cannot depend on $n^{a_{1} \cdots a_{\ell'}}$ with $\ell' > \ell$, e.g., $n^{a_{1} \cdots a_{\ell+1}}$.
Another important condition is that the mass dipole moment tensor must vanish, $M^{a_{1}} = 0$, to ensure the origin of coordinates is located at the center of mass.  

One can re-express Eqs.~\eqref{eq:g00ACMC}--\eqref{eq:g0jACMC} in terms of spherical components. In particular, we impose axial symmetry by aligning the $\hat{k}$ coordinate axis with the star’s rotation axis. This symmetry implies that both the mass and mass-current multipole moments must remain invariant under rotations about $\hat{k}$. Consequently, the tensors $M^{\langle a_{1} \cdots a_{\ell} \rangle}$ and 
$S^{\langle a_{1} \cdots a_{\ell} \rangle}$ must be proportional to the STF products of the vector components
$k^{i}$, i.e., they must be proportional to $ k^{\langle a_{1} \cdots a_{\ell} \rangle}$. 
The proportionality constants for the mass and mass-current multipoles are the scalar quantities $M_{\ell}$ and $S_{\ell}$, respectively, each multiplied by a normalization factor. Therefore, in the axisymmetric case, the moments take the form
\begin{align}
\label{eq:Ml_STF}
M^{\langle a_{1} \cdots a_{\ell} \rangle} &= (2\ell - 1)!! M_{\ell} k^{\langle a_{1} \cdots a_{\ell} \rangle} \,  , \\
S^{\langle a_{1} \cdots a_{\ell} \rangle} &= (2\ell - 1)!! S_{\ell} k^{\langle a_{1} \cdots a_{\ell} \rangle} \, .
\label{eq:Sl_STF}
\end{align}
Using the identities of STF tensors \cite{poisson2014gravity},
\begin{align} k_{ \langle a_{1} \cdots a_{\ell} \rangle}n^{\langle a_{1} \cdots a_{\ell} \rangle } &= \dfrac{\ell!}{(2\ell - 1)!!}P_{\ell}(\mu) \, , \\
k_{ \langle a_{1} \cdots a_{\ell} \rangle}n^{ \langle j a_{1} \cdots a_{\ell} \rangle } &= \dfrac{\ell !}{(2 \ell +1)!!} \left[ \dfrac{d P_{\ell +1}}{d \mu} n^{j} - \dfrac{d P_{\ell}}{d \mu } k^{j} \right] ,  
\end{align}
where $\mu := \hat{n} \cdot \hat{k} = \cos\theta$, and applying Eqs.~\eqref{eq:Ml_STF} and \eqref{eq:Sl_STF}, it follows that
\begin{align}
M^{\langle a_{1} \cdots a_{\ell} \rangle } n^{\langle a_{1} \cdots a_{\ell} \rangle} &= \ell ! M_{\ell} P_{\ell}(\cos \theta) \\
\epsilon^{jka_{\ell}} S^{\langle k a_{1} \cdots a_{\ell-1} \rangle} n^{\langle a_{1} \cdots a_{\ell} \rangle} &= (\ell -1)! \epsilon^{ijk} n^{i} k^{k} S_{\ell} P'_{\ell}(\cos \theta)
\end{align}
with $P'_{\ell}(\cos\theta) \equiv d P_{\ell}(\cos\theta) / d \cos\theta$. By using $\hat{n} = (\sin\theta \cos\phi, \sin\theta \sin\phi, \cos\theta)$ and $\hat{k} = (0,0,1) $, the metric components can be written as\footnote{The vector harmonics in Eq.~\eqref{eq:g0jACMC}, which have parity $(-1)^{\ell}$, vanish when the spacetime is symmetric about the $\hat{k}$ axis~\cite{Cardoso:2016ryw}.}
\begin{align}
\nonumber
g_{00} &= - 1 + \dfrac{2M}{r} + \sum_{\ell \geq 2} \dfrac{1}{r^{\ell +1}} \left[ M_{\ell} P_{\ell}(\cos\theta) + \mathcal{S}_{\ell-1}  \right]  \\
\label{eq:g00multipoles}
&= -1 + \dfrac{2M}{r} + \dfrac{2M_{2}}{r^{3}}P_{2}(\cos\theta) + \cdots \\ \nonumber
g_{0j} &= -2 \sin^{2} \theta \sum_{\ell \geq 1}\dfrac{1}{r^{\ell}} \dfrac{S_{\ell}}{\ell}   \left[ P'_{\ell}(\cos\theta) + \mathcal{S}_{j,\ell-1} \right] \\
&= -\sin^{2} \theta \left[ \dfrac{2S_{1}}{r} + \dfrac{2 S_{3}}{3 r^{3}} P'_{3}(\cos\theta)  + \cdots \right].
\label{eq:g0jmultipoles}
\end{align}
Note that under reflection symmetry (i.e., \( \theta \rightarrow \pi - \theta \)), the Legendre polynomials \( P_{\ell}(\cos\theta) \) transform with a factor of \( (-1)^{\ell} \), while their derivatives \( P'_{\ell}(\cos\theta) \) transform with a factor of \( (-1)^{\ell+1} \). Therefore, imposing reflection symmetry on the metric implies that the mass multipole scalars \( M_{\ell} \) must vanish for odd \( \ell \), and the mass-current multipole scalars \( S_{\ell} \) must vanish for even \( \ell \).

Equations~\eqref{eq:g00multipoles} and~\eqref{eq:g0jmultipoles} can be used to extract the multipole moments of the star. To achieve this, we consider the \( g_{00} \) and \( g_{0j} \) components of the exterior Hartle-Thorne metric solution to $\mathcal{O}(\epsilon^{7})$, expressed in Boyer-Lindquist–type coordinates, and expand the solution about spatial infinity in inverse powers of the radial coordinate $r$. By comparing this asymptotic expansion with Eqs.~\eqref{eq:g00multipoles} and \eqref{eq:g0jmultipoles}, we can identify the scalar mass multipole moments \( M_{0}, M_{2}, M_{4}, M_{6} \) and their corrections, as well as the scalar mass-current multipole moments \( S_{1}, S_{3}, S_{5}, S_{7} \) and their corresponding corrections. 

Let us provide a concrete example. The asymptotic form of \( g_{00} \) and \( g_{0j} \) about spatial infinity is
\begin{align}
\nonumber
g_{00} &= -1 + \dfrac{2}{r} \left( M_{*} + C^{\,(2)}_{0,\, \textsf{ext}} \right) \\
&- \dfrac{2}{r^{3}} \left( \frac{S^2}{M_{\star }} + \frac{8}{5} M_{\star }^3 C^{\, (2)}_{2, \textsf{ext}} \right) P_{2}(\cos\theta) \epsilon^2 \, , \\[1ex] \nonumber
g_{0j} &= - \sin^{2}\theta \Bigg[  \dfrac{2}{r} \left( S \epsilon + C^{\,(3)}_{1, \, \textsf{ext}} \epsilon^{3} \right) \\ 
&+ \dfrac{ C^{\, (3)}_{3, \, \textsf{ext}}}{r^{3}} P'_{3}(\cos\theta)\epsilon^{3} \Bigg] \, ,
\end{align}
where we have left out terms of ${\cal{O}}(1/r^4)$.
Comparing these expressions to Eqs.~\eqref{eq:g00multipoles} and~\eqref{eq:g0jmultipoles}, we find that  
\begin{align}
    M_0 &= M_* + C^{ \,(2)}_{0,\textsf{ext}} \epsilon^2 + {\cal{O}}(\epsilon^4)\,,
    \\
    S_1 &= S \epsilon + C^{\, (3)}_{1,\textsf{ext}} \epsilon^3 + {\cal{O}}(\epsilon^5)\,,
    \\
    M_2 &= - \left( \frac{S^2}{M_{\star }} + \frac{8}{5} M_{\star }^3 C^{\, (2)}_{2, \textsf{ext}} \right)\epsilon^2 + {\cal{O}}(\epsilon^4)\,,
    \\
    S_3 &= \frac{3}{2} C^{\, (3)}_{3, \textsf{ext}} \epsilon^3 \, .
\end{align}
We have not included the higher-order in spin results here for simplicity, but they can be found in App.~\ref{apx:multipoles}.  
 
From expressions like the ones above, we can separate the $s$ spin-order correction to each of the multipole moments. Let us denote the mass multipole moments and their spin-induced corrections using the following expansion,
\begin{align}
M_{0} &= M_{*} + M^{(2)}_{0} \epsilon^{2} + M^{(4)}_{0} \epsilon^{4} + M^{(6)}_{0} \epsilon^{6}, \\
M_{2} &= M^{(2)}_{2} \epsilon^{2} + M^{(4)}_{2} \epsilon^{4} + M^{(6)}_{2} \epsilon^{6}, \\
M_{4} &= M^{(4)}_{4} \epsilon^{4} + M^{(6)}_{4} \epsilon^{6}, \\
M_{6} &= M^{(6)}_{6} \epsilon^{6},
\end{align}
and the mass-current multipole moments as
\begin{align}
S_{1} &= S \epsilon + S^{(3)}_{1} \epsilon^{3} + S^{(5)}_{1} \epsilon^{5} + S^{(7)}_{1} \epsilon^{7}, \\
S_{3} &= S^{(3)}_{3} \epsilon^{3} + S^{(5)}_{3} \epsilon^{5} + S^{(7)}_{3} \epsilon^{7}, \\
S_{5} &= S^{(5)}_{5} \epsilon^{5} + S^{(7)}_{5} \epsilon^{7}, \\
S_{7} &= S^{(7)}_{7} \epsilon^{7},
\end{align}
where the superscript in parentheses indicates the order in spin frequency \( \epsilon \) at which the corresponding correction enters. Explicit expressions for each of the different $s$-order contributions for these multipoles can be found in App.~\ref{apx:multipoles}, calculated up to \( \mathcal{O}(\epsilon^{7}) \)  and expressed in terms of the exterior integration constants \( C^{(n)}_{\ell,\,\textsf{ext}} \).

\subsection{Ryan's Method}

Given the existence of a fixed, global coordinate system, multipole moments can be defined as coefficients in an asymptotic expansion of a field about spatial infinity in terms of spherical harmonics, as done above. However, in GR, the lack of global coordinates makes such definition less useful. The asymptotic form of the metric invariably depends on the choice of coordinates, yet multipole moments should be coordinate-invariant. 
New definitions of multiple moments can be introduced to describe the source from a distance in GR and in a gauge-invariant way.
Geroch \cite{Geroch:1970cd} and Hansen \cite{Hansen:1974zz} provided such a coordinate-invariant multipolar description of asymptotically-flat stationary spacetimes. If additionally the spacetime is axially symmetric, the $\ell$th multipole moment can be represented by a single scalar moment $P_\ell=M_\ell+iS_\ell$. 

The goal of this section is to extract the Geroch-Hansen multipole moments up to $\ell=7$ for an arbitrary body that is isolated, asymptotically flat, stationary, axisymmetric, and possesses reflection symmetry. We do so following Ryan's method \cite{Ryan:1995wh}, which demonstrated that the multipole moments of a massive body of mass $M$ can be derived from the energy change per logarithmic interval of the orbital frequency of a small test particle with mass $\mu$ (i.e.~with $\mu \ll M$) inspiraling into the bigger object. Ryan’s initial work provided results up to $\ell=4$. Here, we extend this result to $\ell=7$, using the framework provided in Secs.~$2$ and $3$ of~\cite{Ryan:1995wh}. 

We begin by expressing the stationary, axially symmetric metric in Papapetrou form, which is defined by
\begin{equation}
ds^2 = -F(dt - \omega d\phi)^2 + \frac{1}{F} \left[ e^{2\gamma}(d\rho^2 + dz^2) + \rho^2 d\phi^2 \right],
\end{equation}
where here we neglect the influence of the much less massive orbiting object in the metric, while $F, \omega, \gamma$ are all functions of $\rho$ and $|z|$. This metric is thus written in cylindrical-like coordinates, where $\rho$ is the radial distance from the rotation axis, $z$ is the distance from the center and up the rotation axis, and $\phi$ is the azimuthal angular coordinate.  
For simplicity, assume that the test particle moves slowly and adiabatically from one circular geodesic in the equatorial plane ($z=0$) to another. On the timescale of one orbital period, the orbital angular frequency of the particle
\begin{equation}
\Omega_{\text{orb}} = \frac{d\phi}{dt} = \frac{-g_{t\phi,\rho} + \sqrt{(g_{t\phi,\rho})^2 - g_{tt,\rho}g_{\phi\phi,\rho}}}{g_{\phi\phi,\rho}}
\label{eq:omega}
\end{equation}
and its energy per mass
\begin{equation}
\frac{E}{\mu} = \frac{-g_{tt} - g_{t\phi}\Omega}{\sqrt{-g_{tt} - 2g_{t\phi}\Omega - g_{\phi\phi}\Omega^2}}
\label{eq:E/mu}
\end{equation}
can be obtained from the geodesic equation by imposing $d\rho/d\tau=0,\ d\rho^2/d\tau=0,\ dz/d\tau=0$. 

As the particle evolves from one geodesic to another, due to the loss of orbital energy to gravitational-wave emission, Eq.~\eqref{eq:E/mu} is not a constant any longer. 
The quantity of interest to us is the change of orbital energy per logarithmic interval of orbital angular frequency
\begin{equation}
\frac{\Delta E}{\mu} \equiv \frac{d(E/\mu)}{d(ln\Omega)}=-\Omega \frac{d(E/\mu)}{d\Omega},
\label{eq:dE}
\end{equation}
which is a coordinate invariant quantity. Inspecting Eq.~\eqref{eq:E/mu} and~\eqref{eq:dE}, one notices the need to find a relation between $\rho$ and $\Omega$, since, at the equator, the metric components are all functions of $\rho$. Fortunately, Eq.~\eqref{eq:omega} offers exactly that. The task then is to find a way to relate the metric components with multipole moments of the gravitational source object, so that $\Delta E/\mu$ can be written as a power series, whose coefficients are functions of the multipole moments.

Fodor et al. \cite{fodor1989multipole, Fodor:2020fnq} introduced an algorithm to express the scalar multipole moments in terms of the coefficients of a power-series expansion of the potential $\tilde{\xi}$, which is related to the Ernst potential by
\begin{equation}
\mathcal{E} = F + i\psi = \frac{\sqrt{\rho^2 + z^2} - \tilde{\xi}}{\sqrt{\rho^2 + z^2} + \tilde{\xi}}
\label{ernpot}
\end{equation}
where $F,\ \rho$ and $\psi$ are related to the metric by
\begin{equation}
g_{tt}=-F,
\label{eqn6}
\end{equation}
\begin{equation}
g_{t\phi} = F \int_{\rho}^{\infty} \frac{\rho'}{F^2} \frac{\partial \psi}{\partial z} d\rho' \bigg|_{\text{constant } z},
\label{eq:gtphi}
\end{equation}
and 
\begin{equation}
g_{\phi \phi} = \dfrac{ 1 } {g_{tt}} \left( g_{t\phi}^{2} - \rho^{2} \right) \, .
\label{eqn8}
\end{equation}
The Fodor et al.~algorithm offers a way to relate metric components to these scalar multipole moments, since the Ernst potential contains all the information necessary to reconstruct the metric. 
We have corrected a minus sign on the right-hand side of Eq.~\eqref{eq:gtphi}, which was present in Ryan's original paper~\cite{Ryan:1995wh}. For more details on the derivation of Eq.~\eqref{eq:gtphi}, we refer the reader to Ref.~\cite{Dietz:1988mg}.

Equations~\eqref{eqn6}-\eqref{eqn8} provide a way to express the relevant metric components in Eq.~\eqref{eq:dE} as functions of $\rho$ and $z$. Since the small object travels along geodesics in the equatorial plane, all three equations are evaluated at $z=0$. We then expand the metric components around spatial infinity, i.e.~$1/\rho \rightarrow 0$. After inserting the power-series expansions into Eqs.\ \eqref{eq:omega}-\eqref{eq:dE}, $\Delta E/\mu$ is now a function of $1/\rho$, and then, it too can be expressed as a power series in $\Omega^{1/3}$, using the relationship between $\rho$ and $\Omega$ given by Eq. \eqref{eq:omega}. At this stage, the expansion coefficients of $\Delta E/\mu$ are functions of $g_{tt},\ g_{t\phi}$ and $g_{\phi\phi}$. What remains to be done is to express these metric components in terms of multipole moments, which is the goal of Ryan's method. This is done using Fodor's algorithm, where $\tilde{\xi}$ serves as the bridge.

Substituting Eq.~\eqref{ernpot} into the Ernst equation, we obtain the field equation for $\tilde{\xi}$
\begin{equation}
\left(r^2 \tilde{\xi} \tilde{\xi}^* - 1\right) \Delta \tilde{\xi} = 2 \tilde{\xi}^* \left[ r^2 (\nabla \tilde{\xi})^2 + 2r \tilde{\xi} \nabla \tilde{\xi} \nabla r + \tilde{\xi}^2 \right].
\label{field eqn}
\end{equation}
The field $\tilde{\xi}$ can be expanded in a power series of the form
\begin{equation}
\tilde{\xi} = \sum_{j,k=0}^{\infty} a_{jk} \tilde{\rho}^j \tilde{z}^k = \sum_{j,k=0}^{\infty} a_{jk} \frac{\rho^j z^k}{(\rho^2 + z^2)^{j+k}}.
\label{xi}
\end{equation}
Substituting Eq. \eqref{xi} into Eq. \eqref{field eqn} yields a recursive relation for the coefficients of $\tilde{\xi}$, namely \cite{Fodor:2020fnq}:
\begin{widetext}
\begin{equation}
\begin{aligned}
(r + 2)^2 a_{r+2,s} &= -(s + 2)(s + 1)a_{r,s+2} 
 + \sum_{\substack{k+m+p=r \\ l+n+q=s}} a_{kl} \bar{a}_{mn} 
[a_{pq}(p^2 + q^2 - 2p - 3q  
 - 2k - 2l - 2pk - 2ql - 2)\\
&\quad + a_{p+2,q-2}(p + 2)(p + 2 - 2k) 
+ a_{p-2,q+2}(q + 2)(q + 1 - 2l)].
\end{aligned}
\label{ajk}
\end{equation}    
\end{widetext}
Using Eq.~\eqref{ajk}, we can express all $a_{jk}$ in terms of the coefficients that contain information of the equatorial plane metric, namely, $a_{j0}$ and $a_{j1}$ ($a_{j1}$ is needed due to the derivative in the integrand of Eq.~\eqref{eq:gtphi}). Expressing $\tilde{\xi}$ purely in terms of $a_{j0}$ and $a_{j1}$ is desirable because $\Delta E/\mu$ is determined by the metric around the equatorial plane, where $z=0$. Equation~\eqref{ajk} implies that $a_{jk}=0$ for odd $j$. Due to reflection symmetry across the equatorial plane, $a_{jk}$ is real for even $k$ and imaginary for odd $k$. The stationarity and axisymmetry of the spacetime is reflected in the absence of $t$ and $\phi$ dependence in Eq.~\eqref{xi}.

The multipole moments are related to $\tilde{\xi}$ in terms of $\tilde{\rho}=\rho/(\rho^2+z^2)$ and $\tilde{z}=z/(\rho^2+z^2)$ through \begin{equation}
M_\ell + iS_\ell = \left. \frac{S_0^{(\ell)}}{(2\ell - 1)!!} \right|_{\bar{\rho}=0, z=0},
\label{Pl}
\end{equation}
where $S_0^{(\ell)}$ can be recusively computed through 
\begin{equation}
S_0^{(0)} = \tilde{\xi}, \quad S_0^{(1)} = \frac{\partial \tilde{\xi}}{\partial \tilde{z}}, \quad S_1^{(1)} = \frac{\partial \tilde{\xi}}{\partial \tilde{\rho}},
\label{S0^n}
\end{equation}
\begin{align}
S_a^{(n)} &= \frac{1}{n} \left[ a \frac{\partial}{\partial \tilde{\rho}} S_{a-1}^{(n-1)} + (n - a) \frac{\partial}{\partial \tilde{z}} S_a^{(n-1)} \right] \nonumber \\
&\quad + a \left( \left[ a + 1 - 2n \right] \gamma_1 - \frac{a - 1}{\tilde{\rho}} \right) S_{a-1}^{(n-1)}\nonumber  \\
&\quad + (a - n)(a + n - 1) \gamma_2 S_a^{(n-1)} + a(a - 1) \gamma_2 S_{a-2}^{(n-2)} \nonumber \\
&\quad + (n - a)(n - a - 1) \left( \gamma_1 - \frac{1}{\tilde{\rho}} \right) S_{a+1}^{(n-1)} \nonumber \\
&\quad - \left[ a(a - 1) \tilde{R}_{11} S_{a-2}^{(n-2)} + 2a(n - a) \tilde{R}_{12} S_{a-1}^{(n-2)} \right. \nonumber \\
&\quad + \left. (n - a)(n - a - 1) \tilde{R}_{22} S_a^{(n-2)} \right] \left( n - \frac{3}{2} \right),
\label{Sa^n}
\end{align}
where $\tilde{R}_{11}$, $\tilde{R}_{12}$, $\tilde{R}_{22}$ can be computed from
\begin{equation}
\tilde{R}_{ij} = \left[ \left( \tilde{\rho}^2 + \tilde{\rho}^2 \right) \left| \tilde{\xi} \right|^2 - 1 \right]^{-2} \left( G_i G_j^* + G_i^* G_j \right),
\label{Rij}
\end{equation}
with
\begin{equation}
G_1 = z \frac{\partial \tilde{\xi}}{\partial\tilde{\rho}} - \tilde{\rho}\frac{\partial \tilde{\xi}}{\partial \tilde{z}}, \quad G_2 = \tilde{\rho} \frac{\partial \tilde{\xi}}{\partial \tilde{\rho}} + \tilde{z} \frac{\partial \tilde{\xi}}{\partial \tilde{z}} + \tilde{\xi},
\label{G}
\end{equation}
and from these,
\begin{equation}
\gamma_1 = \frac{\tilde{\rho}}{2} \left( \tilde{R}_{11} - \tilde{R}_{22} \right), \quad \gamma_2 = \tilde{\rho} \tilde{R}_{12}.
\label{gamma}
\end{equation}

To express $\Delta E/\mu$ up to the $\ell$th multipole moment, one can follow the below steps:
\begin{itemize}
    \item[(1)] Start by writing Eq.~\eqref{xi} with $a_{jk}$ up to $j+k=\ell$.
    \item[(2)] Use the recursive relation in Eq.~\eqref{ajk} to express Eq.~\eqref{xi} in terms of $a_{j0}$ and $a_{j1}$. 
    \item[(3)] Compute the relevant metric components using Eqs.\ \eqref{eqn6}-\eqref{eqn8}, and then expand $\Delta E/\mu$ in powers of $1/\rho$ on the equatorial plane by setting $z=0$. Invert Eq.~\eqref{eq:omega} to find $1/\rho$ as a function of $\Omega$, and then substitute the result into $\Delta E/\mu (1/\rho)$ to obtain the expansion of $\Delta E/\mu$ in powers of $\Omega^{1/3}$, where the coefficients are functions of $a_{j0}$ and $a_{j1}$. 
    \item[(4)] Use Eq.~\eqref{xi} with Eqs.\ \eqref{Pl}-\eqref{gamma} to express $a_{j0}$ and $a_{j1}$ in terms of the scalar multipole moments $S_i$ and $M_i$ where $i\leq \ell$.
    \item[(5)] Substitute $a_{j0}$ and $a_{j1}$ in the $\Delta E/\mu$ expansion to find an expression in terms of  multipole moments.
    \item[(6)] Redo steps (1)--(5) for $\ell+1$.    
\end{itemize}
Through the complex potential $\tilde{\xi}$, the behavior of the metric components in the equatorial plane is related to the multipole moments.

After performing the above steps up to $\ell=7$, we obtained the desired result in terms of $v\equiv (M\Omega)^{1/3}$, namely 
\begin{widetext}
\begin{align}
\nonumber
\frac{\Delta E}{\mu}  &= \frac{1}{3}v^{2} - \frac{1}{2}v^{4} + \frac{20}{9}\frac{S_{1}}{M^{\,2}_{0}} v^{5} + \left(  \frac{M_{2}}{M_{0}^{\, 3}} -\frac{27}{8} \right) v^{6} + \frac{28}{3}\frac{S_{1}}{M_{0}^{\,2}}v^{7} + \left(\frac{80}{27} \frac{S_{1}^{\,2}}{M_{0}^{\, 4}} + \frac{70}{9}\frac{M_{2}}{M_{0}^{\, 3}}  - \frac{225}{16} \right)v^{8} \hspace{0.2cm} \\ \nonumber
&+ \left( \frac{81}{2} \frac{S_{1}}{M_{0}^{\,2}} + 6 \frac{S_{1}M_{2}}{M_{0}^{\,5}} - 6 \frac{S_{3}}{M_{0}^{\, 4}} \right) v^{9} + \left(\dfrac{115}{18}\dfrac{S_{1}^{\, 2}}{M_{0}^{\, 4}} + \dfrac{935}{24} \dfrac{M_{2}}{M_{0}^{\, 3}} + \dfrac{35}{12}\dfrac{M_{2}^{\, 2}}{M_{0}^{\, 6}} - \dfrac{35}{12}\dfrac{M_{4}}{M_{0}^{\, 5}} - \dfrac{6615}{128}  \right)v^{10} \\ \nonumber
&+ \left( 165 \dfrac{S_{1}}{M_{0}^{\, 2}} + \dfrac{1408}{243} \dfrac{S_{1}^{\, 3}}{M_{0}^{\, 6}} + \dfrac{968}{27} \dfrac{S_{1}M_{2}}{M_{0}^{\, 5}} - \dfrac{352}{9}\dfrac{S_{3}}{M_{0}^{\, 4}} \right) v^{11} \\ \nonumber
&+ \left( \dfrac{9147}{56} \dfrac{M_{2}}{M_{0}^{\, 3}} + \dfrac{93}{4} \dfrac{M_{2}^{\, 2}}{M_{0}^{\, 6}} + 24 \dfrac{S_{1}^{\, 2} M_{2} }{M_{7}} - 24 \dfrac{S_{1}S_{3}}{M_{0}^{\, 6}} - \dfrac{99}{4} \dfrac{M_{4}}{M_{0}^{\, 5}} - \dfrac{123}{14} \dfrac{S_{1}^{\, 2}}{M^{4}} - \dfrac{45927}{256} \right)v^{12} \\ \nonumber
&+ \left(\frac{260 M_{2}^2 S_{1}}{9 M_{0}^8}-\frac{65 M_{2} S_{3}}{3 M_{0}^7}-\frac{325 M_{4} S_{1}}{18 M_{0}^7}+\frac{15080 S_{1}^3}{567 M_{0}^6}+\frac{65 S_{5}}{6 M_{0}^6}+\frac{32435 M_{2} S_{1}}{252 M_{0}^5}-\frac{6305 S_{3}}{36 M_{0}^4}+\frac{20475 S_{1}}{32 M_{0}^2}\right) v^{13} \\ \nonumber
&+ \bigg(\frac{385 M_{2}^3}{36 M_{0}^9}-\frac{385 M_{2} M_{4}}{24 M_{0}^8}+\frac{8624 S_{1}^4}{729 M_{0}^8}+\frac{13766 M_{2} S_{1}^2}{81 M_{0}^7}+\frac{385 M_{6}}{72 M_{0}^7}+\frac{100411 M_{2}^2}{864 M_{0}^6}-\frac{161 S_{1} S_{3}}{M_{0}^6}-\frac{38045 M_{4}}{288 M_{0}^5} \\ \nonumber
&-\frac{82027 S_{1}^2}{432 M_{0}^4}+\frac{2160829 M_{2}}{3456 M_{0}^3}-\frac{617463}{1024}\bigg) v^{14} 
+ \bigg(\frac{80 M_{2} S_{1}^3}{M_{0}^9}+\frac{255 M_{2}^2 S_{1}}{M_{0}^8}-\frac{80 S_{1}^2 S_{3}}{M_{0}^8}-\frac{195 M_{2} S_{3}}{M_{0}^7} \\ \nonumber
&-\frac{305 M_{4} S_{1}}{2 M_{0}^7}+\frac{2390 S_{1}^3}{21 M_{0}^6}+\frac{185 S_{5}}{2 M_{0}^6}+\frac{6625 M_{2} S_{1}}{21 M_{0}^5}-\frac{4045 S_{3}}{6 M_{0}^4}+\frac{76545 S_{1}}{32 M_{0}^2} \bigg) v^{15} 
+ \bigg(\frac{4576 M_{2}^2 S_{1}^2}{27 M_{0}^{10}}+\frac{988 M_{2}^3}{9 M_{0}^9} \\ \nonumber
&-\frac{1664 M_{2} S_{1} S_{3}}{9 M_{0}^9}-\frac{2080 M_{4} S_{1}^2}{27 M_{0}^9}-\frac{2951 M_{2} M_{4}}{18 M_{0}^8}+\frac{125840 S_{1}^4}{1701 M_{0}^8}+\frac{520 S_{1} S_{5}}{9 M_{0}^8}+\frac{104 S_{3}^2}{3 M_{0}^8}+\frac{181246 M_{2} S_{1}^2}{231 M_{0}^7} \\ \nonumber
& +\frac{325 M_{6}}{6 M_{0}^7}+\frac{863213 M_{2}^2}{1848 M_{0}^6}-\frac{202384 S_{1} S_{3}}{297 M_{0}^6}-\frac{151073 M_{4}}{264 M_{0}^5}-\frac{83161 S_{1}^2}{63 M_{0}^4}+\frac{4579627 M_{2}}{2016 M_{0}^3}-\frac{4065633}{2048}\bigg) v^{16} \\ \nonumber
&+ \bigg(\frac{7735 M_{2}^3 S_{1}}{54 M_{0}^{11}}-\frac{595 M_{2}^2 S_{3}}{6 M_{0}^{10}}-\frac{2975 M_{2} M_{4} S_{1}}{18 M_{0}^{10}}+\frac{53312 S_{1}^5}{2187 M_{0}^{10}}+\frac{5428253 M_{2} S_{1}^3}{8019 M_{0}^9}+\frac{595 M_{2} S_{5}}{12 M_{0}^9}+\frac{595 M_{4} S_{3}}{12 M_{0}^9} \\ \nonumber
&+\frac{4165 M_{6} S_{1}}{108 M_{0}^9}+\frac{64360181 M_{2}^2 S_{1}}{46332 M_{0}^8}-\frac{7559033 S_{1}^2 S_{3}}{11583 M_{0}^8}-\frac{595 S_{7}}{36 M_{0}^8}-\frac{2572967 M_{2} S_{3}}{2376 M_{0}^7}-\frac{48086353 M_{4} S_{1}}{61776 M_{0}^7} \\
&+\frac{937873 S_{1}^3}{1782 M_{0}^6}+\frac{918323 S_{5}}{1872 M_{0}^6}+\frac{17042483 M_{2} S_{1}}{57024 M_{0}^5}-\frac{45785369 S_{3}}{19008 M_{0}^4}+\frac{2226609 S_{1}}{256 M_{0}^2}\bigg) v^{17}
\label{eq:dE/mu}
\end{align} 
\end{widetext}
As a test of the correctness of our results, we have checked that we recover the well-known no-hair results for a Kerr black hole, namely
\begin{align}
M_{2\ell} &= (-1)^{\ell} M a^{2 \ell} \\
S_{2 \ell + 1} &= (-1)^{\ell} M a^{2 \ell + 1}\,,
\end{align}
up to the mass multipole \( M_6 \) and the mass-current multipole \( S_7 \).

In order to extract the multipole moments of a spacetime, we must now compute $\Delta E/\mu$ for that spacetime and them compare to Eq.~\eqref{eq:dE/mu}. More specifically, if one has a metric, written in any coordinate system, for a spacetime that is stationary, axisymmetric with reflection symmetry across the equator, one can  calculate $\Delta E/\mu$ and compare the result to Eq.\ \eqref{eq:dE} to read out the multipoles $(M_0, S_1, M_2, S_3, M_4, S_5, M_6, S_7)$ order by order in $v$. We applied the above (Ryan's) method to compute the multipole moments of a neutron star, using the exact analytical form of the Hartle-Thorne exterior metric, expanded up to order $\mathcal{O}(\epsilon^{7})$. Given the complexity and size of the expressions for each metric component, and because the multipole extraction requires an expansion at spatial infinity, it is computationally convenient to assume an asymptotic form for the relevant metric components ($g_{tt}$, $g_{\phi \phi}$, and $g_{t \phi}$) of the form
\begin{align}
g_{tt} &= -1 + \sum_{n=1}^{\infty} \dfrac{a_{n}}{r^{n}} \,  \\
g_{\phi \phi} &= r^{2} + \sum_{n=1}^{\infty} \dfrac{b_{n}}{r^{n}} \, \\
g_{t \phi} &= \sum_{n=1}^{\infty} \dfrac{c_{n}}{r^{n}} \, ,
\end{align}
We can then use this functional form to compute $\Omega_{\text{orb}}$ and therefore $\Delta E/\mu$ as a function of $\Omega_{\text{orb}}$ or in powers of $v$. Then, we can compare this result to that in Eq.~\eqref{eq:dE/mu} to extract the multipole moments as functions of the $\{a_n,b_n,c_n\}$. Once that is achieved, one can then insert expressions for the expansion constants $\{a_n,b_n,c_n\}$ in terms of the asymptotic exterior constants of the exterior Hartle-Thorne solution to obtain the multipole moments in terms of the latter. Doing so, we obtain exactly the same result as when working in ACMC coordinates, thus validating our results. 

\section{Discussion and Conclusions}
\label{sec:discussion}

The extension of the Hartle-Thorne approximation to seventh-order in spin has revealed several theoretical and practical insights regarding the structure and observable properties of rotating neutron stars. One of the primary advantages of increasing the spin-order in the perturbative expansion is the ability to extract both higher-order multipole moments and their corrections, as summarized in Table \ref{table:multipoles}. This is closely tied to the harmonic structure of the perturbative equations: at spin-frequency order $n$, only harmonics up to $\ell \leq n$ appear in the spherical-harmonic decomposition, as explained in Sec.~\ref{sec:slow-rot}. Since the extraction of the multipole moments—such as through Thorne's expansion—requires the presence of multipoles of degree $\ell$ to be proportional to the product of $r^{-(\ell+1)}$ and harmonics of mode $\ell$, multipoles with $\ell > n$ are absent from the decomposition unless the expansion is carried to at least $\mathcal{O}(\epsilon^n)$. Therefore, higher-order expansions not only increase the number of accessible moments but also improve the accuracy of lower-order ones by capturing additional corrections.

This spin-frequency extension also sets the stage for exploring several key astrophysical and nuclear physics applications. One promising direction involves computing the corrections to the equatorial radius of the star, which can be achieved via an invariant parameterization of the stellar surface as outlined in~\cite{Hartle:1968si}. This would allow for the extraction of second-, fourth-, and sixth-order spin corrections to the TOV radius $R_{*}$, providing more accurate mass-radius profiles for use in pulse-profile modeling and X-ray observations.
Another application pertains to the determination of the mass-shedding limit within the Hartle-Thorne framework. Following the methodology in \cite{Friedman:1986tx, Benhar:2005gi}, the extended equations could be used to assess the approximation's validity in predicting the maximum spin frequency of neutron stars before mass loss occurs. Such corrections are crucial for interpreting rapidly spinning pulsars and constraining the equation of state.

As illustrated in Tables \ref{table:mass-sensi} and \ref{table:current-sens}, the spin corrections to the multipole moments become sensitive to these higher-order derivatives. This opens up the possibility of probing the microphysics of dense matter—such as the presence of higher-order phase transitions—by examining the sensitivity of certain multipole corrections to the equation of state~\cite{ReinkePelicer:2025vuh}. 

In addition to these theoretical insights, our work paves the way for further follow-up studies. In the second paper of this series, we will numerically solve the interior equations using various realistic equations of state and compute the full set of multipole moments and their corrections for different spin frequencies. The implementation will build on QLIMR \cite{zenodo_qlimr}, a recently released code for computing gravitational neutron star observables up to second-order in spin frequency, developed within the MUSES collaboration and described in \cite{ReinkePelicer:2025vuh}; our work will serve as the foundation for extending the QLIMR solve to seventh order in spin.
This calculation will allow for direct comparison with full numerical relativity and provide concrete estimates of systematic errors in observable quantities, such as the mass, radius, quadrupole moment and higher multipoles. Precise model-to-data comparisons are important for providing accurate constrations for the neutron star equation of state \cite{MUSES:2023hyz}. Such estimates are particularly important for interpreting X-ray timing data from NICER and future missions, where precision measurements demand robust theoretical modeling.
Moreover, this work lays the groundwork for studying quasi-universal or effective no-hair relations between higher multipole moments and the lowest-order ones. These relations, which are largely insensitive to the underlying equation of state, can serve as powerful tools to reduce degeneracies in parameter estimation from both electromagnetic and gravitational-wave observations.

\begin{acknowledgments}
We thank Kent Yagi for his valuable feedback on the Hartle–Thorne metric expanded to fourth order in spin-frequency.
This work was supported by the National Science Foundation (NSF) within the framework of the MUSES collaboration, under grant number OAC-2103680. 
\end{acknowledgments}

\renewcommand{\arraystretch}{1.5}
\newcolumntype{Y}{>{\centering\arraybackslash}X}

\begin{widetext}
    
\begin{table}[h!]
\centering
\begin{minipage}{0.48\textwidth}
\centering
\begin{tabularx}{0.7\textwidth}{|Y|Y|Y|Y|}
\hline
$M^{(s)}_{\ell}$ & $c_s^2$ & $c_s^{\prime 2}$ & $c_s^{\prime\prime 2}$ \\ \hline
\multicolumn{4}{|c|}{Mass monopole moment} \\ \hline
$M^{(0)}_{0}$ & No & No & No \\ \hline
$M^{(2)}_{0}$ & Yes & No & No \\ \hline
$M^{(4)}_{0}$ & Yes & Yes & No \\ \hline
$M^{(6)}_{0}$ & Yes & Yes & Yes \\ \hline
\multicolumn{4}{|c|}{Mass quadrupole moment} \\ \hline
$M^{(2)}_{2}$ & No & No & No \\ \hline
$M^{(4)}_{2}$ & Yes & No & No \\ \hline
$M^{(6)}_{2}$ & Yes & Yes & No \\ \hline
\multicolumn{4}{|c|}{Mass hexadecapole moment} \\ \hline
$M^{(4)}_{4}$ & Yes & No & No \\ \hline
$M^{(6)}_{4}$ & Yes & Yes & No \\ \hline
\multicolumn{4}{|c|}{Mass hexacontatetrapole moment} \\ \hline
$M^{(6)}_{6}$ & Yes & Yes & No \\ \hline
\end{tabularx}
\caption{Dependence of the mass multipole moments and their corrections on derivatives of the squared speed of sound, \( c_s^2 \), with respect to the energy density \( \varepsilon \). The symbol ``  $ ' $ " denotes derivative with respect to energy density $\varepsilon$.}
\label{table:mass-sensi}
\end{minipage}
\hspace{1em}
\begin{minipage}{0.48\textwidth}
\centering
\begin{tabularx}{0.7\textwidth}{|Y|Y|Y|Y|}
\hline
$S^{(k)}_{\ell}$ & $c_s^2$ & $c_s^{\prime 2}$ & $c_s^{\prime\prime 2}$ \\ \hline
\multicolumn{4}{|c|}{Mass-current dipole moment} \\ \hline
$S^{(1)}_{1}$ & No & No & No \\ \hline
$S^{(3)}_{1}$ & Yes & No & No \\ \hline
$S^{(5)}_{1}$ & Yes & Yes & No \\ \hline
$S^{(7)}_{1}$ & Yes & Yes & Yes \\ \hline
\multicolumn{4}{|c|}{Mass-current octupole moment} \\ \hline
$S^{(3)}_{3}$ & Yes & No & No \\ \hline
$S^{(5)}_{3}$ & Yes & Yes & No \\ \hline
$S^{(7)}_{3}$ & Yes & Yes & Yes \\ \hline
\multicolumn{4}{|c|}{Mass-current dotriacontapole moment} \\ \hline
$S^{(5)}_{5}$ & Yes & Yes & No \\ \hline
$S^{(7)}_{5}$ & Yes & Yes & Yes \\ \hline
\multicolumn{4}{|c|}{Mass-current hectoicosaoctapole moment} \\ \hline
$S^{(7)}_{7}$ & Yes & Yes & Yes \\ \hline
\end{tabularx}
\caption{Dependence of the mass-current multipoles and their corrections on derivatives of the squared speed of sound, \( c_s^2 \), with respect to the energy density \( \varepsilon \). The symbol ``  $ ' $ " denotes derivative with respect to energy density $\varepsilon$.}
\label{table:current-sens}
\end{minipage}
\end{table}
\end{widetext}

\appendix

\begin{widetext}
    
\section{Metric form in the \textsf{HT} frame}
\label{apx:HTmetric}

\begin{align}
\nonumber
g^{\textsf{HT}}_{tt}(R,\Theta) &=
%
%
-e^{\nu}
%
%
+ \epsilon ^2 \left[R^{2}\sin^{2}{\Theta} \left( \omega^{(1)} \right)^2-e^{\nu } \left(2 h^{(2)}+\xi^{(2)} \nu' \right)\right] 
%
%
+\epsilon^{4} \bigg\{-\frac{1}{2} e^{\nu } \bigg[\xi^{(2)} \left(4 \partial_{R} h^{(2)}+\xi^{(2)} \left(\nu''+\nu '^2\right)\right) \\ \nonumber 
&+ 4 h^{(2)} \xi^{(2)} \nu '+4 h^{(4)}+2 \xi^{(4)} \nu '\bigg] +  2 R \sin^{2}\Theta \omega^{(1)} \left[\xi^{(2)} \left( R \partial_{R} \omega^{(1)}+\omega^{(1)}\right)+k^{(2)} R \omega^{(1)}+R \omega^{(3)}\right]\bigg\} \\ \nonumber
%
%
&+\epsilon ^6 \bigg\{ -\frac{1}{6} e^{\nu } \bigg[12 \left(\xi^{(2)}\right)^{2} \nu' \partial_{R}h^{(2)} + 6 \left( \xi^{(2)}\right)^{2} \partial^{2}_{R} h^{(2)} + 12 \xi^{(4)} \partial_{R} h^{(2)} + 12 \xi^{(2)} \partial_{R} h^{(4)} +6 h^{(2)} \bigg[2 \xi^{(4)} \nu ' \\ \nonumber
&+ \left(\xi^{(2)}\right)^{2} \left(\nu ''+\nu'^2 \right) \bigg]+12 h^{(4)} \xi^{(2)}\nu' + 12 h^{(6)} + \left( \xi^{(2)} \right)^{3} \left(  \nu''' + \nu'^{3} + 3 \nu' \nu'' \right) + 6\xi^{(2)} \xi^{(4)}  \nu'' + 6 \xi^{(2)} \xi^{(4)} \nu'^{2}\\ \nonumber
& + 6\xi^{(6)} \nu '\bigg] + \sin^{2}\Theta \bigg[2 R^2 \xi^{(2)}  \left(\omega^{(1)}\right)^{2} \partial_{R} k^{(2)} + 4 R k^{(2)}  \omega^{(1)} \left[\xi^{(2)} \left(R \partial_{R} \omega^{(1)} + \omega^{(1)}\right) + R \omega^{(3)}\right] \\ \nonumber
&+ R^{2} \left( \xi^{(2)} \right)^{2}  \left(\partial_{R}\omega^{(1)}\right)^{2}
+ 4 R \left( \xi^{(2)} \right)^{2}  \omega^{(1)} \partial_{R} \omega^{(1)} + R^2 \left(\xi^{(2)}\right)^{2} \omega^{(1)} \partial^{2}_{R} \omega^{(1)} + 2 R^{2} \xi^{(2)}  \omega^{(3)} \partial_{R} \omega^{(1)}  \\ \nonumber
&+ 2 R^{2} \xi^{(2)} \omega^{(1)} \partial_{R}\omega^{(3)} + 2 R^{2} \xi^{(4)}  \omega^{(1)} \partial_{R} \omega^{(1)} + 2 R^{2} k^{(4)}  \left(\omega^{(1)}\right)^{2} + \left(\xi^{(2)}\right)^{2} \left(\omega^{(1)}\right)^{2} + 4 R \xi^{(2)}  \omega^{(1)} \omega^{(3)} \\ 
&+ 2 R \xi^{(4)} \left(  \omega^{(1)} \right)^{2} + 2 R^{2} \omega^{(1)} \omega^{(5)} + R^{2} \left( \omega^{(3)} \right)^{2}  \bigg]\bigg\}
\end{align}

\begin{align}
\nonumber
g^{\textsf{HT}}_{RR}(R,\Theta) &= 
%
%
e^{\lambda } 
%
%
+ \epsilon ^2 e^{\lambda } \left[ \frac{2 m^{(2)}}{R-2 M}+\lambda ' \xi^{(2)} + 2 \partial_{R} \xi^{(2)} \right] 
%
%
+ \frac{1}{2} \epsilon^{4} e^{\lambda } \Bigg\{ \frac{4 \xi^{(2)} }{(R-2 M)^2} \bigg( m^{(2)} \left[2 M'+(R-2 M) \lambda '-1\right] \\ \nonumber
&+ (R-2 M) \partial_{R} m^{(2)}\bigg) +\frac{8 m^{(2)} \partial_{R} \xi^{(2)}}{R-2 M} + \frac{4 m^{(4)}}{R-2 M}+4 \lambda' \xi^{(2)} \partial_{R} \xi^{(2)} +2 \lambda' \xi^{(4)} +  \left(\lambda ''+\lambda '^2\right)\left(\xi^{(2)}\right)^2 \\ \nonumber
&+ 2 \left[ \left( \partial_{R} \xi^{(2)} \right)^{2} + 2 \partial_{R} \xi^{(4)}\right] \Bigg\}
%
%
+ \frac{1}{6} \epsilon ^6 e^{\lambda } \Bigg\{ \frac{24  \xi^{(2)} \partial_{R}\xi^{(2)} }{(R-2 M)^2}\left[ m^{(2)} \left(2 M'+(R-2 M) \lambda '-1\right)+(R-2 M) \partial_{R} m^{(2)}\right] \\ \nonumber
&+ \frac{12 \xi^{(4)} }{(R-2 M)^2} \left[m^{(2)} \left(2 M'+(R-2 M) \lambda '-1\right)+(R-2 M) \partial_{R} m^{(2)}\right] \\ \nonumber
&+ \frac{12 \xi^{(2)} }{(R-2 M)^2} \left[ m^{(4)} \left(2 M'+(R-2 M) \lambda '-1\right)+(R-2 M) \partial_{R} m^{(4)}\right] \\ \nonumber
&+ \frac{6 \left(\xi^{(2)}\right)^2 }{(R-2 M)^3} \bigg[ (R-2 M) \left(2 \partial_{R} m^{(2)} \left(2 M'+(R-2 M) \lambda '-1\right)+(R-2 M) \partial^{2}_{R} m^{(2)}\right) \\ \nonumber
&+ m^{(2)} \left((R-2 M) \left(2 M''+(R-2 M) \lambda ''+\lambda ' \left((R-2 M) \lambda '-2\right)\right)+4 M' \left((R-2 M) \lambda '-2\right)+8 M'^2+2\right) \bigg] \\ \nonumber
&+ \frac{12 m^{(2)} \left(\left( \partial_{R} \xi^{(2)} \right)^{2}+2 \partial_{R}\xi^{(4)}\right)}{R-2 M}+\frac{24 m^{(4)} \partial_{R}\xi^{(2)}}{R-2 M}+\frac{12 m^{(6)}}{R-2 M}+6 \lambda ' \xi^{(2)} \left[\left( \partial_{R} \xi^{(2)} \right)^{2}+2 \partial_{R}\xi^{(4)}\right] \\ \nonumber
&+12 \lambda ' \xi^{(4)} \partial_{R}\xi^{(2)} + 6 \lambda' \xi^{(6)} + 6 \left( \xi^{(2)}\right)^{2} \left(\lambda '' + \lambda'^{2}\right)  \partial_{R}\xi^{(2)} + 6 \xi^{(2)} \xi^{(4)} \left(\lambda ''+\lambda '^2\right) \\ 
&+ \left( \xi^{(2)}\right)^{3} \left(\lambda'''+\lambda '^3+3 \lambda ' \lambda ''\right)+12 \left(\partial_{R}\xi^{(2)} \partial_{R}\xi^{(4)} + \partial_{R} \xi^{(6)} \right)\Bigg\}
\end{align}

\begin{align}
\nonumber
g^{\textsf{HT}}_{\Theta\Theta}(R,\Theta) &=
%
%
R^{2} 
%
%
+ 2 R \epsilon^{2} \left( R k^{(2)} + \xi^{(2)} \right) 
%
%
+ \epsilon^{4} \bigg[ 2 R \left(R \partial_{R} k^{(2)}+2 k^{(2)}\right) \xi^{(2)}+2 R^2 k^{(4)}+e^{\lambda} \left(\partial_{\Theta} \xi^{(2)} \right)^{2} + \left( \xi^{(2)} \right)^{2} \\ \nonumber
&+ 2 R \xi^{(4)} \bigg] 
%
%
+ \epsilon^{6} \bigg\{ \left[R \left(4 \partial_{R} k^{(2)}+R \partial^{2}_{R} k^{(2)} \right)+2 k^{(2)} \right] \left(\xi^{(2)}\right)^{2} + 2 R \left( R \partial_{R} k^{(2)} + 2 k^{(2)}\right) \xi^{(4)} + 2 e^{\lambda} \partial_{\Theta} \xi^{(2)} \partial_{\Theta} \xi^{(4)}\\ 
&+2 R \left(R \partial_{R} k^{(4)} + 2 k^{(4)}\right)\xi^{(2)} + 2 R^2 k^{(6)} + \left[ \frac{2 e^{\lambda} m^{(2)}}{R-2 M} + e^{\lambda} \lambda' \xi^{(2)} \right] \left( \partial_{\Theta} \xi^{(2)} \right)^{2} + 2 \xi^{(2)} \xi^{(4)} + 2 R \xi^{(6)} \bigg\}
\end{align}

\begin{align}
\nonumber
g^{\textsf{HT}}_{\phi \phi}(R,\Theta)  &=
%
%
R^{2} \sin^{2}\Theta
%
%
+ 2 \epsilon ^2 R \sin^{2}\Theta \left( Rk^{(2)} + \xi^{(2)} \right)
%
%
+ \epsilon ^4 \sin^{2}\Theta \bigg[2  R^{2} \xi^{(2)} \partial_{R} k^{(2)} + 4 Rk^{(2)} \xi^{(2)} + 2 R^{2} k^{(4)} + \left( \xi^{(2)}\right)^{2} \\ \nonumber 
&+ 2R \xi^{(4)} \bigg] 
%
%
+\sin^{2}\Theta \epsilon ^6 \bigg\{ 2 \bigg[2 R \left( \xi^{(2)} \right)^{2}  \partial_{R} k^{(2)} + R\left( R \xi^{(4)} \partial_{R}k^{(2)} + \xi^{(6)}\right)+\xi^{(2)} \left(R^{2} \partial_{R} k^{(4)} + \xi^{(4)} \right) \\ 
&+ k^{(2)} \left( \left(\xi^{(2)}\right)^{2} + 2R\xi^{(4)}\right) \bigg] 
+ R^{2} \left( \xi^{(2)} \right)^{2} \partial^{2}_{R} k^{(2)} + 4 R k^{(4)} \xi^{(2)} + 2 R^{2} k^{(6)}  \bigg\}
\end{align}

\begin{align}
\nonumber
g^{\textsf{HT}}_{R\Theta}(R,\Theta) &= \epsilon ^2 e^{\lambda } \left( \partial_{\Theta} \xi^{(2)} \right)
+\epsilon ^4 e^{\lambda } \left[  \left(\frac{2 m^{(2)}}{R-2 M}+\lambda ' \xi^{(2)} + \partial_{R} \xi^{(2)}\right) \partial_{\Theta} \xi^{(2)} + \partial_{\Theta} \xi^{(4)}\right]
+\frac{1}{2} \epsilon ^6 e^{\lambda } \bigg\{ \frac{4 m^{(4)} }{R-2 M}\partial_{\Theta} \xi^{(2)} \\ \nonumber
&+ \frac{4 \xi^{(2)} \partial_{\Theta}\xi^{(2)} }{(R-2 M)^2} \left\{ m^{(2)} \left[ 2 M'+(R-2 M) \lambda '-1\right] + (R-2 M) \partial_{R} m^{(2)}\right\} +\frac{4 m^{(2)} \left(  \partial_{R} \xi^{(2)}\partial_{\Theta} \xi^{(2)} + \partial_{\Theta} \xi^{(4)}\right)}{R-2 M} \\ \nonumber
&+2 \lambda ' \xi^{(2)} \left(  \partial_{R} \xi^{(2)} \partial_{\Theta} \xi^{(2)} + \partial_{\Theta} \xi^{(4)}\right) + 2 \lambda' \xi^{(4)} \partial_{\Theta} \xi^{(2)} + \left(\lambda ''+\lambda '^2\right) \left( \xi^{(2)} \right)^{2}  \partial_{\Theta} \xi^{(2)} \\ 
&+2 \left[ \partial_{R} \xi^{(2)}   \partial_{\Theta} \xi^{(4)}  + \partial_{\Theta} \xi^{(2)}  \partial_{R} \xi^{(4)}  + \partial_{R} \xi^{(6)}\right] \bigg\}
\label{eq:gRTheta}
\end{align}

\begin{align}
\nonumber
g^{\textsf{HT}}_{t\phi}(R,\Theta) &= 
%
%
- \epsilon R^{2}\sin ^2\Theta \omega^{(1)} 
%
%
- \epsilon ^3 R \sin ^2\Theta \left[2 R k^{(2)} \omega^{(1)} + \xi^{(2)} \left(R \partial_{R}\omega^{(1)} + 2 \omega^{(1)}\right) + R \omega^{(3)}\right] \\ \nonumber
%
%
&-\frac{1}{2} \epsilon^{5} \sin^{2}\Theta \Bigg\{ 2 \bigg[R \xi^{(2)} \left(2 R \omega^{(1)} \partial_{R} k^{(2)} +R \partial_{R} \omega^{(3)} + 2 \omega^{(3)} \right) + \left( \xi^{(2)}\right)^{2} \left( 2 R \partial_{R} \omega^{(1)}+\omega^{(1)}\right) \\ \nonumber
&+ R \left( \xi^{(4)} \left(R \partial_{R} \omega^{(1)}+2 \omega^{(1)}\right)+R \omega^{(5)}\right)\bigg] + 4 R k^{(2)} \left[\xi^{(2)} \left(R \partial_{R} \omega^{(1)}+2 \omega^{(1)}\right)+R \omega^{(3)}\right] \\ \nonumber
&+ 4 R^2 k^{(4)} \omega^{(1)} + R^{2} \left( \xi^{(2)}\right)^{2} \partial^{2}_{R} \omega^{(1)}\Bigg\} 
%
%
-\frac{1}{6} \epsilon^{7} \sin^{2}\Theta \Bigg\{ 12 R^{2}  \left( \xi^{(2)} \right)^{2} \partial_{R} k^{(2)} \partial_{R} \omega^{(1)}+6 R^{2} \left( \xi^{(2)} \right)^{2} \omega^{(1)}\partial^{2}_{R} k^{(2)}  \\ \nonumber
&+ 12 R^{2}  \omega^{(3)} \xi^{(2)}  \partial_{R} k^{(2)} + 12 R^{2}  \xi^{(4)} \omega^{(1)}\partial_{R} k^{(2)} + 24 R  \left( \xi^{(2)} \right)^{2} \omega^{(1)}\partial_{R} k^{(2)} + 6 R^{2} k^{(2)} \left( \xi^{(2)} \right)^{2} \partial^{2}_{R}\omega^{(1)}  \\ \nonumber
&+12 R^2 k^{(2)} \xi^{(2)} \partial_{R} \omega^{(3)}+12 R^2 k^{(2)} \xi^{(4)} \partial_{R} \omega^{(1)} +12 R^2 k^{(2)} \omega^{(5)}+24 R k^{(2)} \left( \xi^{(2)} \right)^{2} \partial_{R} \omega^{(1)}\\ \nonumber
&+12 k^{(2)} \left( \xi^{(2)} \right)^{2} \omega^{(1)} + 24 R k^{(2)} \xi^{(2)} \omega^{(3)} + 24 R k^{(2)} \xi^{(4)} \omega^{(1)}+12 R^2  \xi^{(2)} \omega^{(1)}\partial_{R} k^{(4)} + 12 R k^{(4)} \bigg[\xi^{(2)} \Big(R \partial_{R} \omega^{(1)} \\ \nonumber
&+2 \omega^{(1)}\Big)+R \omega^{(3)}\bigg] +12 R^2 k^{(6)} \omega^{(1)}+6 R^2 \xi^{(2)} \xi^{(4)} \partial^{2}_{R}\omega^{(1)} +R^2 \left( \xi^{(2)} \right)^{3} \partial^{3}_{R} \omega^{(1)}+3 R^2 \left( \xi^{(2)} \right)^{2} \partial^{2}_{R} \omega^{(3)}   \\ \nonumber
&+6 R^2 \xi^{(2)} \partial_{R} \omega^{(5)} +6 R^2 \xi^{(4)} \partial_{R} \omega^{(3)}+6 R^2 \xi^{(6)} \partial_{R} \omega^{(1)}+6 R^2 \omega^{(7)}+24 R \xi^{(2)} \xi^{(4)} \partial_{R} \omega^{(1)}+12 \xi^{(2)} \xi^{(4)} \omega^{(1)} \\ \nonumber
&+6 \left( \xi^{(2)} \right)^{3} \partial_{R} \omega^{(1)} +6 R \left( \xi^{(2)} \right)^{3} \partial^{2}_{R}\omega^{(1)} + 12 R \left( \xi^{(2)} \right)^{2} \partial_{R} \omega^{(3)}+6 \left( \xi^{(2)} \right)^{2} \omega^{(3)}+12 R \xi^{(2)} \omega^{(5)} \\ 
&+12 R \xi^{(4)} \omega^{(3)}+12 R \xi^{(6)} \omega^{(1)}\Bigg\}
\end{align}

\end{widetext}

\section{Uniqueness of mode $\ell=1$ at $\mathcal{O}
(\epsilon)$}
\label{apx:unique}

In this appendix, we demonstrate that for a uniformly rotating star, the only non-vanishing mode at first order in the slow-rotation expansion that is both regular at the origin and asymptotically flat at infinity corresponds to $\ell = 1$. At order $\mathcal{O}(\epsilon)$, the only non-zero component of the Einstein equations is the $(t, \phi)$ component. By employing the TOV equations given in Eqs.\ \eqref{eq:enclosed}--\eqref{eq:TOV}, we find
\begin{align}
\begin{split}
\dfrac{\partial^{2}\varpi^{(1)}}{\partial R^{2}} + 4 \dfrac{1 - \pi R^{2}(\varepsilon \, + \, p)e^{\lambda} }{R} \dfrac{ \partial \varpi^{(1)}}{\partial R} \\[1ex]
+ \dfrac{e^{\lambda}}{R^{2}} \left( \dfrac{\partial^{2}\varpi^{(1)}}{\partial \Theta^{2}} + 3 \cot\Theta \dfrac{\partial \varpi^{(1)}}{\partial \Theta}   \right) \\[1ex]
-16 \pi (\varepsilon + p) e^{\lambda} \varpi^{(1)} = 0
\end{split}
\label{eq:tp}
\end{align}
where we have defined
\begin{equation}
    \varpi^{(1)} \equiv \Omega - \omega^{(1)} \, .
\end{equation}
We begin by simplifying Eq.~\eqref{eq:tp} through a vector harmonic decomposition of the form  
\begin{equation}
\varpi^{(1)}(R,\Theta) = \sum_{\ell} \varpi^{(1)}_{\ell} \dfrac{d P_{\ell}(\cos\Theta)}{d\cos\Theta}
\label{eq:wdecomp}
\end{equation}
where \(P_{\ell}(\cos\Theta)\) are the Legendre polynomials. Substituting \eqref{eq:wdecomp} into Eq.~\eqref{eq:tp} and making use of the differential equation satisfied by the Legendre polynomials, we obtain a decoupled ordinary differential equation for each mode \(\varpi^{(1)}_{\ell}(R)\)
\begin{align}
\nonumber
&\dfrac{d^{2} \varpi^{(1)}_{\ell}}{dR^{2}} +  4 \dfrac{1  -  \pi R^{2} (\varepsilon  +   p)e^{\lambda}}{R} \dfrac{d\varpi^{(1)}_{\ell}}{dR} \\[1ex] 
& - \left[ \dfrac{(\ell + 2)(\ell - 1)}{R^{2}} + 16 \pi  (\varepsilon + p)  \right] e^{\lambda} \varpi^{(1)}_{\ell} = 0 \, .
\label{eq:Eqvarpi}
\end{align}
Outside the star $\varepsilon = p = 0$ and therefore the asymptotic behaviour at spatial infinity of the solution is given by
\begin{equation}
\varpi_{\ell}(R \to \infty) = A_{\ell} R^{-\ell-2} + B_{\ell} R^{\ell-1}
\label{eq:inf_w}
\end{equation}
On the other hand, when $R \rightarrow 0$ the asymptotic solution about the origin gives
\begin{equation}
\varpi(R \to 0)  = C_{\ell}R^{S_{+}} + D_{\ell} R^{S_{-}}
\end{equation}
where 
\begin{equation}
S_{\pm} = - \dfrac{3}{2} \pm \sqrt{\dfrac{9}{4} + \dfrac{(\ell+2)(\ell+1)}{j(0)e^{\nu(0)/2}}}
\end{equation}
with \( j \equiv e^{-(\nu + \lambda)/2} \). To ensure regularity at the center \( R = 0 \), we must set \( D_{\ell} = 0 \), which leads to the solution  $\varpi(R) = C_{\ell} R^{S_{+}}$.

Since we are solving a second-order differential equation, we require two independent boundary conditions. One is already fixed by enforcing regularity at the origin. The second condition must be imposed at spatial infinity.
However, the requirement of asymptotic flatness for $\omega^{(1)}$ constrains the behavior of the solution $\varpi^{(1)}$ at large \( R \). Imposing regularity at the origin eliminates one of the two integration constants, leaving a single free constant \( C_{\ell} \). As a result, the constants \( A_{\ell} \) and \( B_{\ell} \) appearing in the asymptotic expansion at infinity cannot be chosen independently. This interdependence implies that neither \( A_{\ell} \) nor \( B_{\ell} \) can vanish unless both do simultaneously~\cite{Hartle:1967he}.

Examining the asymptotic behavior of Eq.~\eqref{eq:Eqvarpi} at spatial infinity, we find that for \(\ell > 1\), the second term in the general solution given by Eq.~\eqref{eq:inf_w}, \( B_{\ell} R^{\ell - 1} \), does not vanish as \( R \to \infty \). To ensure asymptotic flatness, we must therefore set \( B_{\ell > 1} = 0 \). This, in turn, implies that \( A_{\ell > 1} \) must also vanish, due to the interdependence of the asymptotic coefficients. Since the spacetime is asymptotically flat as \( R \to \infty \), \( \omega^{(1)} \) must decrease faster than \( R^{-3} \), ensuring that \( \varpi^{(1)} = \Omega - \omega^{(1)} \) approaches \( \Omega \). Consequently, the only non-vanishing mode that satisfies both regularity at the origin and asymptotic flatness is \(\ell = 1\). Therefore, Eq.~\eqref{eq:Eqvarpi} simplifies to
\begin{align}
\nonumber
&\dfrac{d^{2} \varpi^{(1)}_{1}}{dR^{2}} +  4 \dfrac{1  -  \pi R^{2} (\varepsilon  +   p)e^{\lambda}}{R} \dfrac{d\varpi^{(1)}_{1}}{dR} \\[1ex] 
& -  16 \pi  (\varepsilon + p) e^{\lambda} \varpi^{(1)}_{1} = 0 \, .
\label{eq:Eqvarpi1}
\end{align}

\begin{widetext}

\section{Multipole moments and their corrections}
\label{apx:multipoles}
\begin{align}
\nonumber
%
%
M_{0} &= M_{\star } 
%
+ \textcolor{red}{\epsilon^{2}} C^{ \,(2)}_{0,\textsf{ext}} 
%
%
+ \textcolor{red}{\epsilon^{4}} \left[12 M_{\star } \left(C^{ \,(2)}_{2,\textsf{ext}}\right)^{2} + C^{ \,(4)}_{0,\textsf{ext}} \right] 
%
%
+\textcolor{red}{\epsilon^{6}} \Bigg\{\frac{3 C^{ \,(2)}_{2,\textsf{ext}}}{M_{\star }^{\,4}} \left(8 S^{\,2} C^{ \,(2)}_{0,\textsf{ext}}-C^{\,(4)}_{2,\textsf{ext}}\right) \\ \nonumber
&-\frac{348}{5} C^{ \,(2)}_{0,\textsf{ext}} \left( C^{ \,(2)}_{2,\textsf{ext}}\right)^{2} + \frac{S C^{\,(2)}_{2,\textsf{ext}} }{35 M_{\star }^{\,3}}\left(52799 S C^{\,(2)}_{2,\textsf{ext}}-420 C^{\,(3)}_{1,\textsf{ext}}\right) + \frac{1}{M_{\star }^{\,7}} \bigg[ \frac{1019}{2} S^{\,4} C^{\,(2)}_{2,\textsf{ext}} \\[1ex] \nonumber
&-\frac{315}{256} \left( C^{\,(3)}_{3,\textsf{ext}}\right)^{2} \bigg] +\frac{3057}{4 M_{\star}^{\,5}} S C^{\,(2)}_{2,\textsf{ext}} C^{\,(3)}_{3,\textsf{ext}} -\frac{3384}{35} M_{\star} \left(C^{\,(2)}_{2,\textsf{ext}}\right)^{3} \\[1ex]
&-\frac{105 }{64 M_{\star}^{\,9}} S^{\,3} C^{\,(3)}_{3,\textsf{ext}} -\frac{35 S^{\,6}}{64 M_{\star}^{\,11}} + C^{\,(6)}_{6,\textsf{ext}}  \Bigg\} + \mathcal{O}(\textcolor{red}{\epsilon^{8}})  \\[1.5ex] \nonumber 
M_{2} &=  - \textcolor{red}{\epsilon^2} \left( \frac{S^2}{M_{\star }} + \frac{8}{5} M_{\star }^3 C^{\, (2)}_{2, \textsf{ext}} \right) - \dfrac{\textcolor{red}{\epsilon^{4}}}{175 M_{\star }^5} \bigg\{ 280 M_{\star }^7 C^{\, (2)}_{0, \textsf{ext}} C^{\, (2)}_{2, \textsf{ext}}+35 M_{\star }^3 \left(3 S^2 C^{\, (2)}_{0, \textsf{ext}}-C^{\, (4)}_{2, \textsf{ext}}\right) \\ \nonumber 
&+ 6 S M_{\star }^4 \left(1096 S C^{\, (2)}_{2, \textsf{ext}}+35 C^{\, (3)}_{1, \textsf{ext}}\right)-2340 M_{\star }^8 \left(C^{\, (2)}_{2, \textsf{ext}}\right)^{2}+4095 S M_{\star }^2 C^{\, (3)}_{3, \textsf{ext}}+2730 S^4
\bigg\} \\ \nonumber
&- \frac{\textcolor{red}{\epsilon^{6}}}{504000 M_{\star }^{10}} \bigg\{  2778850580 S^6 M_{\star }-11944800 S M_{\star }^6 C^{\, (2)}_{0, \textsf{ext}} C^{\, (3)}_{3, \textsf{ext}}-73144320 M_{\star }^{13} \left(C^{\, (2)}_{2, \textsf{ext}}\right)^{3} \\ \nonumber
&+ 2621188500 S^3 M_{\star }^3 C^{\, (3)}_{3, \textsf{ext}} + 5 M_{\star }^5 \left[4380422656 S^4 C^{\, (2)}_{2, \textsf{ext}}+45 \left(62976 S^3 C^{\, (3)}_{1, \textsf{ext}}-89719 \left(C^{\, (3)}_{3, \textsf{ext}}\right)^{2}\right)\right] 
\\ \nonumber
& + 57600 M_{\star }^{12} C^{\, (2)}_{2, \textsf{ext}} \left(153 C^{\, (2)}_{0, \textsf{ext}} C^{\, (2)}_{2, \textsf{ext}}+14 C^{\, (4)}_{0, \textsf{ext}}\right)+6871100 S^2 C^{\, (4)}_{4, \textsf{ext}} 
+1000 M_{\star }^4 \bigg(10872 S^4 C^{\, (2)}_{0, \textsf{ext}} \\ \nonumber
&-67 C^{\, (2)}_{2, \textsf{ext}} C^{\, (4)}_{4, \textsf{ext}} 
-4320 S^2 C^{\, (4)}_{2, \textsf{ext}}\bigg) 
+ 64 M_{\star }^9 \bigg[ 941220 S C^{\, (2)}_{2, \textsf{ext}} C^{\, (3)}_{1, \textsf{ext}}+227031596 S^2 \left(C^{\, (2)}_{2, \textsf{ext}}\right)^{2} 
\\ \nonumber
&-1575 \left(S C^{\, (5)}_{1, \textsf{ext}}-3 \left(C^{\, (3)}_{1, \textsf{ext}}\right)^{2}\right) \bigg] 
+960 M_{\star}^7 \bigg[525 S^2 \left(C^{\, (2)}_{0, \textsf{ext}}\right)^{2} -105 C^{\, (2)}_{0, \textsf{ext}} C^{\, (4)}_{2, \textsf{ext}}-365764 S C^{\, (2)}_{2, \textsf{ext}} C^{\, (3)}_{3, \textsf{ext}} 
\\ \nonumber
&-1755 \left(S C^{\, (5)}_{3, \textsf{ext}}-7 C^{\, (3)}_{1, \textsf{ext}} C^{\, (3)}_{3, \textsf{ext}}\right)\bigg] 
- 2880 M_{\star }^8 \bigg[14 S C^{\, (2)}_{0, \textsf{ext}} \left(1056 S C^{\, (2)}_{2, \textsf{ext}} -5 C^{\, (3)}_{1, \textsf{ext}}\right) \\ \nonumber
&-5 \left(117 C^{\, (2)}_{2, \textsf{ext}} C^{\, (4)}_{2, \textsf{ext}}+21 S^2 C^{\, (4)}_{0, \textsf{ext}}-7 C^{\, (6)}_{2, \textsf{ext}}\right)\bigg]
\bigg\} + \mathcal{O}(\textcolor{red}{\epsilon^{8}}) \\[1.5ex] \nonumber
M_{4} &= \dfrac{\textcolor{red}{\epsilon^{4}}}{8820 M_{\star }^4} \bigg[ 
113136 S^2 M_{\star}^5 C^{\, (2)}_{2, \textsf{ext}}+113184 M_{\star }^9 \left(C^{\, (2)}_{2, \textsf{ext}}\right)^{2}-7560 S M_{\star }^3 C^{\, (3)}_{3, \textsf{ext}}+35 C^{\, (4)}_{4, \textsf{ext}}+8988 S^4 M_{\star }
\bigg]  \\ \nonumber
&-\frac{\textcolor{red}{\epsilon^{6}}}{1940400 M_{\star }^8} \bigg\{  
970200 S M_{\star }^6 C^{\, (2)}_{0, \textsf{ext}} C^{\, (3)}_{3, \textsf{ext}} -21605760 M_{\star }^{12} C^{\, (2)}_{0, \textsf{ext}} \left(C^{\, (2)}_{2, \textsf{ext}}\right)^{2} + 384 S M_{\star }^9 C^{\, (2)}_{2, \textsf{ext}} \bigg(25169271 S C^{\, (2)}_{2, \textsf{ext}} \\ \nonumber
&-58520 C^{\, (3)}_{1, \textsf{ext}}\bigg)+518951424 M_{\star }^{13} \left(C^{\, (2)}_{2, \textsf{ext}}\right)^{3} + 1713792900 S^6 M_{\star} 
+15840 M_{\star }^8 C^{\, (2)}_{2, \textsf{ext}} \left(12850 S^2 C^{\, (2)}_{0, \textsf{ext}}+393 C^{\, (4)}_{2, \textsf{ext}}\right) \\ \nonumber
&+420 M_{\star }^3 \left(4666953 S^3 C^{\, (3)}_{3, \textsf{ext}}-55 C^{\, (2)}_{0, \textsf{ext}} C^{\, (4)}_{4, \textsf{ext}}\right)+2160 M_{\star }^7 \left(43113 S C^{\, (2)}_{2, \textsf{ext}} C^{\, (3)}_{3, \textsf{ext}}+770 C^{\, (3)}_{1, \textsf{ext}} C^{\, (3)}_{3, \textsf{ext}}-110 S C^{\, (5)}_{3, \textsf{ext}}\right) \\ \nonumber
&+3397575 S^2 C^{\, (4)}_{4, \textsf{ext}} 
+3 M_{\star }^5 \left[4250102128 S^4 C^{\, (2)}_{2, \textsf{ext}}+1110560 S^3 C^{\, (3)}_{1, \textsf{ext}}-4725 \left(203 \left(C^{\, (3)}_{3, \textsf{ext}}\right)^{2} + 1600 S C^{\, (5)}_{5, \textsf{ext}}\right)\right]  \\ 
&-20 M_{\star }^4 \left(1227600 S^4 C^{\, (2)}_{0, \textsf{ext}}-9931 C^{\, (2)}_{2, \textsf{ext}} C^{\, (4)}_{4, \textsf{ext}}-140514 S^2 C^{\, (4)}_{2, \textsf{ext}}+385 C^{\, (6)}_{4, \textsf{ext}}\right)
\bigg\} + \mathcal{O}(\textcolor{red}{\epsilon^{8}}) \\[1.5ex] \nonumber
M_{6} &= -\frac{\textcolor{red}{\epsilon^{6}}}{416215800 M_{\star }^6} \bigg\{4887865416256 S^2 M_{\star }^9 \left(C^{\, (2)}_{2, \textsf{ext}}\right)^{2} -36247360800 S M_{\star }^7 C^{\, (2)}_{2, \textsf{ext}} C^{\, (3)}_{3, \textsf{ext}} + 75040894464 M_{\star }^{13} \left(C^{\, (2)}_{2, \textsf{ext}}\right)^{3} \\ \nonumber 
& + 2417285740800 S^3 M_{\star }^3 C^{\, (3)}_{3, \textsf{ext}}+1920321740680 S^6 M_{\star } + 42513520 M_{\star }^4 C^{\, (2)}_{2, \textsf{ext}} C^{\, (4)}_{4, \textsf{ext}}  
+ 8 M_{\star }^5 \bigg(1281183889712 S^4 C^{\, (2)}_{2, \textsf{ext}} \\ \nonumber
& +11025 \left[3626 \left(C^{\, (3)}_{3, \textsf{ext}}\right)^{2}+1635 S C^{\, (5)}_{5, \textsf{ext}}\right] \bigg)+175 \left(11786392 S^2 C^{\, (4)}_{4, \textsf{ext}}-693 C^{\, (6)}_{6, \textsf{ext}}\right)\bigg\} + \mathcal{O}(\textcolor{red}{\epsilon^{8}})
\end{align}

\begin{align}
\nonumber
S_{1} &= \textcolor{red}{\epsilon} S \, +  \, \textcolor{red}{\epsilon^{3}} C^{\, (3)}_{1,\textsf{ext}} \, + \, \dfrac{1}{1050 M_{\star }^8} \textcolor{red}{\epsilon^{5}} \bigg\{ 10080 S M_{\star }^{\, 7} C^{ \, (2)}_{0, \textsf{ext}} C^{ \, (2)}_{2, \textsf{ext}}-315 M_{\star }^{\, 3} \left(8 S^{\, 3} C^{ \, (2)}_{0, \textsf{ext}}-S C^{ \, (4)}_{2, \textsf{ext}}\right) \\ \nonumber
&+M_{\star }^{\, 8} \left[17688 S \left(C^{ \, (2)}_{2, \textsf{ext}}\right)^{2}-2520 C^{ \, (2)}_{2, \textsf{ext}} C^{ \, (3)}_{1, \textsf{ext}}-175 C^{ \, (5)}_{1, \textsf{ext}}\right]+6 S^{\, 2} M_{\star }^{\, 4} \bigg(20971 S C^{ \, (2)}_{2, \textsf{ext}} +210 C^{ \, (3)}_{1, \textsf{ext}}\bigg)  \\ \nonumber
&-315 M_{\star }^{\, 6} C^{ \, (2)}_{2, \textsf{ext}} C^{ \, (3)}_{3, \textsf{ext}}+52920 S^{\, 2} M_{\star }^2 C^{ \, (3)}_{3, \textsf{ext}}+35280 S^{\, 5} \bigg\}
-\frac{\textcolor{red}{\epsilon^{7}}}{6048000 M_{\star }^{13}} \bigg\{ 58383328500 S^4 M_{\star }^3 C^{\, (3)}_{3 \, \textsf{ext}}  \\ \nonumber
&+45 M_{\star }^5 \left(6524004256 S^5 C^{\, (2)}_{2 \, \textsf{ext}}-9816960 S^4 C^{\, (3)}_{1 \, \textsf{ext}}+9067695 S \left(C^{\, (3)}_{3 \, \textsf{ext}}\right)^{2} \right) +49104904140 S^7 M_{\star }  \\ \nonumber
&- 1930521600 M_{\star }^{10} C^{\, (2)}_{0 \, \textsf{ext}} C^{\, (2)}_{2 \, \textsf{ext}} C^{\, (3)}_{3 \, \textsf{ext}} -2073600 M_{\star }^{12} C^{\, (2)}_{2 \, \textsf{ext}} \left[C^{\, (2)}_{0 \, \textsf{ext}} \left(1531 S C^{\, (2)}_{2 \, \textsf{ext}}+28 C^{\, (3)}_{1 \, \textsf{ext}}\right)+28 S C^{\, (4)}_{0 \, \textsf{ext}}\right] \\ \nonumber
&+226800 M_{\star }^6 C^{\, (3)}_{3 \, \textsf{ext}} \left(5120 S^2 C^{\, (2)}_{0 \, \textsf{ext}}-C^{\, (4)}_{2 \, \textsf{ext}}\right) +69458300 S^3 C^{\, (4)}_{4 \, \textsf{ext}}-288750 M_{\star }^2 C^{\, (3)}_{3 \, \textsf{ext}} C^{\, (4)}_{4 \, \textsf{ext}} \\ \nonumber 
&+ 1200 M_{\star }^4 \left(25632 S^5 C^{\, (2)}_{0 \, \textsf{ext}}-1157 S C^{\, (2)}_{2 \, \textsf{ext}} C^{\, (4)}_{4 \, \textsf{ext}}+119646 S^3 C^{\, (4)}_{2 \, \textsf{ext}}\right)
+ 384 S M_{\star }^9 \bigg[ 354536279 S^2 \left(C^{\, (2)}_{2 \, \textsf{ext}}\right)^{2} \\ \nonumber
&-6458670 S C^{\, (2)}_{2 \, \textsf{ext}} C^{\, (3)}_{1 \, \textsf{ext}} + 3150 \left(S C^{\, (5)}_{1 \, \textsf{ext}}-9 \left(C^{\, (3)}_{1 \, \textsf{ext}}\right)^{2}\right)\bigg]
-129600 M_{\star }^{11} C^{\, (2)}_{2 \, \textsf{ext}} \bigg( 2 C^{\, (5)}_{3 \, \textsf{ext}}-1120 S \left(C^{\, (2)}_{0 \, \textsf{ext}}\right)^{2} \\ \nonumber
&+ 7833 C^{\, (2)}_{2 \, \textsf{ext}} C^{\, (3)}_{3 \, \textsf{ext}} \bigg)  -5040 S M_{\star }^7 \bigg[14400 S^2 \left(C^{\, (2)}_{0 \, \textsf{ext}}\right)^{2} -1440 C^{\, (2)}_{0 \, \textsf{ext}} C^{\, (4)}_{2 \, \textsf{ext}}+1291759 S C^{\, (2)}_{2 \, \textsf{ext}} C^{\, (3)}_{3 \, \textsf{ext}}+ \\ \nonumber 
 &+ 180 \left(673 C^{\, (3)}_{1 \, \textsf{ext}} C^{\, (3)}_{3 \, \textsf{ext}}-48 S C^{\, (5)}_{3 \, \textsf{ext}}\right)\bigg] + 17280 M_{\star }^8 \bigg[ 4 S^2 C^{\, (2)}_{0 \, \textsf{ext}} \left(31823 S C^{\, (2)}_{2 \, \textsf{ext}}+1050 C^{\, (3)}_{1 \, \textsf{ext}}\right)+1707 S C^{\, (2)}_{2 \, \textsf{ext}} C^{\, (4)}_{2 \, \textsf{ext}} \\ \nonumber 
 &-105 \left(C^{\, (3)}_{1 \, \textsf{ext}} C^{\, (4)}_{2 \, \textsf{ext}}+S C^{\, (6)}_{2, \textsf{ext}}\right)+840 S^3 C^{\, (4)}_{0 \, \textsf{ext}}\bigg] + 5760 M_{\star }^{13} \bigg[ 23280 \left(C^{\, (2)}_{2 \, \textsf{ext}}\right)^{2} C^{\, (3)}_{1 \, \textsf{ext}}+420 C^{\, (2)}_{2 \, \textsf{ext}} C^{\, (5)}_{1 \, \textsf{ext}} \\ 
 &+731024 S \left(C^{\, (2)}_{2 \, \textsf{ext}}\right)^{3}-175 C^{\, (7)}_{1, \textsf{ext}} \bigg]
\bigg\} + \mathcal{O}(\textcolor{red}{\epsilon^{9}}) \\ \nonumber
S_{3} &= \frac{3}{2} \textcolor{red}{\epsilon^{3}} C^{\, (3)}_{3, \textsf{ext}} +\frac{1}{8400 M_{\star }^7}\textcolor{red}{\epsilon^{5}} \bigg\{ 24000 S M_{\star }^8 C^{\, (2)}_{0, \textsf{ext}} C^{\, (2)}_{2, \textsf{ext}}+8 S^2 M_{\star }^5 \left(77341 S C^{\, (2)}_{2, \textsf{ext}}-1140 C^{\, (3)}_{1, \textsf{ext}}\right) \\ \nonumber
&+192 M_{\star }^9 C^{\, (2)}_{2, \textsf{ext}} \left(12747 S C^{\, (2)}_{2, \textsf{ext}}+95 C^{\, (3)}_{1, \textsf{ext}}\right)-416090 S^5 M_{\star} + 25200 M_{\star }^6 C^{\, (2)}_{0, \textsf{ext}} C^{\, (3)}_{3, \textsf{ext}}+120 M_{\star }^4 \left(292 S^3 C^{\, (2)}_{0, \textsf{ext}}-19 S C^{\, (4)}_{2, \textsf{ext}}\right) \\ \nonumber 
&-180 M_{\star }^7 \left(259 C^{\, (2)}_{2, \textsf{ext}} C^{\, (3)}_{3, \textsf{ext}}+10 C^{\, (5)}_{3, \textsf{ext}}\right)+852915 S^2 M_{\star }^3 C^{\, (3)}_{3, \textsf{ext}}+1025 S C^{\, (4)}_{4, \textsf{ext}} \bigg\} + \frac{\textcolor{red}{\epsilon^{7}}}{155232000 M_{\star }^{11}}
\bigg\{  \\ \nonumber
&-1774080 M_{\star }^{12} C^{\, (2)}_{2, \textsf{ext}} \left[C^{\, (2)}_{0, \textsf{ext}} \left(63691 S C^{\, (2)}_{2, \textsf{ext}}-630 C^{\, (3)}_{1, \textsf{ext}}\right)-250 S C^{\, (4)}_{0, \textsf{ext}}\right]+970200 M_{\star }^6 C^{\, (3)}_{3, \textsf{ext}} \left(52204 S^2 C^{\, (2)}_{0, \textsf{ext}}+111 C^{\, (4)}_{2, \textsf{ext}}\right) \\ \nonumber
&+168795314940 S^7 M_{\star } + 350 M_{\star }^3 \left(51877641147 S^4 C^{\, (3)}_{3, \textsf{ext}}-15400 S C^{\, (2)}_{0, \textsf{ext}} C^{\, (4)}_{4, \textsf{ext}}\right)+7711375 M_{\star }^2 C^{\, (3)}_{3, \textsf{ext}} C^{\, (4)}_{4, \textsf{ext}} \\ \nonumber
&+75394053700 S^3 C^{\, (4)}_{4, \textsf{ext}} + 3326400 M_{\star }^{10} \left[ C^{\, (2)}_{0, \textsf{ext}} \left(147 C^{\, (2)}_{2, \textsf{ext}} C^{\, (3)}_{3, \textsf{ext}}-20 C^{\, (5)}_{3, \textsf{ext}}\right)+140 C^{\, (3)}_{3, \textsf{ext}} C^{\, (4)}_{0, \textsf{ext}}\right] \\ \nonumber
&+ 84480 M_{\star }^{13} C^{\, (2)}_{2, \textsf{ext}} \left[ 511035 C^{\, (2)}_{2, \textsf{ext}} C^{\, (3)}_{1, \textsf{ext}}+491348 S \left(C^{\, (2)}_{2, \textsf{ext}}\right)^{2}-665 C^{\, (5)}_{1, \textsf{ext}}\right] -1680 S M_{\star }^7 \bigg[83160 C^{\, (2)}_{0, \textsf{ext}} C^{\, (4)}_{2, \textsf{ext}} \\ \nonumber
&+736560 S^2 \left(C^{\, (2)}_{0, \textsf{ext}}\right)^{2}-3771517733 S C^{\, (2)}_{2, \textsf{ext}} C^{\, (3)}_{3, \textsf{ext}}+18507720 C^{\, (3)}_{1, \textsf{ext}} C^{\, (3)}_{3, \textsf{ext}}-1340295 S C^{\, (5)}_{3, \textsf{ext}}\bigg] \\ \nonumber
&-192 M_{\star }^9 \bigg[712986679847 S^3 \left(C^{\, (2)}_{2, \textsf{ext}}\right)^{2} - 3850 \left[38 S^2 C^{\, (5)}_{1, \textsf{ext}}-2835 \left(C^{\, (2)}_{0, \textsf{ext}}\right)^{2} C^{\, (3)}_{3, \textsf{ext}}-342 S \left(C^{\, (3)}_{1, \textsf{ext}}\right)^{2} \right] \\ \nonumber 
&+280 C^{\, (2)}_{2, \textsf{ext}} \left(601018 S^{2} C^{\, (3)}_{1, \textsf{ext}}-225 C^{\, (5)}_{5, \textsf{ext}}\right)\bigg] -180 M_{\star }^5 \bigg[776913810992 S^5 C^{\, (2)}_{2, \textsf{ext}}+35 S \bigg(5231336 S^3 C^{\, (3)}_{1, \textsf{ext}} \\ \nonumber
&-1131753 \left(C^{\, (3)}_{3, \textsf{ext}}\right){}^2+25736600 S C^{\, (5)}_{5, \textsf{ext}}\bigg)\bigg] - 221760 M_{\star }^8 \bigg[60 S^2 C^{\, (2)}_{0, \textsf{ext}} \left(39461 S C^{\, (2)}_{2, \textsf{ext}}-104 C^{\, (3)}_{1, \textsf{ext}}\right) \\ \nonumber 
&+49829 S C^{\, (2)}_{2, \textsf{ext}} C^{\, (4)}_{2, \textsf{ext}}+190 C^{\, (3)}_{1, \textsf{ext}} C^{\, (4)}_{2, \textsf{ext}}-2920 S^3 C^{\, (4)}_{0, \textsf{ext}}+190 S C^{\, (6)}_{2, \textsf{ext}}\bigg] + 4200 M_{\star}^4 \bigg(12063568 S^5 C^{\, (2)}_{0, \textsf{ext}} \\ \nonumber
&-8653 S C^{\, (2)}_{2, \textsf{ext}} C^{\, (4)}_{4, \textsf{ext}}+4510 C^{\, (3)}_{1, \textsf{ext}} C^{\, (4)}_{4, \textsf{ext}}+326744 S^3 C^{\, (4)}_{2, \textsf{ext}}+4510 S C^{\, (6)}_{4, \textsf{ext}}\bigg) 
-151200 M_{\star }^{11} \bigg[ 32736 S \left(C^{\, (2)}_{0, \textsf{ext}}\right)^{2} C^{\, (2)}_{2, \textsf{ext}} \\ 
&-252819 \left(C^{\, (2)}_{2, \textsf{ext}}\right)^{2} C^{\, (3)}_{3, \textsf{ext}} - 814 C^{\, (2)}_{2, \textsf{ext}} C^{\, (5)}_{3, \textsf{ext}}+220 C^{\, (7)}_{3, \textsf{ext}}\bigg]
\bigg\} + \mathcal{O}(\textcolor{red}{\epsilon^{9}})
\end{align}

\begin{align}
\nonumber
S_{5} &= \textcolor{red}{\epsilon^{5}} \bigg[\frac{21482 }{3465 } \dfrac{S^5}{M_{\star }^4}-\frac{54}{55} M_{\star }^2 C^{\, (2)}_{2, \textsf{ext}} C^{\, (3)}_{3, \textsf{ext}}-\frac{393272  }{3675}S M_{\star }^4\left(C^{\, (2)}_{2, \textsf{ext}}\right)^{2}-\frac{2269466 }{24255}S^3 C^{\, (2)}_{2, \textsf{ext}} + \frac{10639}{462 }  \dfrac{S^2 C^{\, (3)}_{3, \textsf{ext}}}{M_{\star }^2} \\ \nonumber
&- \frac{31 }{504 } \dfrac{S C^{\, (4)}_{4, \textsf{ext}}}{M_{\star }^5}-\frac{5}{22}C^{\, (5)}_{5, \textsf{ext}} \bigg] - \frac{\textcolor{red}{\epsilon ^7}}{3027024000 M_{\star }^9} \bigg\{39936 M_{\star }^{13} \left(C^{\, (2)}_{2, \textsf{ext}}\right)^{2} \left(1196682299 S C^{\, (2)}_{2, \textsf{ext}}+8218965 C^{\, (3)}_{1, \textsf{ext}}\right)  \\ \nonumber
&+199438387200 S M_{\star }^{12} C^{\, (2)}_{0, \textsf{ext}} \left(C^{\, (2)}_{2, \textsf{ext}}\right)^{2} +1996235180905600 S^7 M_{\star }
-60375369600 M_{\star }^{10} C^{\, (2)}_{0, \textsf{ext}} C^{\, (2)}_{2, \textsf{ext}} C^{\, (3)}_{3, \textsf{ext}} \\ \nonumber
&+655200 M_{\star }^6 C^{\, (3)}_{3, \textsf{ext}} \left(63641 S^2 C^{\, (2)}_{0, \textsf{ext}}-567 C^{\, (4)}_{2, \textsf{ext}}\right)-77665000 M_{\star }^2 C^{\, (3)}_{3, \textsf{ext}} C^{\, (4)}_{4, \textsf{ext}}+1794877400500 S^3 C^{\, (4)}_{4, \textsf{ext}} \\ \nonumber
&+14000 M_{\star }^3 \left(37063 S C^{\, (2)}_{0, \textsf{ext}} C^{\, (4)}_{4, \textsf{ext}}+185252206428 S^4 C^{\, (3)}_{3, \textsf{ext}}\right)-80640 M_{\star }^{11} C^{\, (2)}_{2, \textsf{ext}} \left(4371448 C^{\, (2)}_{2, \textsf{ext}} C^{\, (3)}_{3, \textsf{ext}}+5265 C^{\, (5)}_{3, \textsf{ext}}\right) \\ \nonumber
&-480 S M_{\star }^7 \left[44557373147 S C^{\, (2)}_{2, \textsf{ext}} C^{\, (3)}_{3, \textsf{ext}}+390 \left(752668 C^{\, (3)}_{1, \textsf{ext}} C^{\, (3)}_{3, \textsf{ext}}-53195 S C^{\, (5)}_{3, \textsf{ext}}\right)\right] \\ \nonumber
&-249600 M_{\star }^8 \left[C^{\, (2)}_{0, \textsf{ext}} \left(19678108 S^3 C^{\, (2)}_{2, \textsf{ext}}-11025 C^{\, (5)}_{5, \textsf{ext}}\right)+326604 S C^{\, (2)}_{2, \textsf{ext}} C^{\, (4)}_{2, \textsf{ext}}\right] + 80 M_{\star }^5 \bigg(124836882248696 S^5 C^{\, (2)}_{2, \textsf{ext}} \\ \nonumber
&-45 S \left[51387128 S^3 C^{\, (3)}_{1, \textsf{ext}}+115004372 \left(C^{\, (3)}_{3, \textsf{ext}}\right){}^2+68181925 S C^{\, (5)}_{5, \textsf{ext}}\right]\bigg) 
+ 400 M_{\star }^4 \bigg(1150264284 S^5 C^{\, (2)}_{0, \textsf{ext}} \\ \nonumber
&+79961122 S C^{\, (2)}_{2, \textsf{ext}} C^{\, (4)}_{4, \textsf{ext}}+465465 C^{\, (3)}_{1, \textsf{ext}} C^{\, (4)}_{4, \textsf{ext}}-56975178 S^3 C^{\, (4)}_{2, \textsf{ext}}+465465 S C^{\, (6)}_{4, \textsf{ext}}\bigg) - 127945125 S C^{\, (6)}_{6, \textsf{ext}} \\ \nonumber
&+320 M_{\star }^9 \left[ 30 C^{\, (2)}_{2, \textsf{ext}} \left(44031026 S^2 C^{\, (3)}_{1, \textsf{ext}}-390915 C^{\, (5)}_{5, \textsf{ext}}\right)+13655261908268 S^3 \left(C^{\, (2)}_{2, \textsf{ext}}\right)^{2}+2149875 C^{\, (7)}_{5, \textsf{ext}}\right]
\bigg\} + \mathcal{O}(\textcolor{red}{\epsilon^{9}}) \\[1.5ex] \nonumber 
\nonumber
S_{7} &= \frac{\textcolor{red}{\epsilon^{7}}}{27054027000 M_{\star }^7} \bigg\{  136356345854976 S M_{\star }^{13} \left(C^{\, (2)}_{2, \textsf{ext}}\right)^{3}+10706408510686350 S^4 M_{\star }^3 C^{\, (3)}_{3, \textsf{ext}} \\ \nonumber
&+ 8058896560245020 S^7 M_{\star } -135381970080 M_{\star }^{11} \left(C^{\, (2)}_{2, \textsf{ext}}\right)^{2} C^{\, (3)}_{3, \textsf{ext}}+104566977200 S M_{\star }^4 C^{\, (2)}_{2, \textsf{ext}} C^{\, (4)}_{4, \textsf{ext}} \\ \nonumber
& - 510696375 M_{\star }^2 C^{\, (3)}_{3, \textsf{ext}} C^{\, (4)}_{4, \textsf{ext}} + 640 M_{\star }^9 C^{\, (2)}_{2, \textsf{ext}} \left(25824262758599 S^3 C^{\, (2)}_{2, \textsf{ext}}+20336400 C^{\, (5)}_{5, \textsf{ext}}\right) \\ \nonumber 
&+ 980 M_{\star }^5 \left[39640730386702 S^5 C^{\, (2)}_{2, \textsf{ext}}-225 S \left(16491428 \left(C^{\, (3)}_{3, \textsf{ext}}\right){}^2+35195019 S C^{\, (5)}_{5, \textsf{ext}}\right)\right] \\ 
&+ 3675 \left(1737327188 S^3 C^{\, (4)}_{4, \textsf{ext}}-126687 S C^{\, (6)}_{6, \textsf{ext}}\right) + 2100 M_{\star }^7 \left(59033692813 S^2 C^{\, (2)}_{2, \textsf{ext}} C^{\, (3)}_{3, \textsf{ext}}-3006003 C^{\, (7)}_{7, \textsf{ext}}\right)
\bigg\} + \mathcal{O}{(\textcolor{red}{\epsilon^{9}})}
\end{align}

\end{widetext}

\nocite{*} 
\bibliography{references/inspire, references/NOTinspire}

\end{document}